\documentclass[fleqn,usenatbib]{mnras}

\usepackage[T1]{fontenc}
\DeclareRobustCommand{\VAN}[3]{#2}
\let\VANthebibliography\thebibliography
\def\thebibliography{\DeclareRobustCommand{\VAN}[3]{##3}\VANthebibliography}
\usepackage{soul,color}
\usepackage{graphicx}	% Including figure file
\usepackage{amsmath}	% Advanced maths commands
\usepackage{amssymb}	% Extra maths symbols
\usepackage{newtxtext,newtxmath}
\usepackage{physics}
\usepackage{booktabs}
\usepackage{multirow}
\usepackage{soul}
\makeatletter
\renewcommand\paragraph{\@startsection{paragraph}{4}{\z@}%
            {-2.5ex\@plus -1ex \@minus -.25ex}%
            {1.25ex \@plus .25ex}%
            {\normalfont\normalsize\itshape}}
\makeatother
\usepackage{tikz}
\usetikzlibrary{patterns}

\newcommand{\Msun}{\, \mathrm{M}_{\odot}}
\newcommand{\HBT}{{\tt HBT+}}
\newcommand{\HBTHERONS}{{\tt HBT-HERONS}}
\newcommand{\SUBFIND}{{\tt Subfind}}
\newcommand{\ROCKSTAR}{{\tt ROCKSTAR}}
\newcommand{\VELOCIRAPTOR}{{\tt VELOCIraptor}}
\newcommand{\SOAP}{{\tt SOAP}}
\newcommand{\lowres}{{\tt L1\_m10}}
\newcommand{\midres}{{\tt L1\_m9}}
\newcommand{\highres}{{\tt L1\_m8}}

\tikzset{
pattern size/.store in=\mcSize, 
pattern size = 5pt,
pattern thickness/.store in=\mcThickness, 
pattern thickness = 0.3pt,
pattern radius/.store in=\mcRadius, 
pattern radius = 1pt}
\makeatletter
\pgfutil@ifundefined{pgf@pattern@name@_pfdoxm934}{
\pgfdeclarepatternformonly[\mcThickness,\mcSize]{_pfdoxm934}
{\pgfqpoint{0pt}{0pt}}
{\pgfpoint{\mcSize+\mcThickness}{\mcSize+\mcThickness}}
{\pgfpoint{\mcSize}{\mcSize}}
{
\pgfsetcolor{\tikz@pattern@color}
\pgfsetlinewidth{\mcThickness}
\pgfpathmoveto{\pgfqpoint{0pt}{0pt}}
\pgfpathlineto{\pgfpoint{\mcSize+\mcThickness}{\mcSize+\mcThickness}}
\pgfusepath{stroke}
}}
\makeatother

\tikzset{
pattern size/.store in=\mcSize, 
pattern size = 5pt,
pattern thickness/.store in=\mcThickness, 
pattern thickness = 0.3pt,
pattern radius/.store in=\mcRadius, 
pattern radius = 1pt}
\makeatletter
\pgfutil@ifundefined{pgf@pattern@name@_61qwfubkx}{
\pgfdeclarepatternformonly[\mcThickness,\mcSize]{_61qwfubkx}
{\pgfqpoint{0pt}{0pt}}
{\pgfpoint{\mcSize+\mcThickness}{\mcSize+\mcThickness}}
{\pgfpoint{\mcSize}{\mcSize}}
{
\pgfsetcolor{\tikz@pattern@color}
\pgfsetlinewidth{\mcThickness}
\pgfpathmoveto{\pgfqpoint{0pt}{0pt}}
\pgfpathlineto{\pgfpoint{\mcSize+\mcThickness}{\mcSize+\mcThickness}}
\pgfusepath{stroke}
}}
\makeatother

\tikzset{
pattern size/.store in=\mcSize, 
pattern size = 5pt,
pattern thickness/.store in=\mcThickness, 
pattern thickness = 0.3pt,
pattern radius/.store in=\mcRadius, 
pattern radius = 1pt}
\makeatletter
\pgfutil@ifundefined{pgf@pattern@name@_vhaox8mo3}{
\pgfdeclarepatternformonly[\mcThickness,\mcSize]{_vhaox8mo3}
{\pgfqpoint{0pt}{0pt}}
{\pgfpoint{\mcSize+\mcThickness}{\mcSize+\mcThickness}}
{\pgfpoint{\mcSize}{\mcSize}}
{
\pgfsetcolor{\tikz@pattern@color}
\pgfsetlinewidth{\mcThickness}
\pgfpathmoveto{\pgfqpoint{0pt}{0pt}}
\pgfpathlineto{\pgfpoint{\mcSize+\mcThickness}{\mcSize+\mcThickness}}
\pgfusepath{stroke}
}}
\makeatother
\tikzset{every picture/.style={line width=0.75pt}} %set default line width to 0.75pt        

\tikzset{
pattern size/.store in=\mcSize, 
pattern size = 5pt,
pattern thickness/.store in=\mcThickness, 
pattern thickness = 0.3pt,
pattern radius/.store in=\mcRadius, 
pattern radius = 1pt}
\makeatletter
\pgfutil@ifundefined{pgf@pattern@name@_3ztt0290a}{
\pgfdeclarepatternformonly[\mcThickness,\mcSize]{_3ztt0290a}
{\pgfqpoint{0pt}{0pt}}
{\pgfpoint{\mcSize+\mcThickness}{\mcSize+\mcThickness}}
{\pgfpoint{\mcSize}{\mcSize}}
{
\pgfsetcolor{\tikz@pattern@color}
\pgfsetlinewidth{\mcThickness}
\pgfpathmoveto{\pgfqpoint{0pt}{0pt}}
\pgfpathlineto{\pgfpoint{\mcSize+\mcThickness}{\mcSize+\mcThickness}}
\pgfusepath{stroke}
}}
\makeatother

\tikzset{
pattern size/.store in=\mcSize, 
pattern size = 5pt,
pattern thickness/.store in=\mcThickness, 
pattern thickness = 0.3pt,
pattern radius/.store in=\mcRadius, 
pattern radius = 1pt}
\makeatletter
\pgfutil@ifundefined{pgf@pattern@name@_4t081rgbc}{
\pgfdeclarepatternformonly[\mcThickness,\mcSize]{_4t081rgbc}
{\pgfqpoint{0pt}{0pt}}
{\pgfpoint{\mcSize+\mcThickness}{\mcSize+\mcThickness}}
{\pgfpoint{\mcSize}{\mcSize}}
{
\pgfsetcolor{\tikz@pattern@color}
\pgfsetlinewidth{\mcThickness}
\pgfpathmoveto{\pgfqpoint{0pt}{0pt}}
\pgfpathlineto{\pgfpoint{\mcSize+\mcThickness}{\mcSize+\mcThickness}}
\pgfusepath{stroke}
}}
\makeatother

\tikzset{
pattern size/.store in=\mcSize, 
pattern size = 5pt,
pattern thickness/.store in=\mcThickness, 
pattern thickness = 0.3pt,
pattern radius/.store in=\mcRadius, 
pattern radius = 1pt}
\makeatletter
\pgfutil@ifundefined{pgf@pattern@name@_ybtd82z7b}{
\pgfdeclarepatternformonly[\mcThickness,\mcSize]{_ybtd82z7b}
{\pgfqpoint{0pt}{0pt}}
{\pgfpoint{\mcSize+\mcThickness}{\mcSize+\mcThickness}}
{\pgfpoint{\mcSize}{\mcSize}}
{
\pgfsetcolor{\tikz@pattern@color}
\pgfsetlinewidth{\mcThickness}
\pgfpathmoveto{\pgfqpoint{0pt}{0pt}}
\pgfpathlineto{\pgfpoint{\mcSize+\mcThickness}{\mcSize+\mcThickness}}
\pgfusepath{stroke}
}}
\makeatother

\tikzset{
pattern size/.store in=\mcSize, 
pattern size = 5pt,
pattern thickness/.store in=\mcThickness, 
pattern thickness = 0.3pt,
pattern radius/.store in=\mcRadius, 
pattern radius = 1pt}
\makeatletter
\pgfutil@ifundefined{pgf@pattern@_4di8nj5id}{
\pgfdeclarepatternformonly[\mcThickness,\mcSize]{_4di8nj5id}
{\pgfpointorigin}
{\pgfpoint{\mcSize+\mcThickness}{\mcSize+\mcThickness}}
{\pgfpoint{\mcSize}{\mcSize}}{
\pgfsetcolor{\tikz@pattern@color}
\pgfsetlinewidth{\mcThickness}
\pgfpathmoveto{\pgfpointorigin}
\pgfpathlineto{\pgfpoint{0pt}{0.5*\mcSize}}
\pgfpathlineto{\pgfpoint{\mcSize}{0.5*\mcSize}}
\pgfpathmoveto{\pgfpoint{0.5*\mcSize}{0.5*\mcSize}}
\pgfpathlineto{\pgfpoint{0.5*\mcSize}{\mcSize}}
\pgfpathmoveto{\pgfpoint{0pt}{\mcSize}}
\pgfpathlineto{\pgfpoint{\mcSize}{\mcSize}}
\pgfusepath{stroke}}}
\makeatother

\tikzset{
pattern size/.store in=\mcSize, 
pattern size = 5pt,
pattern thickness/.store in=\mcThickness, 
pattern thickness = 0.3pt,
pattern radius/.store in=\mcRadius, 
pattern radius = 1pt}
\makeatletter
\pgfutil@ifundefined{pgf@pattern@_xn8da364r}{
\pgfdeclarepatternformonly[\mcThickness,\mcSize]{_xn8da364r}
{\pgfpointorigin}
{\pgfpoint{\mcSize+\mcThickness}{\mcSize+\mcThickness}}
{\pgfpoint{\mcSize}{\mcSize}}{
\pgfsetcolor{\tikz@pattern@color}
\pgfsetlinewidth{\mcThickness}
\pgfpathmoveto{\pgfpointorigin}
\pgfpathlineto{\pgfpoint{0pt}{0.5*\mcSize}}
\pgfpathlineto{\pgfpoint{\mcSize}{0.5*\mcSize}}
\pgfpathmoveto{\pgfpoint{0.5*\mcSize}{0.5*\mcSize}}
\pgfpathlineto{\pgfpoint{0.5*\mcSize}{\mcSize}}
\pgfpathmoveto{\pgfpoint{0pt}{\mcSize}}
\pgfpathlineto{\pgfpoint{\mcSize}{\mcSize}}
\pgfusepath{stroke}}}
\makeatother

\tikzset{
pattern size/.store in=\mcSize, 
pattern size = 5pt,
pattern thickness/.store in=\mcThickness, 
pattern thickness = 0.3pt,
pattern radius/.store in=\mcRadius, 
pattern radius = 1pt}
\makeatletter
\pgfutil@ifundefined{pgf@pattern@_ttva93p49}{
\pgfdeclarepatternformonly[\mcThickness,\mcSize]{_ttva93p49}
{\pgfpointorigin}
{\pgfpoint{\mcSize+\mcThickness}{\mcSize+\mcThickness}}
{\pgfpoint{\mcSize}{\mcSize}}{
\pgfsetcolor{\tikz@pattern@color}
\pgfsetlinewidth{\mcThickness}
\pgfpathmoveto{\pgfpointorigin}
\pgfpathlineto{\pgfpoint{0pt}{0.5*\mcSize}}
\pgfpathlineto{\pgfpoint{\mcSize}{0.5*\mcSize}}
\pgfpathmoveto{\pgfpoint{0.5*\mcSize}{0.5*\mcSize}}
\pgfpathlineto{\pgfpoint{0.5*\mcSize}{\mcSize}}
\pgfpathmoveto{\pgfpoint{0pt}{\mcSize}}
\pgfpathlineto{\pgfpoint{\mcSize}{\mcSize}}
\pgfusepath{stroke}}}
\makeatother

\tikzset{
pattern size/.store in=\mcSize, 
pattern size = 5pt,
pattern thickness/.store in=\mcThickness, 
pattern thickness = 0.3pt,
pattern radius/.store in=\mcRadius, 
pattern radius = 1pt}
\makeatletter
\pgfutil@ifundefined{pgf@pattern@_phps3flje}{
\pgfdeclarepatternformonly[\mcThickness,\mcSize]{_phps3flje}
{\pgfpointorigin}
{\pgfpoint{\mcSize+\mcThickness}{\mcSize+\mcThickness}}
{\pgfpoint{\mcSize}{\mcSize}}{
\pgfsetcolor{\tikz@pattern@color}
\pgfsetlinewidth{\mcThickness}
\pgfpathmoveto{\pgfpointorigin}
\pgfpathlineto{\pgfpoint{0pt}{0.5*\mcSize}}
\pgfpathlineto{\pgfpoint{\mcSize}{0.5*\mcSize}}
\pgfpathmoveto{\pgfpoint{0.5*\mcSize}{0.5*\mcSize}}
\pgfpathlineto{\pgfpoint{0.5*\mcSize}{\mcSize}}
\pgfpathmoveto{\pgfpoint{0pt}{\mcSize}}
\pgfpathlineto{\pgfpoint{\mcSize}{\mcSize}}
\pgfusepath{stroke}}}
\makeatother

\tikzset{
pattern size/.store in=\mcSize, 
pattern size = 5pt,
pattern thickness/.store in=\mcThickness, 
pattern thickness = 0.3pt,
pattern radius/.store in=\mcRadius, 
pattern radius = 1pt}
\makeatletter
\pgfutil@ifundefined{pgf@pattern@_lzr6ii03l}{
\pgfdeclarepatternformonly[\mcThickness,\mcSize]{_lzr6ii03l}
{\pgfpointorigin}
{\pgfpoint{\mcSize+\mcThickness}{\mcSize+\mcThickness}}
{\pgfpoint{\mcSize}{\mcSize}}{
\pgfsetcolor{\tikz@pattern@color}
\pgfsetlinewidth{\mcThickness}
\pgfpathmoveto{\pgfpointorigin}
\pgfpathlineto{\pgfpoint{0pt}{0.5*\mcSize}}
\pgfpathlineto{\pgfpoint{\mcSize}{0.5*\mcSize}}
\pgfpathmoveto{\pgfpoint{0.5*\mcSize}{0.5*\mcSize}}
\pgfpathlineto{\pgfpoint{0.5*\mcSize}{\mcSize}}
\pgfpathmoveto{\pgfpoint{0pt}{\mcSize}}
\pgfpathlineto{\pgfpoint{\mcSize}{\mcSize}}
\pgfusepath{stroke}}}
\makeatother

\tikzset{
pattern size/.store in=\mcSize, 
pattern size = 5pt,
pattern thickness/.store in=\mcThickness, 
pattern thickness = 0.3pt,
pattern radius/.store in=\mcRadius, 
pattern radius = 1pt}
\makeatletter
\pgfutil@ifundefined{pgf@pattern@_6vaiw07bf}{
\pgfdeclarepatternformonly[\mcThickness,\mcSize]{_6vaiw07bf}
{\pgfpointorigin}
{\pgfpoint{\mcSize+\mcThickness}{\mcSize+\mcThickness}}
{\pgfpoint{\mcSize}{\mcSize}}{
\pgfsetcolor{\tikz@pattern@color}
\pgfsetlinewidth{\mcThickness}
\pgfpathmoveto{\pgfpointorigin}
\pgfpathlineto{\pgfpoint{0pt}{0.5*\mcSize}}
\pgfpathlineto{\pgfpoint{\mcSize}{0.5*\mcSize}}
\pgfpathmoveto{\pgfpoint{0.5*\mcSize}{0.5*\mcSize}}
\pgfpathlineto{\pgfpoint{0.5*\mcSize}{\mcSize}}
\pgfpathmoveto{\pgfpoint{0pt}{\mcSize}}
\pgfpathlineto{\pgfpoint{\mcSize}{\mcSize}}
\pgfusepath{stroke}}}
\makeatother

\tikzset{
pattern size/.store in=\mcSize, 
pattern size = 5pt,
pattern thickness/.store in=\mcThickness, 
pattern thickness = 0.3pt,
pattern radius/.store in=\mcRadius, 
pattern radius = 1pt}
\makeatletter
\pgfutil@ifundefined{pgf@pattern@_epjj7apnb}{
\pgfdeclarepatternformonly[\mcThickness,\mcSize]{_epjj7apnb}
{\pgfpointorigin}
{\pgfpoint{\mcSize+\mcThickness}{\mcSize+\mcThickness}}
{\pgfpoint{\mcSize}{\mcSize}}{
\pgfsetcolor{\tikz@pattern@color}
\pgfsetlinewidth{\mcThickness}
\pgfpathmoveto{\pgfpointorigin}
\pgfpathlineto{\pgfpoint{0pt}{0.5*\mcSize}}
\pgfpathlineto{\pgfpoint{\mcSize}{0.5*\mcSize}}
\pgfpathmoveto{\pgfpoint{0.5*\mcSize}{0.5*\mcSize}}
\pgfpathlineto{\pgfpoint{0.5*\mcSize}{\mcSize}}
\pgfpathmoveto{\pgfpoint{0pt}{\mcSize}}
\pgfpathlineto{\pgfpoint{\mcSize}{\mcSize}}
\pgfusepath{stroke}}}
\makeatother

\tikzset{
pattern size/.store in=\mcSize, 
pattern size = 5pt,
pattern thickness/.store in=\mcThickness, 
pattern thickness = 0.3pt,
pattern radius/.store in=\mcRadius, 
pattern radius = 1pt}
\makeatletter
\pgfutil@ifundefined{pgf@pattern@_9fd7bjsii}{
\pgfdeclarepatternformonly[\mcThickness,\mcSize]{_9fd7bjsii}
{\pgfpointorigin}
{\pgfpoint{\mcSize+\mcThickness}{\mcSize+\mcThickness}}
{\pgfpoint{\mcSize}{\mcSize}}{
\pgfsetcolor{\tikz@pattern@color}
\pgfsetlinewidth{\mcThickness}
\pgfpathmoveto{\pgfpointorigin}
\pgfpathlineto{\pgfpoint{0pt}{0.5*\mcSize}}
\pgfpathlineto{\pgfpoint{\mcSize}{0.5*\mcSize}}
\pgfpathmoveto{\pgfpoint{0.5*\mcSize}{0.5*\mcSize}}
\pgfpathlineto{\pgfpoint{0.5*\mcSize}{\mcSize}}
\pgfpathmoveto{\pgfpoint{0pt}{\mcSize}}
\pgfpathlineto{\pgfpoint{\mcSize}{\mcSize}}
\pgfusepath{stroke}}}
\makeatother

\tikzset{
pattern size/.store in=\mcSize, 
pattern size = 5pt,
pattern thickness/.store in=\mcThickness, 
pattern thickness = 0.3pt,
pattern radius/.store in=\mcRadius, 
pattern radius = 1pt}
\makeatletter
\pgfutil@ifundefined{pgf@pattern@_n06ryn7uc}{
\pgfdeclarepatternformonly[\mcThickness,\mcSize]{_n06ryn7uc}
{\pgfpointorigin}
{\pgfpoint{\mcSize+\mcThickness}{\mcSize+\mcThickness}}
{\pgfpoint{\mcSize}{\mcSize}}{
\pgfsetcolor{\tikz@pattern@color}
\pgfsetlinewidth{\mcThickness}
\pgfpathmoveto{\pgfpointorigin}
\pgfpathlineto{\pgfpoint{0pt}{0.5*\mcSize}}
\pgfpathlineto{\pgfpoint{\mcSize}{0.5*\mcSize}}
\pgfpathmoveto{\pgfpoint{0.5*\mcSize}{0.5*\mcSize}}
\pgfpathlineto{\pgfpoint{0.5*\mcSize}{\mcSize}}
\pgfpathmoveto{\pgfpoint{0pt}{\mcSize}}
\pgfpathlineto{\pgfpoint{\mcSize}{\mcSize}}
\pgfusepath{stroke}}}
\makeatother

\tikzset{
pattern size/.store in=\mcSize, 
pattern size = 5pt,
pattern thickness/.store in=\mcThickness, 
pattern thickness = 0.3pt,
pattern radius/.store in=\mcRadius, 
pattern radius = 1pt}
\makeatletter
\pgfutil@ifundefined{pgf@pattern@_5vnyrfse6}{
\pgfdeclarepatternformonly[\mcThickness,\mcSize]{_5vnyrfse6}
{\pgfpointorigin}
{\pgfpoint{\mcSize+\mcThickness}{\mcSize+\mcThickness}}
{\pgfpoint{\mcSize}{\mcSize}}{
\pgfsetcolor{\tikz@pattern@color}
\pgfsetlinewidth{\mcThickness}
\pgfpathmoveto{\pgfpointorigin}
\pgfpathlineto{\pgfpoint{0pt}{0.5*\mcSize}}
\pgfpathlineto{\pgfpoint{\mcSize}{0.5*\mcSize}}
\pgfpathmoveto{\pgfpoint{0.5*\mcSize}{0.5*\mcSize}}
\pgfpathlineto{\pgfpoint{0.5*\mcSize}{\mcSize}}
\pgfpathmoveto{\pgfpoint{0pt}{\mcSize}}
\pgfpathlineto{\pgfpoint{\mcSize}{\mcSize}}
\pgfusepath{stroke}}}
\makeatother

\tikzset{
pattern size/.store in=\mcSize, 
pattern size = 5pt,
pattern thickness/.store in=\mcThickness, 
pattern thickness = 0.3pt,
pattern radius/.store in=\mcRadius, 
pattern radius = 1pt}
\makeatletter
\pgfutil@ifundefined{pgf@pattern@_w8t7k6z9g}{
\pgfdeclarepatternformonly[\mcThickness,\mcSize]{_w8t7k6z9g}
{\pgfpointorigin}
{\pgfpoint{\mcSize+\mcThickness}{\mcSize+\mcThickness}}
{\pgfpoint{\mcSize}{\mcSize}}{
\pgfsetcolor{\tikz@pattern@color}
\pgfsetlinewidth{\mcThickness}
\pgfpathmoveto{\pgfpointorigin}
\pgfpathlineto{\pgfpoint{0pt}{0.5*\mcSize}}
\pgfpathlineto{\pgfpoint{\mcSize}{0.5*\mcSize}}
\pgfpathmoveto{\pgfpoint{0.5*\mcSize}{0.5*\mcSize}}
\pgfpathlineto{\pgfpoint{0.5*\mcSize}{\mcSize}}
\pgfpathmoveto{\pgfpoint{0pt}{\mcSize}}
\pgfpathlineto{\pgfpoint{\mcSize}{\mcSize}}
\pgfusepath{stroke}}}
\makeatother

\tikzset{
pattern size/.store in=\mcSize, 
pattern size = 5pt,
pattern thickness/.store in=\mcThickness, 
pattern thickness = 0.3pt,
pattern radius/.store in=\mcRadius, 
pattern radius = 1pt}
\makeatletter
\pgfutil@ifundefined{pgf@pattern@_nv34ozx8w}{
\pgfdeclarepatternformonly[\mcThickness,\mcSize]{_nv34ozx8w}
{\pgfpointorigin}
{\pgfpoint{\mcSize+\mcThickness}{\mcSize+\mcThickness}}
{\pgfpoint{\mcSize}{\mcSize}}{
\pgfsetcolor{\tikz@pattern@color}
\pgfsetlinewidth{\mcThickness}
\pgfpathmoveto{\pgfpointorigin}
\pgfpathlineto{\pgfpoint{0pt}{0.5*\mcSize}}
\pgfpathlineto{\pgfpoint{\mcSize}{0.5*\mcSize}}
\pgfpathmoveto{\pgfpoint{0.5*\mcSize}{0.5*\mcSize}}
\pgfpathlineto{\pgfpoint{0.5*\mcSize}{\mcSize}}
\pgfpathmoveto{\pgfpoint{0pt}{\mcSize}}
\pgfpathlineto{\pgfpoint{\mcSize}{\mcSize}}
\pgfusepath{stroke}}}
\makeatother

\tikzset{
pattern size/.store in=\mcSize, 
pattern size = 5pt,
pattern thickness/.store in=\mcThickness, 
pattern thickness = 0.3pt,
pattern radius/.store in=\mcRadius, 
pattern radius = 1pt}
\makeatletter
\pgfutil@ifundefined{pgf@pattern@_89wwvbagr}{
\pgfdeclarepatternformonly[\mcThickness,\mcSize]{_89wwvbagr}
{\pgfpointorigin}
{\pgfpoint{\mcSize+\mcThickness}{\mcSize+\mcThickness}}
{\pgfpoint{\mcSize}{\mcSize}}{
\pgfsetcolor{\tikz@pattern@color}
\pgfsetlinewidth{\mcThickness}
\pgfpathmoveto{\pgfpointorigin}
\pgfpathlineto{\pgfpoint{0pt}{0.5*\mcSize}}
\pgfpathlineto{\pgfpoint{\mcSize}{0.5*\mcSize}}
\pgfpathmoveto{\pgfpoint{0.5*\mcSize}{0.5*\mcSize}}
\pgfpathlineto{\pgfpoint{0.5*\mcSize}{\mcSize}}
\pgfpathmoveto{\pgfpoint{0pt}{\mcSize}}
\pgfpathlineto{\pgfpoint{\mcSize}{\mcSize}}
\pgfusepath{stroke}}}
\makeatother
\tikzset{every picture/.style={line width=0.75pt}} %set default line width to 0.75pt        

\title[Subhalo finding in cosmological simulations]{Assessing subhalo finders in cosmological hydrodynamical simulations}

\author[Victor J. Forouhar Moreno et al.]{Victor J. Forouhar Moreno$^{1}$\thanks{E-mail: forouhar@strw.leidenuniv.nl}, John Helly$^{2}$, Robert McGibbon$^{1}$, Joop Schaye$^{1}$, Matthieu Schaller$^{3,1}$, 
\newauthor 
Jiaxin Han$^{4,5}$, Roi Kugel$^{1}$ and Yannick M.~Bah\'e$^{6,7}$ \\
$^{1}$Leiden Observatory, Leiden University, PO Box 9513, NL-2300 RA Leiden, the Netherlands\\
$^{2}$Institute for Computational Cosmology, Department of Physics, University of Durham, South Road, Durham, DH1 3LE, UK\\
$^{3}$Lorentz Institute for Theoretical Physics, Leiden University, PO Box 9506, 2300 RA Leiden, the Netherlands\\
$^{4}$Department of Astronomy, School of Physics and Astronomy, Shanghai Jiao Tong University, Shanghai 200240, PR China\\
$^{5}$Shanghai Key Laboratory for Particle Physics and Cosmology, Shanghai 200240, China \\
$^{6}$School of Physics and Astronomy, University of Nottingham, University Park, Nottingham NG7 2RD, UK \\ 
$^{7}$Laboratory of Astrophysics, Ecole Polytechnique F\'{e}d\'{e}rale de Lausanne (EPFL), Observatoire de Sauverny, 1290 Versoix, Switzerland
}

\date{Accepted XXX. Received YYY; in original form ZZZ}

\pubyear{2024}

\begin{document}
\label{firstpage}
\pagerange{\pageref{firstpage}--\pageref{lastpage}}
\maketitle 

\begin{abstract}
Cosmological simulations are essential for inferring cosmological and galaxy population properties based on forward-modelling, but this typically requires finding the population of (sub)haloes and galaxies that they contain. The properties of said populations vary depending on the algorithm used to find them, which is concerning as it may bias key statistics. We compare how the predicted (sub)halo mass functions, satellite radial distributions and correlation functions vary across algorithms in the dark-matter-only and hydrodynamical versions of the {\tt FLAMINGO} simulations. We test three representative approaches to finding subhaloes: grouping particles in configuration- ({\SUBFIND}), phase- ({\ROCKSTAR} and {\VELOCIRAPTOR}) and history-space ({\HBTHERONS}). We also present {\HBTHERONS}, a new version of the {\HBT} subhalo finder that improves the tracking of subhaloes. We find 10\%-level differences in the $M_{\mathrm{200c}}$ mass function, reflecting different field halo definitions and occasional miscentering. The bound mass functions can differ by 75\% at the high mass end, even when using the maximum circular velocity as a mass proxy. The number of well-resolved subhaloes differs by up to 20\% near $R_{\mathrm{200c}}$, reflecting differences in the assignment of mass to subhaloes and their identification. The predictions of different subhalo finders increasingly diverge towards the centres of the host haloes. The performance of most subhalo finders does not improve with the resolution of the simulation and is worse for hydrodynamical than for dark-matter-only simulations. We conclude that {\HBTHERONS} is the preferred choice of subhalo finder due to its low computational cost, self-consistently made and robust merger trees, and robust subhalo identification capabilities.
\end{abstract}


%%%%%%%%%%%%%%%%%%%%%%%%%%%%%%%%%%%%%%%%%%%%%%%%%%

%%%%%%%%%%%%%%%%% BODY OF PAPER %%%%%%%%%%%%%%%%%%

\section{Introduction}

The formation of cosmic structure is driven by gravity, which amplifies small deviations from homogeneity in the early Universe. Overdense regions eventually undergo gravitational collapse, leading to the formation of dark matter (DM) haloes. This process is hierarchical in nature, as the smallest haloes form first, and subsequently grow through mass accretion and mergers with neighbouring ones. As the formation of  galaxies is tied to the formation of dark matter haloes, galaxies trace the underlying distribution of matter in the Universe.  

The initial distribution of matter in the Universe, as well as its subsequent evolution, depend on cosmological parameters and the nature of dark matter. The amplitude of density fluctuations depends on $\sigma_{8}$. The subsequent growth of structure is further modulated by the relative amounts of matter and dark energy. As such, measuring the distribution, abundance and properties of structure in the Universe is an essential tool to study its cosmology.

In the past, this approach helped rule out hot models of dark matter \citep{White.1983}, as well as hint at a non-zero cosmological constant \citep[e.g.][]{Efstathiou.1990}. Ongoing surveys aim to constrain key cosmological parameters, such as the amplitude of power fluctuations (e.g. via weak lensing; \citealt{Euclid.2024}), or the nature of dark matter (e.g. using Milky Way satellites; \citealt{Drlica-Wagner.2019}). Making such cosmological inferences requires forward modelling the evolution of structure in the Universe under different sets of parameter choices. This in turn involves running cosmological simulations of increasingly large volumes and higher resolutions. However, simulations follow individual mass elements, rather than the conglomerate structures that they are a part of. As such, an important step in this process is finding where the haloes and the galaxies that they harbour are located.

Identifying (sub)haloes is non-trivial. They are found in a variety of environments and evolutionary stages, with some combinations being inherently more difficult to analyse than others. For instance, field haloes can be found with relative ease using the Friends-of-Friends (FoF) algorithm \citep[e.g.][]{Press.1982}. This method links all particles within a threshold distance, typically chosen to be 0.2 of the mean interparticle separation. This roughly corresponds to an overdensity of 100 times the critical density of the universe \citep{More.2011}. 

Nonetheless, field haloes are often accreted by more massive ones and become satellite subhaloes. If this happens, the simple FoF algorithm is unable to distinguish the satellite from the host. This means that additional steps are required to separate satellite subhaloes from the halo that hosts them. Not doing so would mean that only a subset of the total population of subhaloes in the Universe would be found. 
% The missing satellites, which is an important population in cosmol appear as , are important 

Many algorithms have been developed with the aim of finding subhaloes. They primarily rely on segmenting particles within individual FoF groups into candidate subhaloes. As this can include chance groupings of particles due to discreteness noise, additional checks are done to clean catalogues from spurious subhaloes. For example, by requiring subhaloes to contain a minimum number or fraction of self-bound particles. 

However, differences exist between the subhalo populations that subhalo finders identify. This can often be attributed to how they identify candidate subhaloes, as well as their definition of what a subhalo actually is. There are three main approaches in the literature. Configuration space finders rely on identifying density peaks (or overdensities) in the matter distribution, e.g. by kernel smoothing ({\tt SUBFIND}; \citealt{Springel.2001}, {\tt ADAPTAHOP}; \citealt{Aubert.2004}), adaptive mesh refinement ({\tt AHF}; \citealt{Gill.2004, Knollmann.2009}), or hierarchical 3D FoF finding ({\tt HFoF}; \citealt{Klypin.1999}). Phase-space finders use the position and velocity of particles, such as by doing 6D FoF finding ({\ROCKSTAR}; \citealt{Behroozi.2013}) or finding local peaks in spatial and velocity density ({\VELOCIRAPTOR}; \citealt{Elahi.2011, Elahi.2019}). The additional information provided by the velocity of particles helps overcome several shortcomings of configuration space finders, such as when subhaloes spatially overlap but remain physically distinct entities \citep[e.g.][]{Behroozi.2015}.

The third approach is based on `history'-space, e.g. {\tt SURV} \citep{Giocoli.2010}, {\tt HBT} \citep{Han.2012} and {\HBT} \citep{Han.2018}. The idea is to use information from previous outputs, as that way subhaloes can be first identified when it is easier to do so (e.g. before becoming satellites). The information is then used at later times, so that the algorithm does not have to disentangle the often complicated instantaneous distribution of particles. This alternative approach has been shown to improve the results of subhalo finders and the resulting merger trees \citep{Chandro-Gomez.2025}, and has been gaining traction in recent years \citep[e.g.][]{Springel.2021,Diemer.2024, Mansfield.2024}. 

Each type of subhalo finder has an associated drawback, which can be reflected in the properties of the subhalo population that they (do not) find. As mentioned before, density peak finders struggle to separate subhaloes that are close in space, such as when close pericentre passages or major mergers occur \citep[e.g.][]{Behroozi.2015}. Phase-space finders can also struggle to locate subhaloes that are severely tidally distorted \citep[e.g.][]{Diemer.2024}. Both cases of subhalo misidentification occur even when the number of particles sampling them is large, so the problem is not uniquely driven by a lack of resolution. For history-based finders, the need for a sufficient number of finely time-spaced outputs increases the storage requirements. This also means that the output strategy of simulations needs to be set ahead of time. Additionally, since the catalogues at later times depend on earlier ones, small errors in the tracing of subhaloes, or incorrect assumptions of how they evolve, may amplify over time. Lastly, current history based finders assume hierarchical structure formation, making it difficult to identify subhaloes that formed via fragmentation of a common progenitor, such as tidal dwarf galaxies.  

Understanding how the choice of subhalo finder affects the subhalo population, and hence the predictions to which observations are compared, is essential in the era of precision cosmology. Previous studies have looked at how different subhalo finders perform in a range of idealised test cases, as well as zoom-in and cosmological simulations \citep[e.g.][]{Knebe.2011, Onions.2012, Behroozi.2015}. The comparisons, which were based on only analysing the dark matter particles to find subhaloes, revealed that the properties of subhaloes vary according to where they are spatially located within their host halo, with greater difficulties in recovering the input properties of the test subhaloes at small radii \citep{Knebe.2011}. Subhalo properties that rely on how their edge is defined, such as their masses, can vary by $\approx 20\%$ between subhalo finders \citep{Knebe.2011, Onions.2012}. These differences reflect a combination of how the subhaloes are initially found, and how (and if) the particles are subjected to gravitational unbinding. Overall, phase-space and history-space finders recover the subhalo population properties more robustly than configuration-based finders, although the performance of phase-space finders appears to depend sensitively on their implementation details (e.g. some find similar populations to those in configuration-space finders; \citealt{Onions.2012}).   

However, two important aspects are largely missing from past comparisons. First, how the choice of subhalo finder affects cosmologically-relevant summary statistics, e.g. correlation functions. Although discussed in \citet{Onions.2012} and \citet{Pujol.2014}, they focused on either the $10^{5}$ most massive central subhaloes or on the satellites within a relaxed, Milky Way-mass halo. Neither of these choices is a representative subset of the full variety of cosmic structures, so the actual biases may be larger than those inferred from those studies. Second, how the performance of subhalo finders changes between dark-matter-only (DMO) and hydrodynamical simulations. For example, cold and dense gas can be identified as a separate self-bound density peak within a galaxy disc, the phase-space distribution of baryons is clearly different from that in dispersion-supported dark matter haloes, and structure can form non-hierarchically (e.g. tidal dwarf galaxies). Most finders are first developed with DMO in mind, as reflected in previous comparisons of how subhalo finders perform. Hydrodynamical support is subsequently added without consideration of how the effects resulting from the inclusion of baryons affect the algorithms.

Lastly, an equally important aspect is the computational cost of running (sub)halo finding. As simulations grow larger in volume and reach higher resolutions, the speed and scalability of (sub)halo finding becomes an important consideration. Some algorithms are inherently slower than others, leading to a computational expense comparable to the cost of running the simulation itself.

Here, we explore how the choice of subhalo finder affects key summary statistics relevant to large-scale cosmological studies. For this purpose, we use the {\tt FLAMINGO} simulations \citep[][]{Schaye.2023, Kugel.2023}, which constitute a representative suite of simulations for these kind of studies. We focus our comparison on four subhalo finders: {\HBTHERONS} (a new version of the {\HBT} subhalo finder, the previous version of which was described by \citealt{Han.2018}), {\SUBFIND} \citep{Springel.2005,Springel.2021}, {\VELOCIRAPTOR} \citep{Elahi.2019} and {\ROCKSTAR} \citep{Behroozi.2013}. This set of algorithms is representative of configuration- ({\SUBFIND}), phase- ({\VELOCIRAPTOR}, {\ROCKSTAR}) and history-space finding ({\HBTHERONS}). 

In this comparison, we find $10\%$ differences in even basic summary statistics like the $M_{\mathrm{200c}}$ mass functions. The magnitude of the disagreement between subhalo finders worsens for more detailed properties, like the radial distribution of satellite subhaloes around haloes. Part of the differences is caused by a poorer performance of the algorithms when running in hydrodynamical simulations. Altogether, the aforementioned differences in subhalo finding result in different two-point correlation functions, both in the two- and one-halo regimes. Overall, we find that {\HBTHERONS} provides a robust and computationally cheap way of identifying subhaloes in simulations. Its history-based approach means that it is able to identify subhaloes in instantaneous configurations that are difficult to analyse using alternative algorithms. The resulting merger trees contain fewer catastrophic failures (e.g. severe mass swapping and massive transients; \citealt{Chandro-Gomez.2025}), substantially reducing the need to `prune' merger trees in post-processing. Given this, we make the catalogues made by {\HBTHERONS} the fiducial {\tt FLAMINGO} ones.

This paper begins by presenting the simulations (\S\ref{section:simulations}) and the (sub)halo finders we use (\S\ref{section:halo_finders}). As we present a new version of {\HBT}, {\HBTHERONS}, we discuss how it works in more detail than the other subhalo finders. Each change we made to {\HBT} is described in detail in Appendix \ref{section:hbt_improvements}. The main results of the comparisons across subhalo finders are shown in \S\ref{section:results}, where we examine the halo \& subhalo mass functions (\S\ref{section:halo_mass_functions}), the subhalo radial distributions within centrals (\S\ref{section:radial_distributions}) and the subhalo correlation functions (\S\ref{section:correlation_functions}). The key takeaway points are summarised in \S\ref{section:conclusions}.

\section{Simulations}\label{section:simulations}

We use the {\tt FLAMINGO} simulations \citep{Schaye.2023,Kugel.2023} to compare subhalo finders. The {\tt FLAMINGO} simulations are a suite of large-scale cosmological hydrodynamical simulations calibrated to reproduce the observed $z = 0$ stellar mass function \citep{Driver.2022} and cluster gas fractions \citep{Kugel.2023}. The simulations, which also include neutrinos through the $\delta f$ method \citep{Elbers.2021}, were run using the open source code Swift \citep{Schaller.2024} using the SPHENIX SPH scheme \citep{Borrow.2022}.

{\tt FLAMINGO} includes subgrid models for element-by-element radiative cooling and heating \citep{Ploeckinger.2020}, star formation \citep{Schaye.2008}, stellar mass loss \citep{Wiersma.2009, Schaye.2015}, feedback energy from supernova \citep{DallaVecchia.2008, Chaikin.2022,Chaikin.2023}, seeding and growth of black holes, and feedback from active galactic nuclei \citep{Springel.2005, Booth.2009, Bahe.2022}. AGN feedback is either thermal \citep{Booth.2009} or jet-based \citep{Husko.2022}. The subgrid parameters were fit to observations using emulators \citep{Kugel.2023}, but simulation variations in the feedback strength span the observational uncertainties. In this work, we only analyse the simulations that use the fiducial subgrid model and their dark-matter-only counterparts.

Several {\tt FLAMINGO} box sizes, resolutions and cosmology variations exist. For this study, we analyse the $1\, \mathrm{Gpc}^{3}$ volumes that use the cosmological parameters from the ‘3x2pt + all external constraints’ dark energy survey year 3 results of \citet{Abbott.2022}  ($\Omega_{\mathrm{m}} = 0.306$, $\Omega_{\mathrm{b}} = 0.0486$, $\sigma_{8} = 0.807$, $H_{0} = 68.1$, $n_{\mathrm{s}}$ = 0.967). Their initial conditions were generated using a modified version of {\tt Monofonic} \citep{Hahn.2021, Elbers.2022}. 

The identifiers of the three resolution levels of the cosmological simulations we analyse, and their corresponding dark matter particle mass ($m_{\mathrm{dm}}$), baryonic particle mass ($m_{\mathrm{b}}$) and the maximum physical softening length ($\epsilon$) are listed in Table \ref{table:simulation_summary}. The simulations have 79 outputs, with an average spacing in expansion factor of $\langle \Delta \ln a\rangle = 0.015$. The output strategy was not tuned to work with the {\HBTHERONS} subhalo finder.

\begin{table}
\centering
\caption{Summary of the initial mass for dark matter ($m_{\mathrm{dm}}$) and baryonic ($m_{\mathrm{b}}$) particles in the cosmological simulation suite we use in this work, as well as the maximum physical softening length ($\epsilon$). Note that we do not use the hydrodynamical version of the {\highres} simulation due to the large computational cost required to analyse it with all of the subhalo finders we compare.}
\begin{tabular}{@{}cclll@{}}
\toprule
\multicolumn{1}{l}{Simulation Identifier} & \multicolumn{1}{l}{Type} & $m_{\mathrm{DM}}$ {[}$\Msun${]} & $m_{\mathrm{bar}}$ {[}$\Msun${]} & $\epsilon$ {[}kpc{]} \\ \midrule
\multirow{2}{*}{L1\_m10}       & DMO                      & $5.38\times10^{10}$             & \multicolumn{1}{c}{-}            & 11.40                \\ \cmidrule(l){2-5} 
                               & Hydro                    & $4.52\times10^{10}$             & $8.56\times10^{9}$               & 11.40                \\ \midrule
\multirow{2}{*}{L1\_m9}        & DMO                      & $6.72\times10^{9}$              & \multicolumn{1}{c}{-}            & \phantom{1}5.70                 \\ \cmidrule(l){2-5} 
                               & Hydro                    & $5.65\times10^{9}$              & $1.07\times10^{9}$               & \phantom{1}5.70                 \\ \midrule
\multirow{2}{*}{L1\_m8}        & DMO                      & $8.40\times10^{8}$              & \multicolumn{1}{c}{-}            & \phantom{1}2.85                 \\ \cmidrule(l){2-5} 
                               & Hydro                    & $7.06\times10^{8}$              & $1.34\times10^{8}$               & \phantom{1}2.85                 \\ \bottomrule
\end{tabular}

\label{table:simulation_summary}
\end{table}
\section{Halo and subhalo finding}\label{section:halo_finders}

In this section, we discuss how we identify (sub)haloes within the simulations. Although the {\tt FLAMINGO} simulations include neutrinos, we omit them when finding (sub)haloes because they only bind to the most massive structures and they account for a small fraction of the mass ($\Omega_{\nu} / \Omega_{\mathrm{m}} \approx 4 \times 10^{-3}$). We start by explaining in \S\ref{section:fof_parameters} how the Friends-of-Friends (FoF) percolation algorithm works, as it is a required step for each of the subhalo finders we use. We then discuss the working principles behind the four subhalo finders that are compared in this work, which use three distinct approaches: configuration- ({\SUBFIND}, \citealt{Springel.2001, Springel.2021}), phase- ({\ROCKSTAR}, \citealt{Behroozi.2013}; {\VELOCIRAPTOR}; \citealt{Elahi.2019}) or history-space ({\HBTHERONS}). We conclude this section by discussing how we compute subhalo properties in a consistent manner across finders (\S\ref{section:soap}).

Throughout the remainder of the paper, we use the following nomenclature. Haloes are collections of particles found using the 3D FoF particle clustering algorithm, and they contain one or more subhaloes. A subhalo is a self-bound collection of particles distinct from the background in which they are embedded, according to the particular metric that each subhalo finder uses. Subhaloes can be either centrals or satellites, with the position of central subhaloes used as the position of field haloes. The defining property of central subhaloes varies, but they are typically chosen based on comparing a characteristic property (e.g. mass) to other subhaloes within the FoF they are part of ({\SUBFIND}, {\VELOCIRAPTOR}, {\HBTHERONS}) or within a given spherical aperture ({\ROCKSTAR}). 

\subsection{Friends-Of-Friends}\label{section:fof_parameters}

All of the subhalo finders we use rely on initially grouping particles using the Friends-of-Friends percolation algorithm. Each distinct FoF group is composed by particles that are within a critical distance from each other, measured using a spatial or phase-space metric. In {\HBTHERONS}, {\SUBFIND} and {\ROCKSTAR}, the metric is spatial and set to be a fraction of the mean dark matter interparticle separation. {\VELOCIRAPTOR} also starts with a 3D FoF, but it can subsequently trim it by using a 6D FoF algorithm. This optional step is done with the aim of removing spurious particle bridges, which can boost the halo mass function  based on the virial overdensity of \citet{Bryan.1998} \citep[e.g.][]{Euclid.2023}. In this study, we choose not to use the 6D trimming to find field haloes in {\VELOCIRAPTOR}, as we want to find them as consistently as possible across subhalo finders. This also reflects the same choice as was made for the {\tt FLAMINGO} halo catalogues presented in \citet{Schaye.2023}.

The fiducial linking length used by the 3D FoF catalogues in {\HBTHERONS}, {\SUBFIND} and {\VELOCIRAPTOR} is set to be 0.2 times the mean interparticle separation. For {\ROCKSTAR}, we use instead the recommended value of 0.28. In the DMO simulations, the mean interparticle separation is defined as:
\begin{equation}
    \langle l \rangle_{\mathrm{DMO}} = \Big[\dfrac{m_{\mathrm{dm}}}{(\Omega_{\mathrm{cdm}} + \Omega_{\mathrm{b}})\rho_{\mathrm{crit}}}\Big]^{1/3}\, ,
\end{equation}
where 
\begin{equation}
    \rho_{\mathrm{crit}}(z) = \dfrac{3H^{2}(z)}{8\pi G}\, .
\end{equation}
The baryon density parameter is $\Omega_{\mathrm{b}}$, and the dark matter (minus the contribution of neutrinos) density parameter is $\Omega_{\mathrm{cdm}}$. In the hydrodynamical simulations, we use the mean dark matter interparticle separation:
\begin{equation}
    \langle l \rangle_{\mathrm{HYDRO}} = \Big[\dfrac{m_{\mathrm{dm}}}{\Omega_{\mathrm{cdm}}\rho_{\mathrm{crit}}}\Big]^{1/3}\, .
\end{equation}
For {\HBTHERONS} and {\SUBFIND}, the linking in hydrodynamical simulations is first done using only the DM particles. All other particle types are subsequently attached to their nearest dark matter-defined FoF group, and those with fewer than 32 total particles are subsequently discarded. In {\VELOCIRAPTOR}, the FoF links pairs of particles as long as one of them is a DM particle. Similar to {\HBTHERONS} and {\SUBFIND}, all FoF groups with fewer than 32 total particles are removed from the catalogues. {\ROCKSTAR} runs the FoF algorithm on all particle types directly and by default retains those with more than 10 particles. Each subhalo finder ran its own implementation of FoF, except for {\HBTHERONS}, which used the FoF output of {\tt Swift}. We verified that the FoF algorithms implemented within {\tt GADGET-4} and {\tt Swift} produce the same results. 

\subsection{HBT-HERONS}\label{HBT_section}

\begin{figure}
    \centering
    \input{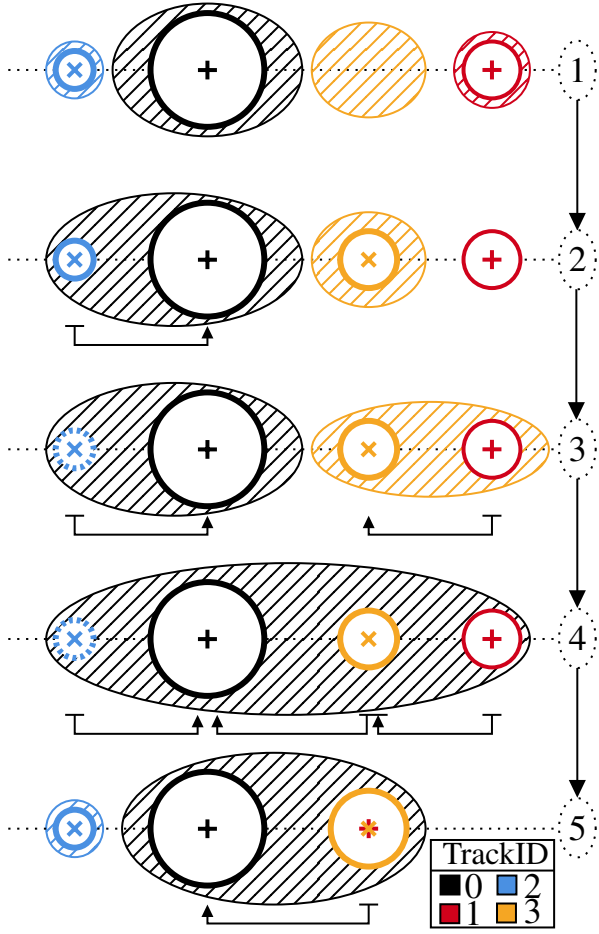}
    \caption{Schematic of how {\HBTHERONS} tracks subhaloes, shown across five output times. Each column shows the time evolution of a unique subhalo, whose bound component at a given time is represented by a circle. If no bound component is identified, the edge of the circle is dashed. Different colours correspond to different evolutionary branches (`Tracks'), which have unique IDs (`TrackIDs') given by the legend at the bottom. As each TrackID is unique and time-persistent, the evolution of individual subhaloes is easy to retrieve. The shaded ellipses are FoF groups, coloured as their corresponding central subhalo. The lines connecting different subhaloes at a fixed time indicate the current subhalo hierarchy. The crosses and pluses represent the position of its most bound tracer particle. \textit{Output 1}: four FoF groups exist, but one of them does not contain a self-bound subhalo. Only three Tracks (0,1,2) are created. \textit{Output 2}: The FoF without a previous bound subhalo now has one, so a new Track (3) is created. Track 0 and 2 are assigned to the same FoF, based on the weighted FoF membership of their tracer particles. In this case, Track 0 is much more massive than Track 2, so Track 2 is automatically assigned as its satellite. Track 1 is not found in any FoF group, but the past particle membership is used to identify that there is still a bound (hostless) subhalo. \textit{Output 3}: Tracks 1 and 3 are in the same FoF. As they have similar masses, the central subhalo is the one with the minimum kinetic energy in the centre of mass frame of the host FoF (Track 3). Track 2 is now unresolved, so its position and velocity is found using the most bound tracer from output 2. \textit{Output 4}: All subhaloes are found in the same FoF group. Since Tracks 0 and 3 were centrals in the previous output, they are the only candidates to be centrals. Based on the relative mass, Track 0 is chosen as the central, making Track 1 a satellite of a satellite. \textit{Output 5}: The cores of Tracks 1 and 3 overlap in phase-space, so they are merged into a single subhalo. The tracer of Track 2 is found in a separate FoF, i.e. it is a splashback object. The FoF hosting Track 2 has no resolved subhaloes at that time, meaning that its particles are assigned to the orphaned Track 2. The particles are then confirmed to be self-bound, making Track 2 appear as a resolved subhalo again.}
    \label{figure:hbt_general_diagram}
\end{figure}

The {\HBTHERONS}\footnote{Available at \href{https://github.com/SWIFTSIM/HBT-HERONS}{https://github.com/SWIFTSIM/HBT-HERONS}} (\textbf{H}ierarchical \textbf{B}ound \textbf{T}racing - \textbf{H}ydro-\textbf{E}nabled \textbf{R}etrieval of \textbf{O}bjects in \textbf{N}umerical \textbf{S}imulations) algorithm is an updated version of the history-based subhalo finder {\HBT} \citep{Han.2018}. This new version collects several new additions that improve the identification and tracking of subhaloes in both DMO and hydrodynamical simulations, which are presented in detail in Appendix \ref{section:hbt_improvements}\footnote{The {\HBTHERONS} catalogues for {\tt FLAMINGO} were made without including the symmetric merging criterion (\S\ref{appendix:merging_calculations}), the center refinement change described in \S\ref{appendix:unbinding_subsampling_effects} and the gas re-attachtment step (\S\ref{appendix:gas_reattaching}). The need for the last two modifications only became clear when running on higher resolution simulations. We have verified that the effect of these changes is negligible for all of the results presented here.}. We additionally modified the way in which particles are partitioned within the code to speed it up. This allowed us to run {\HBTHERONS} using 256 nodes (128 cores per node) on the {\tt FLAMINGO-10k} DMO simulation (\citealt{Pizzati.2024}; Schaller et al. in prep), which contains over $10080^{3}$ ($\approx10^{12}$) particles. We also used it to analyse the largest hydrodynamical simulation run to $z = 0$ to date, the 2.8~Gpc {\tt FLAMINGO} box with $2\times5040^{3}$ particles. Given its small computational cost and robust identification of structures, both at fixed and across time \citep{Chandro-Gomez.2025}, we make {\HBTHERONS} our fiducial choice of subhalo finder. In this subsection, we present a general overview of how {\HBTHERONS} works, with an accompanying diagram to illustrate important steps in Fig.~\ref{figure:hbt_general_diagram}.

{\HBTHERONS} assumes that structure forms hierarchically, i.e. that every present-day subhalo has been a central subhalo in the past. In practice, this means that particles belonging to a given halo become associated to its central subhalo in what is called a `source' subhalo (see \citealt{Han.2012}). The source subhalo is the set of particles that constitute a tentative subhalo within the simulation, whose existence needs confirming after a `cleaning' step to identify whether it is indeed a physical subhalo or just spurious noise. Our choice to determine when a subhalo is resolved is the same as the choice made by {\SUBFIND} and {\VELOCIRAPTOR}, which is to check whether the subhalo contains a minimum number of self-bound particles (\S\ref{unbinding_step}). However, alternative criteria used to deem a subhalo as physical exist, like requiring a minimum fraction of self-bound particles in a given 6D FoF subhalo in {\ROCKSTAR} or by discarding subhaloes whose concentrations are non-physical in {\tt Bloodhound} \citep{Kong.2025}.

The way in which the source subhalo is defined differs across subhalo finders, and it is closely related to the performance with which they are able to identify subhaloes. For example, the source subhalo in {\SUBFIND} corresponds to all particles enclosed within a density peak in configuration space, excluding additional density peaks within. In contrast, {\HBTHERONS} defines it based on the past evolution of subhaloes: the source subhalo of satellite subhaloes is built from the source subhalo when they were last a central, excluding its resolved substructure. This approach circumvents the difficulties involved when trying to separate heavily overlapping subhaloes in phase-space. Indeed, variations of this history-based approach have recently been used to augment other structure finding algorithms, e.g. {\tt Subfind-HBT}\footnote{Despite the similarity in the name, the implementation details differ from those in {\HBT}. This leads to differences that we discuss in Appendix \ref{Appendix:VersionsOfHBT}.} \citep{Springel.2021}, {\tt ROCKSTAR+MORIA} \citep{Diemer.2024}, {\tt ROCKSTAR+SYMFIND} \citep{Mansfield.2024} and {\tt  ROCKSTAR+Bloodhound} \citep{Kong.2025}.

{\HBTHERONS} requires a series of simulation outputs saved at sufficiently fine time spacing\footnote{\citet{Han.2012} found that using a time spacing of $\Delta\ln a = 0.2$ leads to a bound subhalo mass function whose amplitude is 5\% of the $z = 2.1$ value obtained using a finer time spacing. The convergence improves at lower redshifts and with more frequent outputs. We independently confirmed that {\tt FLAMINGO} has sufficient outputs to provide "time"-converged mass functions, as there are 79 available outputs with an average spacing of $\langle \Delta \ln a\rangle = 0.015$.}, each containing information on the phase-space distribution and FoF group memberships of particles. Each output is then analysed consecutively, from early to late times. As structures form, they are assigned unique IDs used throughout the remainder of the simulation, whose time-persistence facilitates the retrieval of the evolution of the subhalo along its main progenitor branch. At each output, every pre-existing subhalo is assigned to a given FoF group (\S\ref{fof_host_finding}), and a central subhalo is chosen per FoF group based on a mass and orbital kinetic energy criterion (\S\ref{central_identification}). The classification of whether a subhalo is a central or not determines which physical mechanisms are eligible to affect its mass from one output to the next (\S\ref{diffuse_mass_accretion}, \S\ref{unbinding_step}, \S\ref{merging_step} and \S\ref{non_hierarchical_step}), which reflects the expected evolutionary differences between central and satellite subhaloes. 

\subsubsection{Assigning host haloes}\label{fof_host_finding}

The first step consists of assigning a host FoF group to every pre-existing subhalo. Each subhalo has an associated set of $N_{\mathrm{tracer}}$ (10 by default) tracer particles, whose type can be constrained to prevent issues when tracking subhaloes in hydrodynamical simulations. For example, allowing gas to be the tracer of a subhalo makes it susceptible to issues arising from gas blowouts (see Appendix \ref{Appendix:incorrect_tracers}). By default, {\HBTHERONS} uses stars and dark matter particles as subhalo tracers, because of their collisionless and time-persistent nature in most cosmological simulations. 

The FoF membership of each tracer particle is used to score a list of candidate hosts. Each individual tracer particle is weighted as follows \citep[e.g.][]{Springel.2021}:
\begin{equation}\label{host-finding-equation}
    w_{i} = \dfrac{1}{1 + \sqrt{r_{i}}}\,,
\end{equation}
where $r_{i}$ is its bound ranking in the previous output. The sum is done over the $N_{\mathrm{host-finding}}$ (10 by default) most bound particle tracers. The chosen host halo is the FoF group with the highest score out of the candidate list. Once all the pre-existing subhaloes have been assigned to hosts, three distinct possibilities can occur.

First, FoF groups without any assigned subhaloes, and hence no previously existing structures associated to it, are considered potential seeds for new subhaloes. In other words, all of the particles belonging to that FoF group are grouped into a single source subhalo. If the candidate subhalo is identified as self-bound, a new central subhalo with a unique ID (TrackID) is created.

Second, the groups that host one or more previously-existing central subhaloes require a decision concerning which subhalo remains as the central (see \S\ref{central_identification}).

Third, some subhaloes may not be found in any FoF group at all. This occurs primarily near the resolution limit of the simulation, as FoF groups can fragment due to noise and no longer be above the particle number threshold to be found. However, as the particles were previously tagged as being part of a subhalo, {\HBTHERONS} is able to check if the particles of the now-fragmented FoF remain self-bound. If so, these subhaloes are denoted as hostless, and typically have fewer than 50 bound particles in {\tt FLAMINGO}. 

\subsubsection{Identification of the central subhalo}\label{central_identification}

During the assignment of host FoF groups, two or more subhaloes that were centrals in the previous output can be found within the same host halo. This indicates an accretion event has occurred, and a choice about which subhalo is the central needs to be made. Making a sensible choice is important, as it determines which subhalo acquires the particles that have been accreted by its host FoF and that do not already belong to satellites, i.e. diffuse mass accretion at a given resolution (see \S\ref{diffuse_mass_accretion}). 

To decide which subhalo is the central, all subhaloes in the current FoF group are ranked according to their previous bound mass. The previous bound mass of the most massive subhalo in the previous output, $M_{0}$, is used to define the following threshold: 
\begin{equation}
    M_{\mathrm{candidate}} = f_{\mathrm{major}}M_{0} \,,
\end{equation}
where $f_{\mathrm{major}} = 0.8$ by default. Only subhaloes whose bound masses in the previous output were above this threshold are considered central candidates. Thus, the choice is trivial if only minor accretions occur.

If more than one subhalo satisfies the above mass threshold, the specific orbital kinetic energy of each candidate is computed in the centre of mass reference frame of the host FoF group. The subhalo candidate with the lowest value, and hence moving most like its host FoF group, is chosen as central\footnote{Note that if the subhalo chosen to be the central is not the most massive subhalo in the FoF group after unbinding (\S\ref{unbinding_step}), the most massive subhalo of the group is changed to be the central. However, diffuse mass accretion (\S\ref{diffuse_mass_accretion}) at that output will reflect the original central choice.}. Updating which subhaloes are assigned to be satellites also propagates to the hierarchy of structures that they brought with them, e.g. their satellites become satellites of satellites.
 
We note that sensible values for $f_{\rm major}$ are of $\mathcal{O}(0.1)$. Values much lower than 0.1 make low mass subhaloes become candidate centrals. As they dominate in number over massive subhaloes, and the central is chosen out of the candidates based on their \textit{specific} orbital kinetic energy, it is very likely that a low-mass subhalo is incorrectly chosen as a central instead of a more suitable massive subhalo. Consequently, large mass fluctuations may occur, as a low-mass subhalo suddenly acquires most of the mass in the FoF group. On the other hand, using $f_{\rm major} = 1$ is too restrictive, as only the most massive subhalo in the previous output is a candidate central. Since mass is assigned in an exclusive manner, it then becomes a matter of which subhaloes have a more massive satellite population prior to being in the same FoF, even if they have similar spherical overdensity masses.

\subsubsection{Diffuse mass accretion}\label{diffuse_mass_accretion}

We make the physically-motivated assumption that only central subhaloes can grow in mass through the accretion of surrounding diffuse mass, as the evolution of satellites is more constrained and most will have ceased growing or are even losing mass. Consequently, the source subhalo of central subhaloes can grow by accreting particles that have been newly acquired by their host FoF group, and which are not already associated to the source subhalo of resolved subhaloes. For the simplest case of a single (central) subhalo in a FoF group, all of the particles in the FoF group are initially part of its source. 

\subsubsection{Unbinding and mass stripping}\label{unbinding_step}

The hierarchy of structures within each FoF group is analysed recursively from the bottom of the subhalo hierarchy. In other words, the subhaloes without any children are subject to unbinding first, at which point their parents can be analysed. The process is repeated until the central subhalo of the FoF is analysed. This depth-first\footnote{In {\HBTHERONS}, depth refers to the number of hierarchical connections that a given subhalo is away from the central subhalo of its host FoF group. A central subhalo thus has a depth of 0, its satellites a depth of 1, the satellites of satellites a depth of 2, etc.} approach allows for the possibility of parents gaining bound mass that is not bound to its children subhaloes. In other words, any central or satellite subhalo can accrete mass that is stripped from the subhaloes in their own satellite system. 

The unbinding of a single subhalo is done iteratively, whereby its centre of mass position and velocity are updated based on the particles deemed to be bound in any given iteration. To speed up the process of unbinding, which is generally the most expensive step in the subhalo finding process, the particle distribution is subsampled if the subhalo contains more than $N_{\mathrm{subsample}}$ particles. By default, $N_{\mathrm{subsample}}$ = 1000, which provides results very close to not subsampling the particles at all whilst reducing the computational cost (by a factor of $\approx10$ when using the default parameters for the {\midres} simulation). While the exact value of $N_{\mathrm{subsample}}$ is user-adjustable, we caution against using values lower than the default, as the code becomes increasingly likely to choose a subset of particles that do not represent the true mass distribution of the subhalo, leading to its premature disruption. Once the random subsample of particles is chosen, the particle masses are appropriately upscaled to ensure mass conservation, either on a total subhalo mass or a total particle-type mass level. The subsampling of specific particles types can also be disabled, e.g. for black hole particles given their low numbers and already large masses compared to the rest of the particles. 

In practice, particle subsampling can lead to a few-percent differences in the subhalo bound mass functions at the high mass end between different runs of {\HBTHERONS} on the same simulation. This reflects different random numbers being generated when doing the particle subsampling. Additionally, the threshold used to determine whether the iterative unbinding has converged can be changed to reduce the number of iterations. The threshold used to stop the iterations is defined based on the ratio of bound particles ($N_{\mathrm{bound}}$) between two consecutive iterations: $N^{i+1}_{\mathrm{bound}} / N^{i}_{\mathrm{bound}} \geq f_{\mathrm{converge}}$, where $f_{\mathrm{converge}} = 0.995$ by default. We recommend using values in the range of $0.9 \leq f_{\mathrm{converge}} < 1$, as the condition is strict enough to provide similar results to $f_{\mathrm{converge}} = 1$, but cuts down on the number of unbinding iterations and therefore computational cost. Using a value that is too small results in too few iterations to have converged bound masses, meaning that some physically unbound particles are misclassified as bound. Since mass is assigned exclusively and in a depth-first manner, satellite subhaloes would then become too massive, while central subhaloes would artificially miss some of their mass. 

As mentioned previously, each satellite subhalo has its own source subhalo that reflects the source subhalo it had when it was last a central subhalo. Despite the possibility of accreting mass from its own children subhaloes, more bound mass is typically lost than gained. Therefore, the source component is generally larger than the set of bound particles. Although tracking particles in this manner is crucial to separate satellite subhaloes from the background of their host, it can lead to failures in subhalo finding without adequate care. For example, if the source subhalo is dominated by particles that are truly unbound, the centre of mass estimate can be offset from the true centre of the subhalo. Consequently, the kinetic energies in its core will be biased high, as the Hubble flow contribution (proportional to the distance to the centre of mass) is larger. This can lead to an apparently unbound core, even though it is purely driven by an incorrect choice of reference frame. 

To prevent such instabilities during unbinding, {\HBTHERONS} limits the number of particles that any source subhalo can have ($N^{\mathrm{max}}_{\mathrm{source}} \equiv f^{\mathrm{max}}_{\mathrm{source}}N_{\mathrm{bound}}$). By default, it equals three times the number of bound particles, i.e. $f^{\mathrm{max}}_{\mathrm{source}} = 3$. After the unbinding step, the most unbound particles of the source subhalo are removed until $N_{\mathrm{source}} \leq N^{\mathrm{max}}_{\mathrm{source}}$. The particles removed from the source subhalo are accreted by the source of its parent subhalo, if the subhalo has one. The default value we choose is sufficiently large to allow for the re-accretion of particles after they have been momentarily found as unbound. Using too small a value, e.g. $f^{\mathrm{max}}_{\mathrm{source}} \approx 1$, results in the secular loss of mass from satellites due to the continuous removal of particles that are found to be unbound from one snapshot to the next (e.g. as shown for {\tt Subfind-HBT} in Fig.~\ref{figure:subfind_hbt}).

Lastly, if a previously resolved subhalo no longer has enough bound particles to be above the threshold to be classified as self-bound ($N^{\mathrm{min}}_{\mathrm{bound}}$; 20 by default) or has an insufficient number of tracer particles ($N^{\mathrm{min}}_{\mathrm{tracer}}$; 10 by default), it is deemed disrupted and becomes an orphan. Based on our choice to determine when subhalos disrupt, we recommend to keep the value of $N^{\rm min}_{\rm bound}$ small, in the range $10 \leq N^{\rm min}_{\rm bound} \leq 20$. Higher values result in the `disruption' of subhaloes that are truly physical (i.e. they remain self-bound for many snapshots after their supposed `disruption').

The position and velocity of orphan subhaloes, which are often used in the context of semi-analytical models of galaxy formation \citep[e.g.][]{Kitzbichler.2008, Guo.2014}, are subsequently tracked by their most bound tracer particle when the subhalo was last resolved. Note that if the tracer particle is found in a FoF host without currently resolved subhaloes, the particles are acquired by the source of the orphan subhalo. This means that if these particles are found to be bound, orphans can re-appear as fully resolved subhaloes in the simulation.

\subsubsection{Merging}\label{merging_step}

Pairs of structures with similar masses can merge as a consequence of dynamical friction, whereby the least massive subhalo loses orbital energy and eventually becomes deeply embedded in the core of its host. This process is sometimes insufficient to fully disrupt the sinking subhalo \footnote{If the sinking process does not sufficiently disturb the subhalo so as to disrupt it, checking its self-boundness is similar to checking the self-boundness of a subset of the most bound particles of its host halo. Hence, it is likely to be found as self-bound even though it no longer represents a separate physical structure.}, which for a tracking subhalo finder like {\HBTHERONS} can result in persisting subhaloes that heavily overlap in phase-space. This is not a problem for other methods of subhalo finding, since density peaks would coalesce into one and configuration and phase-space finders would in principle not find two separate subhaloes. 

To prevent this unwanted behaviour, which would lead to artificially high numbers of objects in the central regions of subhaloes, {\HBTHERONS} merges subhaloes that are within a threshold phase-space offset. This offset is defined for two pairs of subhaloes within the same hierarchical tree branch, $i,j$, as: 
\begin{equation}
    \Delta_{ij} = \dfrac{|\Vec{x}_{i} - \Vec{x}_{j}|}{\sigma_{i, x}} + \dfrac{|\Vec{v}_{i} - \Vec{v}_{j}|}{\sigma_{i,v}}
    \label{equation:phase_space_offset}
\end{equation}
Here, $\sigma_{i,x}$ and $\sigma_{i,v}$ are, respectively, the position and velocity dispersion of the $N^{\rm min}_{\rm core}$ (20 by default) most bound tracer particles of subhalo $i$, which are also used to compute its position $\Vec{x}_{i}$ and velocity $\Vec{v}_{i}$. The position and velocity of subhalo $j$ are also estimated using its $N^{\rm min}_{\rm core}$ most bound tracer particles. The number of tracer particles used to estimate the properties of the subhalo core should be $N^{\rm min}_{\rm core} \approx N^{\rm min}_{\rm bound}$. Using a value much larger than $N^{\rm min}_{\rm bound}$ means that the measured phase-space dispersion is no longer representative of the core of the subhalo, but of the whole subhalo. This can lead to premature merging between pairs of well resolved subhaloes that are neither nearby in space nor stationary with respect to each other.

We compute the phase-space offset in both directions, i.e. one time using the measured dispersion of subhalo $i$ and another time using the dispersion of subhalo $j$, and take the minimum of the two values: $\Delta_{\rm min} = \mathrm{min}(\Delta_{ij},\Delta_{ji})$. Usually, the more massive subhalo of the pair has a larger phase-space dispersion than the less massive one, but this is not necessarily the case when both subhaloes are relatively poorly resolved ($N_{\mathrm{bound}}\lesssim 100$). 

This condition is checked between pairs of subhaloes contained in the same hierarchical tree branch, e.g. TrackIDs 0, 3 and 1 in output 4 of Fig.~\ref{figure:hbt_general_diagram}. If $\Delta_{\rm min} \leq 2$, the subhalo located deeper in the hierarchy is marked as merged, and all of the particles in its source subhalo are acquired by the more massive one. This is followed by a new unbinding of the subhalo that accreted the mass, to update its bound mass based on the gravitational potential that includes the newly accreted mass. Consequently, both central and satellite subhaloes are eligible to grow in mass if they overlap with a subhalo contained in their own satellite system. 

As merger checks can only happen between subhaloes that share an explicit hierarchical connection, not all potential satellite-satellite merger configurations are handled by {\HBTHERONS}. For example, the current approach is unable to account for mergers between satellite subhaloes that were independently accreted as separate centrals, but subsequently encounter each other at sufficiently low velocities so as to form a bound system. The fact that the subhaloes were accreted separately means that no hierarchical relation links them, and hence no merger checks occur between them. Nonetheless, this is minute fraction of satellite-satellite mergers, which are themselves not very common \citep{Bahe.2019}. Future work will modify {\HBTHERONS} to be able to handle such encounters.

\subsubsection{Non-hierarchical transfer of mass}\label{non_hierarchical_step}

An important caveat to the mass transfer channels discussed above is that they are implicitly designed to be hierarchical in nature, e.g. low mass subhaloes can only transfer mass to more massive subhaloes, and not vice versa. This underlying assumption works well for particles that are collisionless in nature, like dark matter or stars. However, gas is inherently collisional, meaning that the assumption of purely hierarchical mass transfer may not work as well in simulations that contain gas particles. Not accounting for the possibility of non-hierarchical gas exchange would entail that gas particles (or eventually star particles formed from them) will (incorrectly) remain assigned to their original subhalo, rather than the one in which they now physically reside.

To avoid this problem, we perform a re-attachment of gas particles that does not assume anything about the underlying hierarchical connections between subhaloes, and hence allows for the transfer of mass between arbitrary subhaloes within a FoF groups (as long as the local physical conditions are conducive for it to happen). In short, for each gas particle in a FoF group we find their 10 nearest tracer particles. We then check whether any of those neighbours belong to the same TrackID as the gas particle itself. If none do, we remove the gas particle from the source subhalo of its original TrackID, and re-attach it to the source subhalo of the TrackID with the largest representation among the 10 nearest tracer particles. This approach means that only gas particles that are truly far away from their supposed subhalo are allowed to be re-attached. In practice, this step is done \textit{before} before any subhalo is subject to unbinding, so that the re-attached gas can become bound to their newly assigned TrackID. More details about the re-attachment are provided in \S\ref{appendix:gas_reattaching}

\subsection{Subfind}

In this work, we use the {\SUBFIND} version that is available within the public version of {\tt GADGET-4} \citep{Springel.2021}. To identify subhalo candidates within a FoF group, its member particles are used to estimate the local total mass density field via adaptive kernel estimation. The resulting density peaks are seeds for potential subhaloes, which grow through an excursion set algorithm until they reach saddle points that connect them to neighbouring peaks.

The self-boundness of each candidate subhalo is checked by an iterative unbinding process. All particles are considered initially, with subsequent iterations only using those deemed as bound for gravitational potential calculations. In each step, the centre of mass reference frame of bound particles is used, and only a fraction of particles can be removed at a time. Iterations stop whenever the subhalo is deemed as disrupted (20 particles by default), or the number of bound particles has converged ($N^{i+1}_{\mathrm{bound}} / N^{i}_{\mathrm{bound}} = 1$).

\subsection{ROCKSTAR}

The position and velocity dispersion of particles ($\sigma_{x}$, $\sigma_{v}$) associated to the initial 3D FoF group defines a length and velocity scale used to define a phase-space distance between two particles $i,j$:
\begin{equation}\label{distance_metric_rockstar_1}
    d(\vec{x}_{i},\vec{v}_{i};\vec{x}_{j},\vec{v}_{j}) = \sqrt{\dfrac{|\vec{x}_{i} - \vec{x}_{j}|^{2}}{\sigma^{2}_{x}} + \dfrac{|\vec{v}_{i} - \vec{v}_{j}|^{2}}{\sigma^{2}_{v}}}\,.
\end{equation}
Particles are linked if they are within a certain phase-space distance. The threshold value is found by requiring that a minimum specified fraction of particles (0.7 by default) have at least one neighbour. This process is repeated for each subsequent 6D FoF group that has more than a predefined number of associated particles (10 by default), to find substructures within substructures. Particles in the 3D FoF are assigned to the closest seed subhalo in phase-space based on a distance metric similar to equation (\ref{distance_metric_rockstar_1}), where the position normalisation $\sigma_{x}$ is replaced by a characteristic radius $r_{0} \equiv V_{\mathrm{max}}t_{\mathrm{vir}} = V_{\mathrm{max}}(4\pi G\rho_{\mathrm{vir}} /3)^{-1/2}$. Here, $\rho_{\mathrm{vir}}$ is the virial density as defined in \citet{Bryan.1998}. 

Contrary to the other subhalo finders used in this work, {\ROCKSTAR} only does a single unbinding iteration, and keeps all subhaloes whose 6D FoF group contains more than a specified fraction of bound particles (0.5 by default). This approach is computationally cheap, although an iterative unbinding approach is able to cut down the number of spurious subhaloes and identify otherwise unresolved ones, as discussed in \citet{Griffen.2016}. In this work, we use the default of a single unbinding pass, to reflect the choice made in other large-scale cosmological simulations, such as the Euclid flagship simulations \citep{EuclidCollaboration.2024}. Another difference relative to the other three subhalo finders is that {\ROCKSTAR} includes particles that belong to parent subhaloes in the calculation of particle gravitational potentials if the subhalo being analysed is undergoing a major merger. The most important difference for the purposes of this study is that {\ROCKSTAR} allows particles to be bound to multiple subhaloes at once (i.e. it uses an \textit{inclusive} mass assignment), whereas the other three subhalo finders only allow any given particle to be bound to one subhalo (i.e. \textit{exclusive} mass assignment). We outline how we address this difference in \S\ref{section:soap}.

Another choice concerns how to run {\ROCKSTAR} on hydrodynamical simulations. Some studies do the 6D FoF linking using all particle types, which is the default behaviour of {\tt ROCKSTAR-galaxies} \citep[e.g.][]{Hahn.2017, VillaescusaNavarro.2023}. Others restrict the FoF linking to be done on dark matter particles first, and subsequently rely on a post-processing step to assign stellar masses to each subhalo \citep[e.g.][]{Necib.2019, Jung.2024}. In the main analysis of study, we use all particle types except black holes\footnote{The inclusion of black holes resulted in structures that spanned over 10~Mpc in length, which no longer happened once we removed them from the FoF linking.} during linking and unbinding, as per the recommended way of using {\tt ROCKSTAR-galaxies}  (Behroozi; private communication). The alternative would neglect the contribution of baryons to the gravitational potential, which could lead to identifying baryon-dominated objects as less massive or even disrupted. Nonetheless, we provide a brief comparison in Appendix \ref{Appendix:LinkingParticleTypes} of how basic summary statistics change between these two analysis choices.

\subsection{VELOCIraptor}

{\VELOCIRAPTOR} uses a different approach to identify structure in phase-space compared to {\ROCKSTAR}. The core idea behind its subhalo finding implementation is that the centres of subhaloes are not only peaks in the density field, but also have colder velocity distributions than the surrounding background particles. This means their cores appear as peaks in both the spatial and velocity density fields.

To identify the velocity peaks, {\VELOCIRAPTOR} first estimates the expected \textit{background} velocity density distribution for each particle, $f_{\mathrm{bg}}$. For this purpose, it assumes that the background velocity density follows a multivariate Gaussian distribution \citep[e.g.][]{Vogelsberger.2009}:
\begin{equation}
    f_{\mathrm{bg}} = \dfrac{\exp\Big[-\dfrac{1}{2}\Delta\vec{v}\,\langle\Sigma_{v}\rangle^{-1} \Delta\vec{v}\Big]}{(2\pi)^{3/2}\,|\langle\Sigma_{v}\rangle|^{1/2}}\,.
\end{equation}
Here, $\langle \Sigma_{v} \rangle$ is the mean local velocity dispersion tensor, and $\Delta \vec{v}$ is the velocity of the particle relative to the mean background velocity, $\vec{v}_{i} - \langle{\vec{v}\rangle}$. Both $\langle{\vec{v}\rangle}$ and  $\langle \Sigma_{v} \rangle$ are computed using a mass-weighted average of all particles within a rectangular cuboid built using a kD-tree algorithm. The values between adjacent cells are subsequently interpolated onto the positions of individual particles using inverse distance weighting. 

The \textit{local} velocity density of an individual particle, $f_{i}$, is computed using Epanechnikov kernel smoothing \citep{Sharma.2006} with its closest neighbouring particles in velocity, themselves chosen from its closest spatially neighbouring particles. The ratio between local and background velocity densities, $\mathcal{R} = \ln f_{i} / f_{\mathrm{bg}}$, is proportional to the likelihood that the particle is associated to a dense and cold structure, e.g. a subhalo. The actual likelihood is computed as:
\begin{equation}
    \mathcal{L}_{i} \equiv \dfrac{\mathcal{R}_{i} - \langle \mathcal{R}\rangle}{\sigma_{\mathcal{R}}} \,,
\end{equation}
where $\langle \mathcal{R}\rangle$ and $\sigma_{\mathcal{R}}$ are the mean and standard deviation of a skewed Gaussian fitted to the binned distribution of $\mathcal{R}$.

Particles are then grouped by first applying a cut on the minimum required value of $\mathcal{L}_{i}$, by default 2.5, and then linking them in phase space using a FoF-like algorithm. The spatial linking is done in a similar way as the 3D FoF algorithm, but the linking length is halved and hence targets higher overdensities. The linking in velocity space is based on the velocity ratio between particles and the angle between velocity vectors. Finally, all candidate subhaloes are subject to unbinding. If the number of particles within a subhalo is sufficiently high, this whole process is iteratively repeated on its particle members until the resulting subhaloes have fewer than a threshold number of particles (800 by default).

\subsection{Halo and subhalo properties}\label{section:soap}

Different subhalo finders calculate (sub)halo properties under different sets of assumptions and choices. For example, spherical overdensities are computed using only bound particles in {\ROCKSTAR} and {\HBTHERONS}, whereas {\SUBFIND} and {\VELOCIRAPTOR} include all particles regardless of subhalo membership. Moreover {\ROCKSTAR} assigns mass in an \textit{inclusive} manner, whereas all the other subhalo finders used in this work assign mass \textit{exclusively}. This drives differences in population statistics purely due to operational definitions, rather than the performance of each (sub)halo finder. Therefore, in order to compare different structure finding algorithms, the definition of mass and the computation of properties need to be as consistent as possible.

To do so, we modified {\ROCKSTAR} so that it assigns masses to subhaloes exclusively, similar to what is done in the other subhalo finders we compare in this work. To emphasise the difference in mass assignment of this modified version of {\ROCKSTAR} from its default, we refer to it as {\tt ROCKSTAR (Exc.)} whenever it is relevant to the plots being shown. We also compute properties for {\HBTHERONS}, {\SUBFIND} and {\VELOCIRAPTOR} subhaloes using {\SOAP}\footnote{Available at \href{https://github.com/SWIFTSIM/SOAP}{https://github.com/SWIFTSIM/SOAP}} (\textbf{S}pherical \textbf{O}verdensity and \textbf{A}perture \textbf{P}rocessor; \citealt{McGibbon.2025}). {\SOAP} is a Python package that computes a variety of subhalo properties in different spherical and projected apertures, both with and without taking into account the bound membership of particles. In short, {\SOAP} loads the centres of subhaloes and memberships of particles as given by each subhalo finder\footnote{{\ROCKSTAR} does not provide the bound membership of particles in its outputs, meaning that we only use {\SOAP} for the values of $M_{\mathrm{200c}}$. However, bound memberships are only required for bound mass and maximum circular velocity calculations. We have verified that the relevant routines in {\SOAP} and {\ROCKSTAR} are identical.}, and computes the properties of interest using the same routines, hence circumventing additional choices made by each particular halo finding algorithm. We explore how the changes we made to the mass assignment in {\ROCKSTAR} and how {\SOAP}-computed $M_{200\mathrm{c}}$ values differ from those provided by subhalo finders in Appendix \ref{Appendix:MassDefinitions}. 

In this study, we calculate three different subhalo properties: their bound mass ($M_{\mathrm{bound}}$), the maximum circular velocity ($V_{\mathrm{max}}$) and the virial mass ($M_{\mathrm{200c}}$) if the subhalo is a central. The last two properties require passing a subhalo centre to {\SOAP}, as they use spherical apertures to compute their values. The definition of a subhalo centre differs among codes. {\HBTHERONS} uses the position of the most bound particle. {\SUBFIND} and {\VELOCIRAPTOR} use the position of the particle with the largest value of the gravitational potential energy. Lastly, {\ROCKSTAR} uses the centre of mass position of the particles that are bound to the subhalo.

The bound mass is simply the sum of the mass of all particles bound to a subhalo. We note that we include the thermal energy of gas when subjecting subhaloes to unbinding across every subhalo finder. The value of $V_{\mathrm{max}}$ corresponds to the peak value of the circular velocity rotation curve, which is defined as: 
\begin{equation}\label{equation:maximum_circular_velocity}
    V_{\mathrm{circ}}(r) = \sqrt{\dfrac{GM(<r)}{r} }\,.
\end{equation}
Here, only particles that are bound to the subhalo contribute to the sum at a given radius. We also apply the constraint that no particles can be less than a gravitational softening length ($\epsilon$) away from the centre of the subhalo. In other words, we constrain $R_{\mathrm{max}} \geq \epsilon$. This removes unrealistically high values of $V_{\mathrm{max}}$ driven by particle sampling noise in the inner core of subhaloes.
 
We define the value of $M_{\mathrm{200c}}$ by summing up the total mass within a spherical region whose mean density is 200 times the critical density of the Universe, i.e.
\begin{equation}
    M_{\mathrm{200c}} = \dfrac{4\pi R^{3}_{\mathrm{200c}}}{3}\times200\rho_{\mathrm{crit}}.
\end{equation}
The value is found by iteratively increasing the radius of a sphere placed on the subhalo centre until the mean density is less than $200\rho_{\mathrm{crit}}$. The value of $R_{\mathrm{200c}}$ is interpolated between the particles adjacent to where the drop below $200\rho_{\mathrm{crit}}$ occurs, to obtain the exact target overdensity. Note that we only compute $M_{\mathrm{200c}}$ for central subhaloes, and include the mass contribution of all particles within the aperture, regardless of whether the enclosed particles are bound to the central or not. This means that, despite not using neutrinos when identifying (sub)haloes, we account for their mass contribution enclosed within the $R_{200\mathrm{c}}$ of each halo.

 \begin{figure*}
    \centering
    \includegraphics[width=1\textwidth,keepaspectratio]{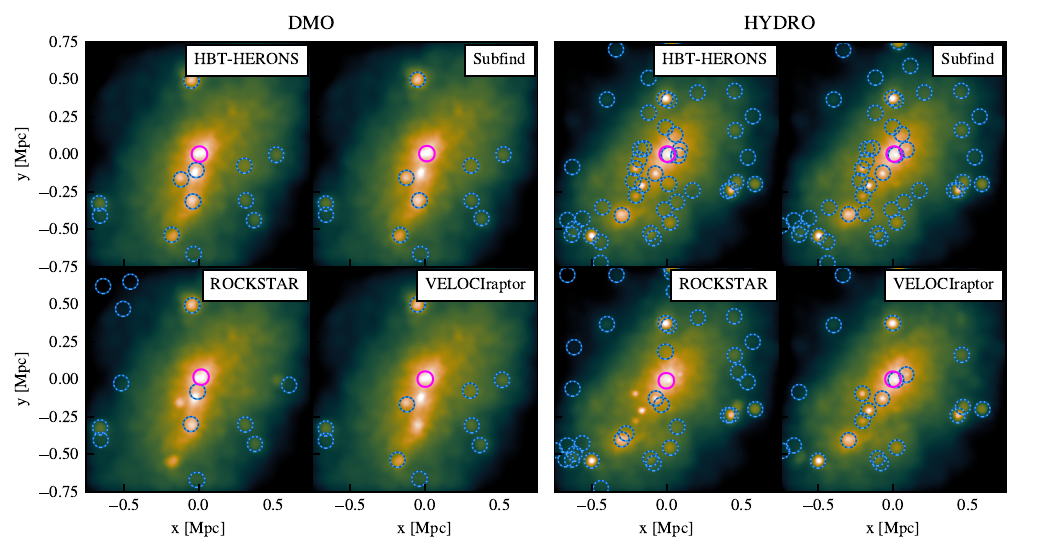}
    \caption{Projection of the dark matter density along a volume of side-length $1.5~\mathrm{Mpc}$ and depth $0.4~\mathrm{Mpc}$. The centre was chosen to be the {\HBTHERONS}-identified centre of a $M_{\mathrm{200c}} = 2.3 \times 10^{15} \Msun$ halo, the second most massive of the {\midres} DMO simulation. Its counterpart in the hydrodynamical version of the box is shown in the right set of images. The colour map scale is consistent across all panels. The circles indicate the location of resolved subhaloes found by each subhalo finder, as indicated in the top right corner of each image. The colour of each circle indicates whether they are centrals (magenta) or satellites (blue dashed). }
    \label{figure:image_example}
\end{figure*}

\section{Results}\label{section:results}

We start this section by illustrating how the choice of subhalo finder affects the structures found in the vicinity of a cluster-mass halo, in its DMO and hydrodynamical versions (\S\ref{Section:example}). We then proceed to discuss how the differences propagate to three summary statistics commonly used in cosmology: halo and subhalo mass functions (\S\ref{section:halo_mass_functions}), radial distributions of subhaloes around haloes (\S\ref{section:radial_distributions}) and the correlation functions of subhaloes based on a $V_{\mathrm{max}}$ selection (\S\ref{section:correlation_functions}).

Before discussing the results, we note how long it took to run each subhalo finder on the $z = 0$ snapshot of the DMO {\midres} box. In our tests, the slowest were {\SUBFIND} and {\VELOCIRAPTOR}, which took $\mathcal{O}(1000)$~CPU-h to run. The fastest were {\HBTHERONS}\footnote{The time for {\HBTHERONS} is for a single time output, but it needs to analyse all preceding ones. Nonetheless, its speed-up relative to the slowest subhalo finders means that it is sometimes cheaper to process all outputs with {\HBTHERONS} instead of just one with another subhalo finder. Additionally, if one is interested in the time evolution of subhaloes, all of the other subhalo finders would need to be run on all outputs too, as well as running a merger tree algorithm.} and {\ROCKSTAR}, taking only $\mathcal{O}(10)$~CPU-h to run. For for {\HBTHERONS}, {\SUBFIND} and {\VELOCIRAPTOR} the times taken to run on the hydrodynamical versions are about two to three times slower than for the corresponding DMO analysis. {\ROCKSTAR} actually took less time to run on the hydrodynamical simulation, likely due to the problems it suffers from when trying to find subhaloes when linking all particle types, as discussed below.

\subsection{Representative example}\label{Section:example}

We select a dark matter halo of  $M_{\mathrm{200c}} = 2.3 \times 10^{15} \Msun$ from the {\midres} box ($m_{\mathrm{dm}} = 5.65\times 10^{9} \Msun$ in the hydrodynamical version) that is undergoing a three-way merger at $z = 0$. The projected density of dark matter within a thin slice of its central region is shown in ~Fig.~\ref{figure:image_example}. Its clearly identifiable merging companions provide a good test-bench for how well the subhalo finders are able to decompose heavily overlapping subhaloes. Note that this test makes the implicit assumption that every visible density peak is caused by the presence of a subhalo.

The positions of all subhaloes with at least 20 bound particles are highlighted using circles. Their colours encode whether the subhaloes are centrals (magenta) or satellites (blue dashed). Focusing on the image from the DMO simulation, it is clear that differences in the satellite population exist across different subhalo finders. {\VELOCIRAPTOR} and {\SUBFIND} do not find the subhalo associated to the second most prominent density peak, which is located just below the central subhalo. Additionally, {\VELOCIRAPTOR} also misses a companion subhalo just below it. {\ROCKSTAR} finds the two subhaloes that {\VELOCIRAPTOR} and {\SUBFIND} miss, but fails to find two subhaloes in the lower left quadrant that are found by all other subhalo finders. {\HBTHERONS} is the only finder capable of identifying all of the visually detectable subhaloes in the central regions of this example.

Several subhaloes without visual counterparts are found in the outskirts of the image of the DMO version. These subhaloes are near the resolution limit of the simulation, but their existence and location is generally found to be consistent across subhalo finders. The outlier is {\ROCKSTAR}, which finds four additional subhaloes in the upper left quadrant that are not found by the other finders. Additionally, the subhalo it finds in the right hand side of the imaged region is clearly offset from the location that other finders identify. {\ROCKSTAR} also finds a centre for the central companion that is evidently offset from its associated density peak.

Turning to the hydrodynamical counterpart of the same halo, we see that the number of resolved subhaloes increases for all subhalo finders. The associated density peaks in the projected DM surface density maps are more prominent, suggesting that the central dark matter distribution has contracted and become denser in response to the presence of galaxies at their centres. Differences remain between subhalo finders, with {\HBTHERONS} and {\SUBFIND} finding the most subhaloes. There is a distinct lack of {\VELOCIRAPTOR} subhaloes in the outskirts, most of which are poorly sampled. {\ROCKSTAR} misses clearly identifiable subhaloes in the central regions, though we note that catalogue is likely influenced by our choice to run its 6D FoF group finding on all particle types. {\ROCKSTAR} provides the most extreme example of how the performance of subhalo finders can change between DMO and hydrodynamical simulations.

\subsection{Mass functions}\label{section:halo_mass_functions}

The (sub)halo mass function measures the number density of dark matter (sub)haloes as a function of a given mass definition. For many applications, the mass enclosed within a given spherical overdensity is used for central subhaloes. Alternative definitions rely on the mass that is self-bound, and are particularly useful when dealing with satellite subhaloes, which are embedded within a more massive central subhalo.

In this subsection, we explore how sensitive alternative definitions of masses are to the choice of (sub)halo finder. Since $V_{\mathrm{max}}$ is another commonly used metric to quantify the mass of subhaloes, we also include it in this analysis. Given that we are interested in the differences across finders, we always express the measured mass functions relative to what is found using {\HBTHERONS}.  We also discuss the convergence of these quantities across resolution levels when relevant.

\subsubsection{Halo mass functions}\label{Section:M200_mass_functions}

\begin{figure}
    \centering
    \includegraphics{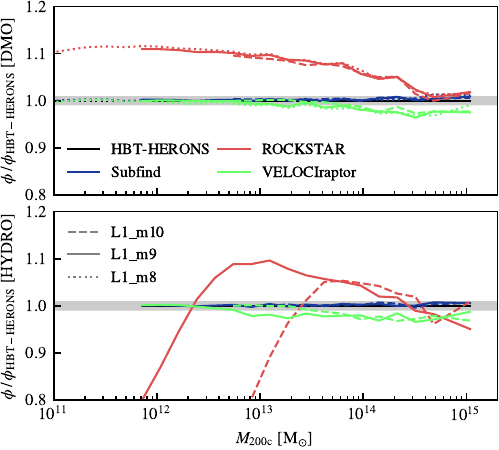}
    \caption{Number density of central subhaloes as a function of their associated $M_{\mathrm{200c}}$ mass. The top panel is for the DMO version of the box, and the bottom panel for the HYDRO version. Each coloured line indicates the mass function found using a different subhalo finder, which are expressed relative to the one found using {\HBTHERONS} at the same resolution. Different resolutions are indicated by the styles of the lines. Only the mass range corresponding to masses greater than the equivalent of 100 DMO dark matter particles (at a given resolution) is shown. The shaded region highlights differences smaller than 1\%.}
    \label{figure:M200_mass_function}
\end{figure}

We show the $z = 0$ $M_{\mathrm{200c}}$ mass functions across subhalo finders and expressed relative to {\HBTHERONS} in Fig.~\ref{figure:M200_mass_function}. The $M_{\mathrm{200c}}$ is computed by {\tt SOAP}, and it corresponds to the total mass enclosed within a spherical region that is 200 times denser than the critical density of the Universe. The only inputs required from each subhalo finder are which subhaloes are centrals, and where their centres are located. Therefore, one might expect this statistic to be less sensitive to the choice of subhalo finder, compared to those based on mass that rely on identifying which particles are bound to a given subhalo.

Focusing first on the DMO simulation in the top panel of Fig.~\ref{figure:M200_mass_function}, we see good agreement between {\HBTHERONS} and {\SUBFIND} irrespective of the resolution of the simulation. The median difference is much less than 1\% across the mass range we consider here. As both subhalo finders rely on the same spatial 3D FoF group catalogue, the similarities between their mass functions suggest that the centre locations of the centrals are, on average, similar. However, we note that there are still some differences in the location of individual haloes between {\HBTHERONS} and {\SUBFIND}, simply due to the choice of which subhalo within a FoF group is the central (see Appendix \ref{Appendix:VirialMassDifferences}). Despite this, the differences are not systematic, and are averaged out when measuring the $M_{\mathrm{200c}}$ mass function of the whole population of haloes.

Larger differences appear when comparing to {\VELOCIRAPTOR}. The mass function is systematically suppressed at the high-mass end by up to 3\% regardless of the resolution of the simulation. The difference relative to {\HBTHERONS} becomes progressively smaller at lower masses, and it is within 1\% below $M_{\mathrm{200c}} \approx 8\times10^{13}\,\Msun$. The aforementioned suppression at large masses is somewhat surprising, as {\VELOCIRAPTOR} assigns a single central subhalo per individual FoF group, similar to what {\HBTHERONS} and {\SUBFIND} do. As the FoF groups for these three finders are all built using a spatial FoF with a linking length of 0.2 times the mean interparticle separation, the difference likely stems from where the centres of the spherical overdensity apertures are placed.

We investigate the discrepancy with {\VELOCIRAPTOR} in more detail in Appendix \ref{Appendix:VirialMassDifferences}. In short, we bijectively match central subhaloes between subhalo finders, and measure the spatial offset and $M_{\mathrm{200c}}$ ratio relative to the values found in {\HBTHERONS}. We find that a fraction of high-mass central subhaloes are miscentred in {\VELOCIRAPTOR}. In such cases, the centre is incorrectly identified with the location of a satellite, leading to lower values of $M_{\mathrm{200c}}$ relative to all the other subhalo finders (Fig.~\ref{figure:mass_ratio_M200_matched}). Some examples have an $M_{\mathrm{200c}}$ whose value is lowered by an order of magnitude.

Despite the miscentering of {\VELOCIRAPTOR}, the most discrepant mass function is the one found using {\ROCKSTAR}. It is within 1\% of the value found by {\HBTHERONS} above $M_{\mathrm{200c}} \approx 4 \times 10^{14} \, \Msun$, but the differences quickly exacerbate towards lower masses. This results in a disagreement in the halo mass functions at the 10\%-level relative to all other subhalo finders considered in this work. Contrary to {\VELOCIRAPTOR}, the differences are not due to the positioning of central subhaloes (see Appendix \ref{Appendix:VirialMassDifferences}). The cause is the definition of central subhaloes, which differs from the other subhalo finders, as was also discussed in \citet{Euclid.2023}. 

\begin{figure}
    \centering
    \includegraphics{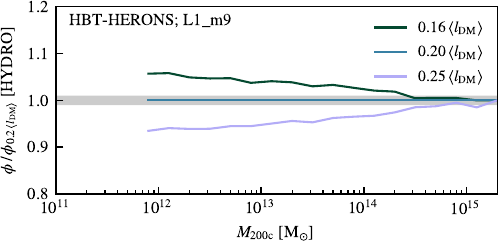}
    \caption{Similar to Fig.~\ref{figure:M200_mass_function}, but showing how the $M_{\mathrm{200c}}$ mass function changes for the hydrodynamical version of the {\tt L1\_m9} simulation when using different FoF linking lengths. The results obtained using shorter and longer linking lengths, which target higher and lower overdensities respectively,  are expressed relative to the fiducial value of $0.2\langle l_{\mathrm{DM}}\rangle$.}
    \label{figure:fof_linking_length_effect}
\end{figure}

{\HBTHERONS}, {\SUBFIND} and {\VELOCIRAPTOR} assign a single central subhalo per FoF group. This is done by identifying the most massive one ({\HBTHERONS} \& {\SUBFIND}) or the one found before doing refined searches in velocity space ({\VELOCIRAPTOR}). {\ROCKSTAR} does not use FoF information to decide which subhaloes are centrals. Instead, it relies on building a hierarchy of subhalo parent-children connections based on their spatial distribution and measured properties. Starting from the subhalo with the highest $V_{\mathrm{max}}$ in the simulation, it searches for all subhaloes within its virial radius ($R_{\mathrm{vir}}$). The virial radius is computed for all subhaloes, using only the particles bound to them and based on the spherical overdensity of \citet{Bryan.1998}. All neighbouring subhaloes whose $R_{\mathrm{vir}}$ and $V_{\mathrm{max}}$ values are smaller are assigned as its children. This process is repeated for all subhaloes in a decreasing $V_{\mathrm{max}}$ order, and the subhaloes which have no parents at the end of this process are deemed centrals.

In practice, this choice of central definition leads to many subhaloes that are identified as satellites in the other subhalo finders being classified as centrals in {\ROCKSTAR}. This in turn leads to more spherical overdensities being computed, and hence boosts the number density of subhaloes at fixed $M_{\mathrm{200c}}$. Of course, one may argue about which central definition is the more appropriate one. On the one hand, FoF-based central identification can be sensitive to spurious particle bridges. On the other hand, physical overdensities do not distinguish between bound and unbound material, and should arguably only be defined for virialised subhaloes that have not been stripped. However, bound-only variants are required to assign `virial' radii to satellite subhalos, as not excluding unbound particles may lead to their `virial' radii being much larger than their actual physical extent. In practice, the approach that {\ROCKSTAR} uses means that the assignment of virial radii becomes strongly dependent on the bound mass of a given subhalo, and hence how its particles are subject to unbinding.

Leaving aside the discussion of which definition of central is more appropriate, which in any case will depend on the application, the differences we see highlight the current freedom in choosing the definition of what constitutes a central. As we show here, this propagates directly to predicted summary statistics, and will bias inferences based on comparisons to observations. 

\begin{figure}
    \centering
    \includegraphics{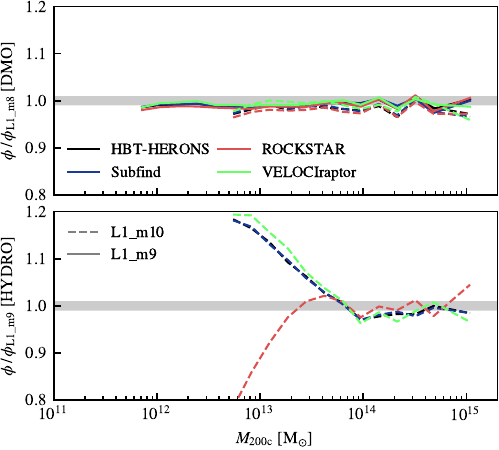}
    \caption{Similar to Fig.~\ref{figure:M200_mass_function}, but showing how the $M_{\mathrm{200c}}$ mass function converges across resolution levels using different subhalo finders. The mass functions for a given subhalo finder are normalised to the mass function obtained from the same subhalo finder using the highest resolution simulation available for that box ({\highres} for DMO and {\midres} for hydro).}
    \label{figure:M200_mass_function_convergence}
\end{figure}

We show the $M_{\mathrm{200c}}$ mass functions measured from the hydrodynamical simulations in the bottom panel of Fig.~\ref{figure:M200_mass_function}. We only computed them for the {\lowres} and {\midres} versions due to the increasingly high computational cost of running {\SUBFIND}. The overall trends are similar to those we found for the DMO simulation. However, the suppression of the halo mass function of {\VELOCIRAPTOR} now extends to even lower masses ($\approx 5\times10^{12}\,\Msun$). This happens because the miscentering worsens when baryons are included.  In essence, it results from more satellites missing from the {\VELOCIRAPTOR} catalogues, which means that the satellite particles are assigned to be part of the central subhalo instead. The relatively dense distribution of particles that are located around the centres of missing satellites are conducive to having the centre of the field halo being placed there.

The {\ROCKSTAR} mass function in the hydrodynamical simulation now has a suppression in the number density of both low- and high-mass haloes relative to {\HBTHERONS}. As discussed in \S\ref{Section:example}, this is likely caused by the issues that arise in halo finding when using all particle types for 6D FoF finding. To explore whether this behaviour remains when excluding baryons from the FoF calculations, we ran it using only DM particles. As shown in Appendix \ref{Appendix:LinkingParticleTypes}, in that case the $M_{\mathrm{200c}}$ functions show a similar behaviour to what was found in the DMO version, indicating that the linking of different particle types is the culprit of the differences.

Returning to the way in which centrals are defined, beyond the details of how to assign radii to subhaloes, the way in which spherical exclusivity is defined can affect the resulting mass functions by up to 6\% within {\ROCKSTAR} \citep{Garcia.2019}. Presumably, the measured $M_{\mathrm{200c}}$ mass function using subhalo finders that assign one central per FoF group can also be influenced by the choice of linking length used within the FoF algorithm. Changing the linking length will change the number of FoF groups, and hence the number of central subhaloes within the same simulation box. The need to choose a linking length constitutes an uncertainty in comparisons with observations, given that observed objects are obviously not identified using a FoF algorithm run on dark matter particles.

As the fiducial FoF linking length of 0.2 times the mean dark matter interparticle separation is not obviously the best a priori choice when comparing against observations, we test the effect of different linking lengths on the $M_{\mathrm{200c}}$ mass function. We do this by changing its value to be 0.16 and 0.25 of the mean dark matter interparticle separation $\langle l_{\mathrm{DM}}\rangle$. These values roughly correspond to a change of a factor of 2 in overdensity relative to the fiducial linking length of $0.2\langle l_{\mathrm{DM}}\rangle$. After constructing these new FoF catalogues for the hydrodynamical version of the {\midres} simulation, we run {\HBTHERONS} to identify the $z = 0$ central subhaloes and measure their $M_{\mathrm{200c}}$ masses using {\SOAP}. 

The resulting mass functions are shown in Fig.~\ref{figure:fof_linking_length_effect}. We see a clear mass-dependence on the measured mass function relative to the fiducial FoF linking choice. The shorter linking length (0.16$\langle l_{\mathrm{DM}}\rangle$), which targets higher overdensities, results in a steeper mass function than the fiducial linking length (0.2$\langle l_{\mathrm{DM}}\rangle$). This in turn yields an increase of 5.8\% in the number density of $\approx 10^{12}\,\Msun$ haloes, which is not as large of a difference compared to that caused by the central definition used in {\ROCKSTAR}. Conversely, the longer linking length of $0.25\langle l_{\mathrm{DM}}\rangle$ that selects lower overdensities, results in shallower mass functions. The abundance of $\approx 10^{12}\,\Msun$ haloes is instead lowered by $6.5\%$ relative to the fiducial linking length. The differences decrease with increasing halo mass, but only become smaller than 1 per cent for $M_{200\mathrm{c}} \geq 5\times 10^{14}\,\Msun$.  We note that, although this test was only done using {\HBTHERONS}, we expect our findings to be applicable to any subhalo finders that assign a single central subhalo per 3D FoF group. In this study, this applies to {\SUBFIND} and {\VELOCIRAPTOR}.

Another important aspect when considering the effect of subhalo finders on the measured $M_{\mathrm{200c}}$ mass functions is how well they converge across resolution levels. In principle, there is no guarantee that different subhalo finders converge to the result they find at the high resolution simulation at the same rate. In other words, one algorithm may perform particularly poorly in high resolution simulations relative to lower resolution counterparts, or vice versa. Thus, we conclude this subsection by exploring the convergence of the $M_{\mathrm{200c}}$ mass functions for different subhalo finders in Fig.~\ref{figure:M200_mass_function_convergence}. For this purpose, we normalise the mass functions found by each subhalo finder using the mass function that the same subhalo finder obtained for the highest resolution simulation with available catalogues ({\highres} for DMO and  {\midres} for the hydrodynamical version).

For the DMO simulations, the difference between {\lowres} (\midres) and {\highres} is about 2\% (1\%) across the explored mass range. All subhalo finders show essentially the same convergence rate, although {\VELOCIRAPTOR} and {\ROCKSTAR} bracket the extremes: {\VELOCIRAPTOR} agrees the best between different resolutions and {\ROCKSTAR} the worst. While differences are small for DMO, the picture changes dramatically for hydrodynamical simulations. There is now a substantial difference in the $M_{\mathrm{200c}}$ mass function below $\approx 10^{14}\,\Msun$ between the {\lowres} and {\midres} simulations, regardless of the subhalo finder. 

One caveat to consider when discussing convergence of the $M_{\mathrm{200c}}$ mass functions in the hydrodynamical {\tt FLAMINGO} simulations is whether the actual properties of galaxies and the circumgalactic medium remain consistent across resolutions. The differences could be not due to issues in the subhalo finding, but rather due to changing properties of the haloes themselves. As a matter of fact, the gas fractions within $R_{500c}$ are higher in {\lowres} (Fig. 10 of \citealt{Schaye.2023}) than in {\midres} below $M_{500c} \approx 3\times10^{13}\,\Msun$. This entails that more gas and dark matter is enclosed within the circumgalactic medium of the haloes below the aforementioned mass scale. As $M_{200} > M_{500}$, the mass scale where the differences start to appear is higher. Hence, the fact that the subhalo finders find an enhancement in {\lowres} relative to {\midres} is not entirely unexpected. 

Nonetheless, two differences unrelated to the above discussion are worth highlighting. The first one is that {\ROCKSTAR} finds fewer low mass haloes at higher resolution, a trend opposite to the one found for the other three subhalo finders. This behaviour is likely caused by the difficulties it encounters when running on hydrodynamical simulations. The second difference is that {\VELOCIRAPTOR} now appears to converge at a slightly slower rate than {\HBTHERONS} and {\SUBFIND}, both of which exhibit similar convergence properties. 

We conclude this subsection by noting that the differences we find in $M_{200\mathrm{c}}$ also extend to other spherical overdensity apertures, like $M_{\mathrm{500c}}$. As the choice of which halo mass function to use has important consequences for predictions relevant to ongoing and upcoming cosmological probes \citep[e.g. cluster counts;][]{Kugel.2025}, choosing one subhalo finder over the other will likely affect cosmological inferences. We also stress that the convergence properties of a given subhalo finder are insufficient to determine which subhalo finder provides `better' results. As we showed here, even though all subhalo finders share similar convergence properties for the DMO simulations, they converge to different answers. As we will show later on, the convergence to different answers also affects the subhalo bound mass functions and their radial distribution within a halo.

\begin{figure}
    \centering
    \includegraphics{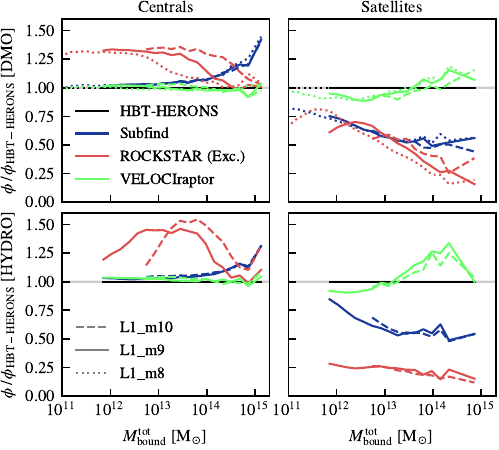}
    \caption{Number density of central (left column) and satellite (right column) subhaloes as a function of their bound total mass. The top panels show the results for the DMO versions of the simulation, and the bottom ones for the hydrodynamical versions. Each coloured line indicates the mass function found using a different subhalo finder, which are expressed relative to the one found using {\HBTHERONS} at the same resolution. Different resolutions are indicated by the styles of the lines. Only the mass range corresponding to masses greater than the equivalent of 100 DMO dark matter particles (at a given resolution) is shown. Note that for the purposes of this comparison, we modified {\ROCKSTAR} to assign masses exclusively, which is not the default way in which it is run. See Fig.~\ref{figure:bound_mass_function_inclusive} for the plot corresponding to an inclusive mass assignment, in which the differences become much larger between {\ROCKSTAR} and the other subhalo finders.
}
    \label{figure:bound_mass_function}
\end{figure}

\subsubsection{Bound subhalo mass functions}\label{section:bound_mass_functions}

We now shift our attention to the mass bound to subhaloes. We compute it by summing the masses of all particles bound to a given subhalo, which is information provided separately by each structure finder. As the bound mass is measurable for satellite and central subhaloes, we show both of their bound mass functions in Fig.~\ref{figure:bound_mass_function}.

The differences across subhalo finders in the bound mass functions are larger than for the $M_{\mathrm{200c}}$ mass functions. Instead of differing by 5\% to 10\%, differences now reach upwards of 50\% for both centrals and satellites. The larger changes are expected due to this mass definition depending more strongly on the assumptions and methodology that each finder uses, e.g. how well they separate satellites from centrals, whether they use an iterative unbinding scheme, etc.

Focusing first on the differences between {\SUBFIND} and {\HBTHERONS}, {\SUBFIND} finds more massive central subhaloes at $M^{\mathrm{tot}}_{\mathrm{bound}} \geq 10^{13}\Msun$, with the most massive being $40\%$ more massive in {\SUBFIND}. As the differences between them are much smaller for the $M_{\mathrm{200c}}$ mass functions, the differences are  driven by how much bound mass each central subhalo is assigned. Indeed, as we saw qualitatively in Fig.~\ref{figure:image_example}, {\HBTHERONS} is better at separating satellites from centrals than {\SUBFIND}. This is both because {\SUBFIND} misses more satellites, and because it is more likely to remove mass from their outskirts \citep{Han.2012}. This means that some of the mass which {\HBTHERONS} finds to be bound to satellites is instead assigned to centrals in {\SUBFIND}. Thus, for the same central, one typically expects to have more bound mass in {\SUBFIND}. Conversely, the {\SUBFIND} satellite bound mass functions are suppressed by up to 50\% at fixed $M_{\mathrm{bound}}$ relative to {\HBTHERONS}. The differences for centrals and satellites remain consistent with resolution and  whether or not the simulation uses hydrodynamics.

\begin{figure}
    \centering
    \includegraphics{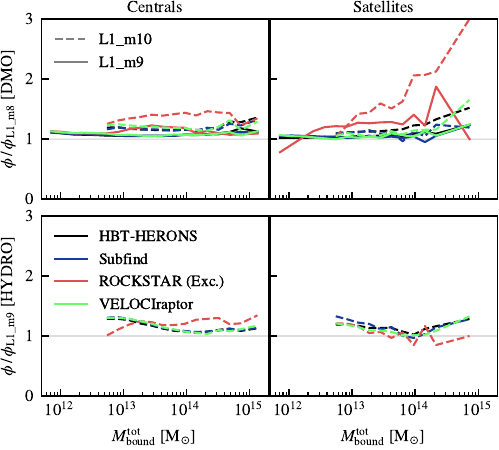}
    \caption{Similar to Fig.~\ref{figure:bound_mass_function}, but showing how the $M_{\mathrm{bound}}$ mass function converges across resolution levels using different subhalo finders. The mass functions for a given subhalo finder are normalised to the mass function obtained from the same subhalo finder using the highest resolution simulation available for that box ({\highres} for DMO and {\midres} for hydro). Note that for the purposes of this comparison, we modified {\ROCKSTAR} to assign masses exclusively, which is the way mass is assigned by the other subhalo finders, but it is not {\ROCKSTAR}'s default behaviour.}
    \label{figure:convergence_bound_mass_function}
\end{figure}

The bound mass function of centrals in {\VELOCIRAPTOR} is very similar the one found by {\HBTHERONS}. There is a minor trend with bound mass where, relative to {\HBTHERONS}, {\VELOCIRAPTOR} overestimates the mass function by 5\% at the low mass end and underestimates it by the same amount at the high mass end. For satellites, larger differences appear relative to {\HBTHERONS}, but they remain the most consistent out of the subhalo finders used in this work. Similar to the central bound mass function, there is a trend with mass, but it is now reversed and more evident. The suppression at the low mass end is partially caused by {\VELOCIRAPTOR} not identifying small subhaloes and assigning less mass to them. At higher masses, the disagreement is caused by the definition of central. In {\HBTHERONS}, the central is the most massive subhalo within a FoF group. This is not necessarily the case in {\VELOCIRAPTOR}, as the central is the top level structure in the 3D FoF before doing any phase-space searches. This means that satellites can be more massive than centrals in certain cases.  

{\ROCKSTAR} differs strongly in its central bound mass relative to the other subhalo finders, but the interpretation is complicated by its definition of central subhaloes differing from the other subhalo finders. A more consistent approach would modify the central population that {\ROCKSTAR} finds to be more in line with the other three finders \citep[e.g.][]{Gomez.2022}. With this difference in mind, the boost relative to {\HBTHERONS} at low masses is likely driven by satellites in {\HBTHERONS} being counted as centrals by {\ROCKSTAR}. Comparing the high mass end is more informative, as the contributing number of satellites is expected to be low for all subhalo finders. We see that it agrees well with {\HBTHERONS}, although the range of masses where both agree is sensitive to the resolution of the simulation. 

Shifting to the satellite population, {\ROCKSTAR} results in the most suppressed bound mass functions relative to {\HBTHERONS}. The satellite high mass end is the least populated out of the four finders, although this is in part due to {\ROCKSTAR} classifying more subhaloes as centrals. The number densities of low mass satellites are comparable to {\SUBFIND}, and both are lower by $\approx 25\%$ compared to {\HBTHERONS} and {\VELOCIRAPTOR} in the DMO simulations. For the hydrodynamical simulations, the suppression relative to {\HBTHERONS} is as severe as $\approx 75\%$, a consequence of including gas and stars to run 6D FoF. Overall, both the central and satellite bound mass functions depend more strongly on resolution for {\ROCKSTAR} than is the case for the rest of finders. Since this analysis is based on a modified version of {\ROCKSTAR} to enable exclusive mass assignment (i.e. a particle can only be bound to a single subhalo at a time), we re-do the same analysis based on the default version of {\ROCKSTAR} in Appendix \ref{Appendix:MassDefinitions}. Using {\ROCKSTAR}'s inclusive mass definition results in much larger differences in the number densities of subhaloes at a fixed $M_{\mathrm{bound}}$ relative to the other subhalo finders. The differences in the number densities increase as the bound mass of the subhalo increases, with $10^{15}\,\Msun$ subhaloes reaching number densities that can be higher by a factor of 10 in {\ROCKSTAR} than in the other subhalo finders.

Given the strong changes in the ratio of mass functions as a function of resolution for {\ROCKSTAR}, we proceed to look at the convergence of the bound mass functions. In other words, how well each subhalo finder approaches the results it obtains at the highest resolution we have available for each box. We show this convergence test in Fig.~\ref{figure:convergence_bound_mass_function}.

Generally, the bound mass of centrals is larger in lower resolution simulations for all subhalo finders. This trend is caused by the fact that fewer satellite subhaloes are resolved at lower resolutions. As the resolution increases, they are found as resolved satellites and hence the central loses out on that bound mass. For satellites, the trend appears to be similar, except for satellites in the hydrodynamical simulation found by {\ROCKSTAR}. We believe this is because unresolved substructures (e.g. satellites of satellites) become resolved, and hence remove bound mass from its parent satellite. 

{\ROCKSTAR} differs much more across resolution levels than the other subhalo finders. For example, for the other three subhalo finders the {\lowres} central bound mass function differs by no more than 25\% from {\highres}. In contrast, {\ROCKSTAR} is a factor of three higher than the values it finds at {\highres}. We remind the reader that this analysis is based on our implementation of exclusive mass definition within {\ROCKSTAR}. We also explored the convergence of $M_{\mathrm{bound}}$ using the default (inclusive mass) version of {\ROCKSTAR}. We found that its convergence properties are comparable to the results of other subhalo finders, if not marginally better. The slight improvement when using inclusive masses is due to the fact that centrals do not lose mass as satellites become resolved, whereas using an exclusive mass definition leads to the reassignment of mass from central to satellite subhaloes. 

The hydrodynamical versions may appear to converge better at first glance. However, the ratios are between two consecutive resolution levels ({\lowres} to {\midres}), whereas the DMO ones are always relative to {\highres}. Still, we observe the same trends as we did in the DMO versions, although satellites in {\ROCKSTAR} remain consistently low regardless of resolution.

We conclude this subsection by noting that differences in the mass function do not only reflect the performance of individual subhalo finders in distinguishing satellites from centrals, but also the definition of what a subhalo is. This is caused by different criteria for how resolved subhaloes are selected. {\HBTHERONS}, {\SUBFIND} and {\VELOCIRAPTOR} do iterative unbinding and keep those subhaloes with more than a threshold number of particles. {\ROCKSTAR} only does one unbinding iteration and retains in the catalogue those subhaloes whose bound particle fraction relative to the initial 6D FoF groups is above a threshold. Changing the threshold will change the type of objects included in {\ROCKSTAR} catalogues. For example, a stream may remain coherent in 6D phase space, despite having few bound particles. Thus, if the value of the threshold is set low enough, these types of objects will be included in {\ROCKSTAR} catalogues. By virtue of the definition used by other subhalo finders, these structures will generally always be excluded, although {\VELOCIRAPTOR} can also be used to find streams \citep[e.g.][]{Elahi.2011}. 

\subsubsection{Maximum circular velocity functions}\label{section:vmax_functions}

Besides the subhalo mass function, it is also of interest to quantify the number of subhaloes as a function of their maximum circular velocity, as given by equation (\ref{equation:maximum_circular_velocity}). The maximum circular velocity is expected to be less sensitive to the choice of subhalo finder the than the bound mass of a subhalo \citep[e.g.][]{Knebe.2011}, since its value is less reliant on how the edge of the subhalo is defined. As a large fraction of mass is located around the outskirts of subhaloes, even if the subhalo core is found consistently across subhalo finders, slightly changing the edge of the subhalo may propagate to larger changes in $M_{\mathrm{bound}}$ than the corresponding changes in $V_{\mathrm{max}}$. 

\begin{figure}
    \centering
    \includegraphics{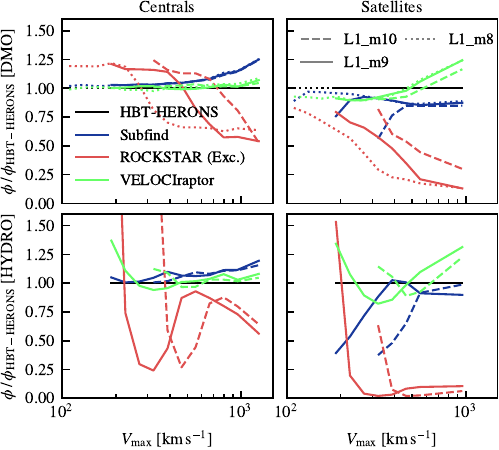}
    \caption{Similar to Fig.~\ref{figure:bound_mass_function}, but the distributions are based on the $V_\mathrm{max}$ values of subhaloes. The lowest $V_{\mathrm{max}}$ we show at a given resolution corresponds to the 99th percentile of $V_{\mathrm{max}}$ values for subhaloes with bound masses that correspond to fewer than 100 DMO dark matter particles. Note that for the purposes of this comparison, we modified {\ROCKSTAR} to assign masses exclusively, which is the way mass is assigned by the other subhalo finders, but it is not {\ROCKSTAR}'s default behaviour. See Fig.~\ref{figure:vmax_function_inclusive} for the plot corresponding to an inclusive mass assignment, which leads to smaller differences than the version presented here. 
}
    \label{figure:vmax_mass_function_exclusive}
\end{figure}

To explore if the differences we discussed in \S\ref{section:bound_mass_functions} only reflect differences in the subhalo outskirts, or if they are also attributable to changes in the central mass distribution, we show in Fig.~\ref{figure:vmax_mass_function_exclusive} the subhalo number densities as a function of their $V_{\mathrm{max}}$. In order to study a similar mass range as shown in Fig. \ref{figure:bound_mass_function}, $M_{\mathrm{bound}} \in [10^{11},10^{15}] \,\Msun $, we measure the mean $V_{\mathrm{max}}$ of subhaloes whose bound masses are near the upper and lower limits of $M_{\mathrm{bound}}$. We then round the values to reach the upper and lower limits we use, $V_{\mathrm{max}} \in [100,1500] \,\mathrm{km}\,\mathrm{s}^{-1}$. To identify the lowest value of $V_{\mathrm{max}}$ that we show for each resolution level, we use the 99th percentile of $V_{\mathrm{max}}$ values for subhaloes whose bound masses are less than the equivalent mass of 100 DMO dark matter particles.

The clearest difference is the improvement in the agreement between {\SUBFIND} and {\HBTHERONS}, but differences remain. The abundance of the most massive central subhaloes at a fixed $V_{\mathrm{max}}$ is within $\approx 30\%$, whereas differences based on $M_{\mathrm{bound}}$ were within $\approx50\%$. On the other hand, the agreement between {\VELOCIRAPTOR} and {\HBTHERONS} does not substantially improve for central subhaloes. In fact, the differences increase for satellite subhaloes, so using $V_{\mathrm{max}}$ instead of $M_{\mathrm{bound}}$ does not automatically result in better agreement. 

{\ROCKSTAR} has systematically suppressed number densities for the centrals with the highest $V_{\mathrm{max}}$, despite the relatively good agreement for the highest $M_{\mathrm{bound}}$ central subhaloes (see the top right panel of Fig.~\ref{figure:bound_mass_function}). One possibility that could explain these lower $V_{\mathrm{max}}$ values is that {\ROCKSTAR} uses the centre of mass of a subhalo as the centre of the spherical aperture used to compute $V_{\mathrm{max}}$. If the centre of mass is offset from the true density peak of its centre, $V_{\mathrm{max}}$ might be underestimated. Another explanation is that there is more substructure in the core of central subhaloes compared to all of the other subhalo finders, which we explore in \S\ref{section:radial_distributions}. If the core of a central subhalo has less bound mass because it is instead assigned to its substructure, it will lower the central subhalo $V_{\mathrm{max}}$. Similar to what we find for central subhaloes, the satellite $V_{\mathrm{max}}$ values in {\ROCKSTAR} are also suppressed compared to all other subhalo finders. 

We note that these findings are based on our implementation of exclusive mass assignment, but we show the equivalent plot that uses the default (inclusive mass) implementation of {\ROCKSTAR} in Fig.~\ref{figure:vmax_function_inclusive}. Using an inclusive mass definition results in less discrepant number densities between {\ROCKSTAR} and {\HBTHERONS} at the high $V_{\mathrm{max}}$ end of satellite subhaloes, although they still differ by $\approx 50\%$. The better agreement is due to satellites having larger $V_{\mathrm{max}}$ values in the inclusive {\ROCKSTAR} mass definition, compared to those found using an exclusive mass assignment. Hence, the number densities are not as suppressed as when using an exclusive mass assignment, which reached differences of up to $\approx 75\%$. 

The hydrodynamical versions show the same qualitative trends as the DMO simulations. {\SUBFIND} and {\HBTHERONS} agree better for the highest $V_{\mathrm{max}}$ subhaloes than for the  corresponding $M_{\mathrm{bound}}$ function, although the number density of the lowest $V_{\mathrm{max}}$ satellites are more suppressed in {\SUBFIND} relative to {\HBTHERONS} than in the DMO simulations. {\VELOCIRAPTOR} and {\HBTHERONS} agree less well than a comparison based on $M_{\mathrm{bound}}$. The number densities based on {\ROCKSTAR} are heavily suppressed relative to other subhalo finders, due to its difficulties in analysing hydrodynamical simulations. This last difference is present regardless of whether we use inclusive or exclusive masses (see the bottom right panel of Fig.~\ref{figure:vmax_function_inclusive}).

To summarise the findings discussed within this subsection, we have shown that a variety of commonly used mass metrics can result in widely different mass functions depending on the chosen subhalo finder. Halo masses based on a spherical overdensity, e.g. $M_{200\mathrm{c}}$, are the most consistent out of the three we have explored here. However, differences of up to 10\% appear in the number densities of low $M_{\mathrm{200c}}$ haloes, which are largely driven by the purely theory-based definition of what constitutes a central subhalo. Choices concerning whether to assign a single central per FoF group ({\HBTHERONS}, {\SUBFIND} and {\VELOCIRAPTOR}) or based on a spherical exclusion criteria (\ROCKSTAR), alongside  additional choices involved in either of these two approaches (e.g  FoF linking length and definition of spherical exclusivity), all contribute to different definitions of central subhalo populations. Quantifying subhalo masses using $M_{\mathrm{bound}}$ or $V_{\mathrm{max}}$ results in even more discrepant mass functions compared to using $M_{\mathrm{200c}}$. We also find that using $V_{\mathrm{max}}$ as a mass proxy instead of $M_{\mathrm{bound}}$ does not necessarily result in a better agreement across finders. Lastly, the inclusion of hydrodynamics can have severe effects on the capabilities of finding subhaloes, with the most extreme example being {\ROCKSTAR}. 

\subsection{Abundance and radial distribution of satellite subhaloes}\label{section:radial_distributions}

One of the main use cases of subhalo finders is to disentangle satellite subhaloes from the central subhalo that hosts them. As we saw from the example discussed in \S\ref{Section:example}, this does not always work as intended. The reason why subhalo finders fail to identify satellites is varied, and reflects the underlying algorithm they use. Density-based finders are known to struggle when there is insufficient contrast between adjacent density peaks \citep[e.g.][]{Behroozi.2015}. Phase space finders also struggle if the subhalo is severely tidally distorted \citep[e.g.][]{Mansfield.2024}. As these properties correlate with the environment the subhalo is in, e.g. when undergoing a close pericentre passage, there may be a bias in finding subhaloes that varies as a function of the distance to its host and their relative masses.

Given the above, a natural test of how well subhalo finders perform is to compare the abundance and distributions of satellites. In this subsection, we explore several relations that concern both aspects. We begin by looking at the subhalo number density in the outskirts of central subhaloes (\S\ref{section:number_haloes_r200}). We then examine how the shape of the subhalo number density distribution varies as a function of host $M_{\mathrm{200c}}$ (\S\ref{Section:radial_distribution_m200_bins}) and satellite-to-host bound mass ratio  (\S\ref{Section:radial_distribution_mass_ratio_bins}). As we measure the last two statistics at a fixed resolution level, haloes with lower $M_{\mathrm{200c}}$ or satellites with lower satellite-to-host bound mass ratios are less well sampled by particles than those with correspondingly larger values. In other words, resolution trends may masquerade as mass trends. To disentangle one from the other we also present convergence tests.

The central subhalo population is significantly different for {\ROCKSTAR} compared to the other subhalo finders (\S\ref{Section:M200_mass_functions}). To prevent changes in the definition of central from driving differences in the radial distributions, we count all subhaloes at a given distance regardless of whether they are classified as centrals or satellites. Throughout this subsection, we only take into account subhaloes with a bound mass that is at least 100 times the mass of a dark matter particle in the DMO simulations.

\subsubsection{Number of subhaloes in the outskirts of central subhaloes}\label{section:number_haloes_r200}

We start the comparison by looking at how many subhaloes there are near $R_{\mathrm{200c}}$ of central subhaloes, measured using a bin of width 0.1 dex. This corresponds to approximately $[0.8 - 1]\times R_{\mathrm{200c}}$. Our focus on the outskirts of haloes reflects the fact that we expect the difficulties associated with heavily overlapping or disturbed subhaloes to be less important than in the inner regions. Thus, this test will reveal whether there are already systematic differences in the number of subhaloes in regions expected to be relatively well resolved.

\begin{figure}
    \centering
    \includegraphics{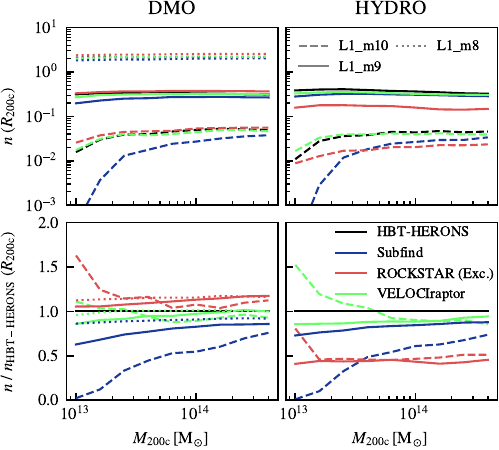}
    \caption{\textit{Top panels}: Mean number density of subhaloes in the outskirts of haloes, as a function of their central $M_{\mathrm{200c}}$ mass. The number density is measured by counting subhaloes with a total bound mass greater than 100 DMO dark matter particles in a spherical shell spanning $[0.8 - 1] \times R_{\mathrm{200c}}$ of the corresponding host. All subhaloes within the shell are included regardless of whether they are a central or a satellite, in order to minimise differences that are driven purely by the definition of a central. Results from the DMO and hydrodynamical versions of the boxes are shown in the left and right columns, respectively. \textit{Bottom panels}: The same as above, but expressed relative to the number density found by {\HBTHERONS}. Note that for the purposes of this comparison, we modified {\ROCKSTAR} to assign masses exclusively, which is the way mass is assigned by the other subhalo finders, but it is not {\ROCKSTAR}'s default behaviour.}
    \label{figure:subhalo_number_R200}
\end{figure}

The mean number density of subhaloes near $R_{\mathrm{200c}}$ as a function of the $M_{\mathrm{200c}}$ of the host is shown in the top panels of Fig.~\ref{figure:subhalo_number_R200}. We show this relation for all resolutions levels for which subhalo catalogues are available. For the {\midres} and {\highres} simulations, there is little to no dependence of $n({R}_{\mathrm{200c}})$ on $M_{\mathrm{200c}}$. This is because $n({R}_{\mathrm{200c}})\propto N_{\mathrm{sub}} R^{-3}_{\mathrm{200c}} \propto N_{\mathrm{sub}} M^{-1}_{\mathrm{200c}}$, where $N_{\mathrm{sub}}$ is the number of subhaloes enclosed within the spherical shell. Since the number of subhaloes within $R_{200}$ scales approximately linearly with $M_{\mathrm{200c}}$ \citep[e.g.][]{Wang.2012}, the number of subhaloes in the outskirts also scales linearly with mass, and hence $n(R_{200})$ remains approximately constant. Comparing {\midres} to {\highres}, we see an upwards shift in the number density of subhaloes, corresponding to more of them being resolved thanks to the increased resolution.

A caveat worth adding is that, when selecting subhaloes according to a fixed subhalo mass limit, a linear relation between the number of subhaloes and $M_{200\mathrm{c}}$ is only expected when the subhalo mass limit is far away from the exponential cut-off mass scale of the corresponding subhalo mass function. As we lower the resolution, the mass limit (100 DMO particles) used to select subhaloes increases, resulting in limits that shift closer to the exponential cut-off scale. For the lowest $M_{200\mathrm{c}}$ haloes, the mass limit can become comparable to the exponential cut-off scale, thus making them the most susceptible to experience a departure from linearity between $M_{200\mathrm{c}}$ and $n({R}_{\mathrm{200c}})$. Indeed, this could be one of the reasons behind why the subhalo number density no longer remains constant with $M_{200\mathrm{c}}$ in the {\lowres} simulation across subhalo finders. As the subhalo mass functions can also differ across subhalo finders, e.g. with the exponential cut off in the subhalo mass function occurring at a lower mass scale or with a systematic shift in the assigned masses, the departure from $N_{\mathrm{sub}} \propto M_{200\mathrm{c}}$ will generally depend on the employed subhalo finder. 

To explore the differences in the measured number density of subhaloes across subhalo finders, we show in the bottom panels of Fig.~\ref{figure:subhalo_number_R200} their $n(R_{\mathrm{200c}})$ relative to that of {\HBTHERONS}. For all of the DMO simulations, {\SUBFIND} always finds the lowest number density out of the four subhalo finders. The suppression relative to {\HBTHERONS} is resolution dependent. At the high mass end, the ratio increases from 0.75 for {\lowres} to 0.87 for {\highres}. For lower masses, the number varies from virtually zero to 0.87 that of {\HBTHERONS}. At the other end of the spectrum, {\ROCKSTAR} consistently finds the largest subhalo number density across all resolutions of the DMO simulations, around $20\%$ more than the results based on {\HBTHERONS}. Out of the three alternative subhalo finders, {\VELOCIRAPTOR} finds values of $n(R_{\mathrm{200c}})$ that are most consistent with {\HBTHERONS}, regardless of the resolution. 

\begin{figure*}
    \centering
    \includegraphics{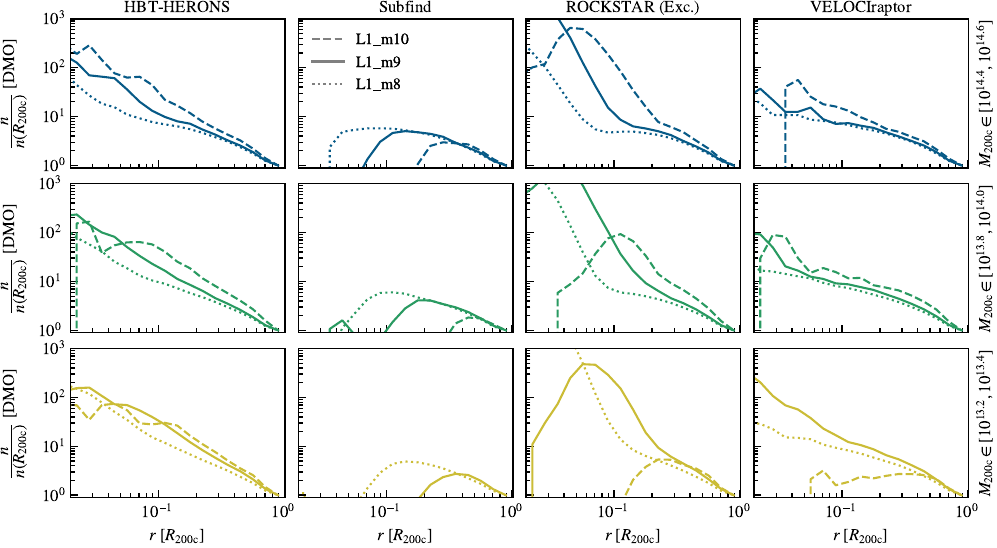}
    \caption{Normalised subhalo number density as a function of normalised distance to the central subhalo for different DMO resolutions (different line styles) and averaged in different bins of $M_{\mathrm{200c}}$ (different line colours and rows). The values of $M_{\mathrm{200c}}$ that bracket each bin are indicated on the right hand size of each corresponding row, and encompass 0.2 dex in $M_{\mathrm{200c}}$ and are spaced 0.6 dex from each other. Each column corresponds to the results found by a different subhalo finder. The number densities are normalised to the value found by each subhalo finder in the outermost radial shell for each resolution and mass bin. We count all subhaloes whose bound mass is at least 100 times the DMO particle mass, regardless of whether they are classified as centrals or satellites. This choice is made to minimise differences that are driven purely by the definition of a central. Note that for the purposes of this comparison, we modified {\ROCKSTAR} to assign masses exclusively, which is the way mass is assigned by the other subhalo finders, but it is not {\ROCKSTAR}'s default behaviour.}
    \label{figure:subhalo_number_density_distribution_convergence}
\end{figure*}

We now shift our attention to how the subhalo number densities change in the hydrodynamical versions of {\tt FLAMINGO}, which are shown in the top right panel of Fig.~\ref{figure:subhalo_number_R200}. The clearest change is that {\ROCKSTAR} is no longer the subhalo finder that consistently finds the highest subhalo number density. In fact, for {\midres}, it finds the lowest densities out of all subhalo finders. As discussed previously, this is driven by the fact that {\ROCKSTAR} struggles to find subhaloes when linking particles regardless of their type. Another difference relative to the DMO versions is that the number density of subhaloes tends to be higher across $M_{\mathrm{200c}}$ for the {\midres} hydrodynamical version, according to {\HBTHERONS}, {\SUBFIND} and {\VELOCIRAPTOR}.

The increase in the number of satellites may be surprising. At Milky Way- (MW) mass scales, which many radial distribution studies focus on, the net change caused by the addition of hydrodynamics is a decrease in the number of satellite subhaloes \citep[e.g.][]{Samuel.2020}. This is attributed to a combination of their hindered growth due to early mass loss \citep[e.g.][]{Sawala.2017} and stronger tides due to the presence of a baryonic disc in the host \citep[e.g.][]{Garrison-Kimmel.2017}. Evidently, extrapolating those results to the mass scales we study here would yield an incorrect expectation.

What could cause the difference between the outcome resulting from the inclusion of baryons for MW scales and cluster scales? The efficiency of galaxy formation varies with mass, so satellites in clusters correspond to the mass range expected to harbour sufficient baryons to induce a contraction of their DM subhalo. Indeed, as we already saw in Fig~\ref{figure:image_example}, subhaloes in the hydrodynamical simulations clearly have  denser cores than in the DMO simulation. As denser centres make subhaloes more resilient to gravitational tides \citep[e.g.][]{Penarrubia.2010}, they are less likely to be stripped to masses below the resolution of the simulation. On MW-mass scales, the masses of satellites correspond to the regime of inefficient galaxy formation, where contraction is expected to less important (and expansion may even occur).

The level of agreement between subhalo finders changes when looking at the hydrodynamical simulations. In particular, {\SUBFIND} and {\HBTHERONS} agree to within $\approx10\%$ for the {\midres} hydrodynamical simulation, i.e. as well as for the {\highres} DMO simulation. On the other hand, {\ROCKSTAR} now underpredicts the number of subhaloes relative to {\HBTHERONS}. As before, we note that this is likely driven by our choice of linking all particle types for {\ROCKSTAR}. This highlights once again the importance of taking proper care when analysing hydrodynamical simulations using subhalo finders that were developed for DMO simulations.

As discussed earlier, relative to the DMO simulation, the number of subhaloes increases for all subhalo finders but {\ROCKSTAR}. However, the increase is not uniform across finders. In particular, the increase in the number of {\SUBFIND} subhaloes at $R_{\mathrm{200c}}$ is the largest. The net effect is better agreement between {\HBTHERONS} and {\SUBFIND} for {\midres} hydro than for the highest resolution DMO simulation we have ({\highres}). The improvement is likely caused by the fact that subhaloes become denser in the hydrodynamical simulations, which helps {\SUBFIND} identify density peaks more easily.  

\subsubsection{Radial distributions as a function of host halo mass}\label{Section:radial_distribution_m200_bins}

Another important property of subhalo populations is how they are distributed within their host haloes. In this subsection, we measure their distribution by counting how many subhaloes are present within a given spherical shell. We use 20 shells whose edges are logarithmically spaced between $[0.01 - 1]\times R_{\mathrm{200c}}$. A subhalo is assigned to a shell based on where its centre is located. We then divide by the volume of the shell to obtain the subhalo number density as a function of distance to the centre of the host. 

As the satellite distribution of an individual halo can be noisy and not representative of the mean trend, we average the number density distributions in nine $M_{\mathrm{200c}}$ bins between $10^{13}$ and $10^{15}\,\Msun$. All but the highest $M_{\mathrm{200c}}$ bin span 0.2 decades in mass. The highest mass bin is enlarged to to 0.4 decades to increase the number of centrals it contains to about 755\footnote{There are small differences in the number of centrals across subhalo finders in the highest $M_{\mathrm{200c}}$ bin due to different $M_{\mathrm{200c}}$ being measured for the same haloes (e.g. Fig~\ref{figure:M200_mass_function}).}. 

In the preceding subsection, we found that the number density of subhaloes differs between subhalo finders even near $R_{\mathrm{200c}}$. This means that differences in the distributions will appear due to differences in the normalisation. In this subsection, we are more interested in how the shape of the distribution changes between finders. Therefore, to decouple the effect of different numbers of subhaloes found across subhalo finders, we normalise the subhalo number densities by the number density found by each subhalo finder in the outermost spherical shell (spanning $[0.8 - 1] \times R_{\mathrm{200c}}$). 

\begin{figure*}
    \centering
    \includegraphics{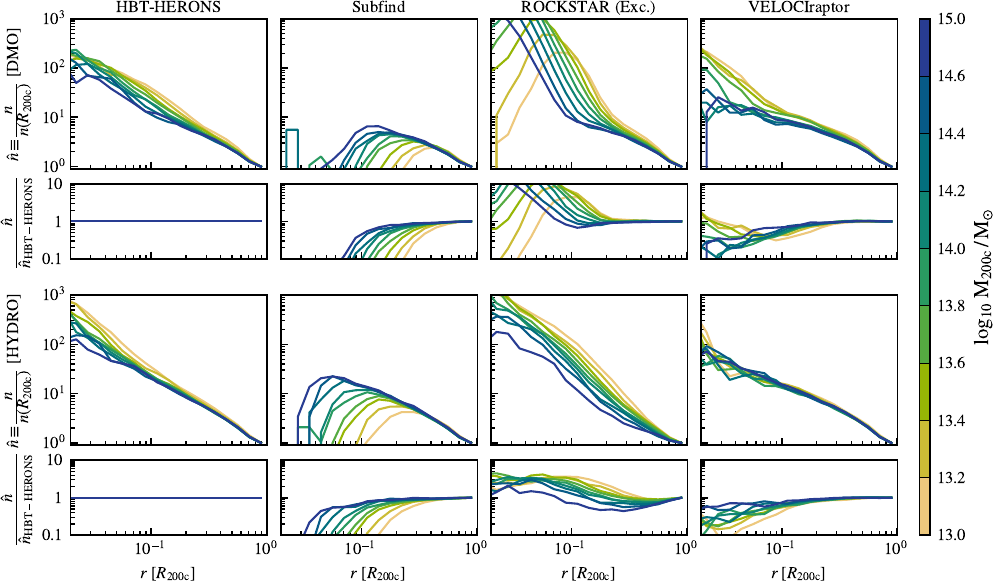}
    \caption{Normalised subhalo number density as a function of normalised distance to the central subhalo for the {\midres} DMO and hydrodynamical simulations averaged in different bins of $M_{\mathrm{200c}}$ (different line colours). \textit{Top panels}: the radial distributions found in the DMO version of the simulations. The secondary panel shows the ratio of the normalised subhalo number densities of each subhalo finder relative to the values found in {\HBTHERONS} \textit{Bottom panels}: Same as the top panels, but for the hydrodynamical version of the simulations.}
    \label{figure:subhalo_number_density_distribution}
\end{figure*}

An important caveat is that, since we are comparing the radial distribution of subhaloes in a simulation using a fixed particle mass, bins with lower values of $M_{\mathrm{200c}}$ correspond to less well resolved systems. Therefore, an apparent trend with $M_{\mathrm{200c}}$ may just be a trend with resolution. To help us interpret the trends and differences between subhalo finders, we first examine how the DMO radial distributions change across the {\lowres}, {\midres} and {\highres} resolutions. The corresponding distributions are shown in Fig.~\ref{figure:subhalo_number_density_distribution_convergence}, which are measured in the manner described above for a subset of the $M_{\mathrm{200c}}$ bins we will explore. 

Focusing first on the normalised subhalo number densities from the {\lowres} simulation, we see that the shapes are very different at a fixed $M_{\mathrm{200c}}$ depending on the subhalo finder used to derive them. The shapes for ${\SUBFIND}$ are non-monotonic and peak at a radius that decreases as $M_{\mathrm{200c}}$ (or the number of particles within $R_{\mathrm{200c}}$) increases. On the other end of the spectrum, using {\HBTHERONS} generally results in a monotonic number density for the plotted radial range across all $M_{200\mathrm{c}}$. {\ROCKSTAR} and {\VELOCIRAPTOR} show a behaviour that can be more similar to either {\SUBFIND} (e.g. $M_{200\mathrm{c}} \in [10^{13.2},10^{13.4}]\Msun$ in {\lowres}) or {\HBTHERONS} (e.g. $M_{200\mathrm{c}} \in [10^{14.4},10^{14.6}]\Msun$ in {\lowres}). 

Looking at how the normalised subhalo number density distributions change as the resolution increases, we see that the distributions converge differently across subhalo finders. For {\SUBFIND}, the radius below which no substructure is found shifts to smaller distances. The normalised subhalo number densities found by {\SUBFIND} always agree well across resolutions beyond the radius at which the normalised number density peaks. This scale provides a natural `convergence' radius. On the other hand, the other three subhalo finders do not exhibit a clear  `convergence' radius, as increasing the resolution results instead in increasingly flatter shapes for $n(r)/n(R_{\mathrm{200c}})$. We can clearly see that this change from {\midres} to {\highres} is primarily driven by a decrease in the normalised number density of subhaloes in the central regions.

How can one make sense of the fact that subhalo finders that find more subhaloes, e.g. {\HBTHERONS}, exhibit a decrease in the normalised number density of subhaloes as resolution increases? It all has to do with how massive (relative to the host halo) the largest resolvable satellites are. At a fixed host $M_{\mathrm{200c}}$, increasing the resolution increases the number of particles that sample its virial region. Hosts with $M_{200\mathrm{c}} \in [10^{13.2},10^{13.4}]\Msun$ have approximately 300 to 500  particles within $R_{\mathrm{200c}}$ at the {\lowres} resolution. We can estimate the minimum resolvable satellite-to-central mass ratio if we assume all particles within $R_{\mathrm{200c}}$ are bound the central or a single satellite subhalo. For our selection cut of 100 bound particles, the satellite-to-central mass ratio is $N_{\mathrm{sat}} / (N(\leq R_{\mathrm{200c}}) - N_{\mathrm{sat}})\approx 100 / (500 -100) = 0.25$.  Thus, lowering the resolution increasingly biases the radial distributions to be more dominated by major mergers. As we show in the following subsection, the high mass end of the satellite population is more concentrated towards the centre in {\HBTHERONS}. This explains why the shapes are steeper for a lower resolution simulation than for a higher resolution one. 

We also show in \S\ref{Section:radial_distribution_mass_ratio_bins} that {\SUBFIND} struggles to identify satellites with masses above $\approx 0.1\%$ that of their host, either by not finding the subhalo itself or by removing so much mass from it that it falls below our 100 DMO dark matter particle mass cut. This makes its trend with resolution different to the other subhalo finders. Since changing $M_{\mathrm{200c}}$ at a fixed resolution is equivalent to changing the number of particles within $R_{\mathrm{200c}}$, we expect that the resolution trends we just discussed will also appear as mass trends. In other words, we expect that lowering $M_{\mathrm{200c}}$ increasingly biases the radial distributions to be more dominated by major mergers and hence more radially concentrated for {\HBTHERONS}, {\ROCKSTAR} and {\VELOCIRAPTOR}. For {\SUBFIND}, the difficulty in finding massive satellites will result in more suppressed subhalo number densities as $M_{\mathrm{200c}}$ is lowered.

\begin{figure*}
    \centering
    \includegraphics{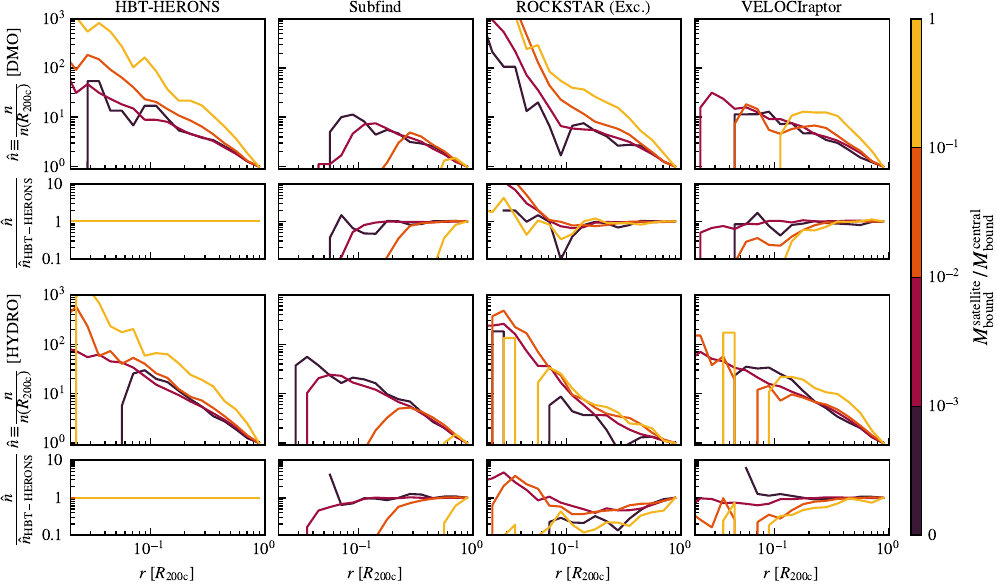}
    \caption{Normalised subhalo number density as a function of normalised distance to central subhalos with $M_{200\mathrm{c}} \in [10^{14.6},10^{15.0}]$, averaged in different satellite-to-host bound mass ratios (different line colours) for the {\midres} DMO and hydrodynamical simulations \textit{Top panels}: the radial distributions found in the DMO version of the simulations. The secondary panel shows the ratio of the normalised subhalo number densities of each subhalo finder relative to the values found in {\HBTHERONS}. \textit{Bottom panels}: Same as the top panels, but for the hydrodynamical version of the simulations.}
    \label{figure:radial_distribution_satellite_mass_ratio_bins}
\end{figure*}

In Fig.~\ref{figure:subhalo_number_density_distribution}, we can now compare the performance of different subhalo finders at the same resolution level and different $M_{\mathrm{200c}}$. The number density of subhaloes generally increases towards the centre, except when there is a clear cut-off due to resolution-dependent issues related to subhalo finding. Only {\SUBFIND} and {\ROCKSTAR} show a drop at small radii, suggesting they perform worse in the inner regions than {\HBTHERONS} and {\VELOCIRAPTOR} at the same resolution level. Interestingly, {\ROCKSTAR} exhibits a peak in the number density of subhaloes at radii slightly larger than where their density sharply decreases. We attribute this increase to resolution-dependent spurious subhaloes being identified in poorly resolved regions, which are prone to noise in the particle distribution. The chosen bound fraction threshold (0.5) was therefore unable to remove them. Further post-processing steps could help remove them if they are indeed spurious \citep[e.g.][]{Behroozi.2013_consistent_trees}, but we intentionally do not use any additional tools external to the subhalo finders themselves.

The level of agreement between subhalo finders in the shapes worsens for lower $M_{\mathrm{200c}}$, i.e. in more poorly sampled haloes. This is particularly prominent for {\SUBFIND}, as it agrees well with {\HBTHERONS} beyond $0.2R_{\mathrm{200c}}$ for the highest $M_{\mathrm{200c}}$ bin we consider. For the lowest $M_{\mathrm{200c}}$ bin, it only agrees with {\HBTHERONS} beyond $0.6R_{\mathrm{200c}}$. This trend is less prominent for {\VELOCIRAPTOR}, as the agreement with {\HBTHERONS} is always good beyond $0.3R_{\mathrm{200c}}$. {\ROCKSTAR} agrees well beyond $0.3R_{\mathrm{200c}}$ but within $r \ll 0.1R_{\mathrm{200c}}$, the subhalo number density in {\ROCKSTAR} is more than a factor 10 larger than for {\HBTHERONS}. The distance below which the normalised number densities of subhaloes disagree is likely driven by how well different subhalo finders perform in poorly resolved regions. Recall that even if the shapes agree, the overall normalisation generally does not (see Fig.~\ref{figure:subhalo_number_R200}).

{\HBTHERONS}, {\ROCKSTAR} and {\VELOCIRAPTOR} do not exhibit good agreement between the normalised subhalo number density distribution across haloes of different $M_{\mathrm{200c}}$ (i.e. resolution). However, the shapes of the distributions and trends with $M_{\mathrm{200c}}$ differ between them. Generally, they find more substructure at smaller distances than {\SUBFIND}, which is expected since density-based finders can struggle to cleanly separate (or identify) subhaloes that are in close proximity, leading to a substantial change in their properties and hence whether they are selected at a given cut or not. {\HBTHERONS} finds the highest number density across all the shown $M_{\mathrm{200c}}$ bins, as it is the only finder that measures an approximately monotonic relation in the plotted radial range. Both {\ROCKSTAR} and {\VELOCIRAPTOR} exhibit a cut-off in the number density that shifts to larger normalised distances as $M_{\mathrm{200c}}$ decreases. Similarly to {\SUBFIND}, this occurs because the subhalo finders are unable to find substructure interior to this radius.

\begin{figure*}
    \centering
    \includegraphics{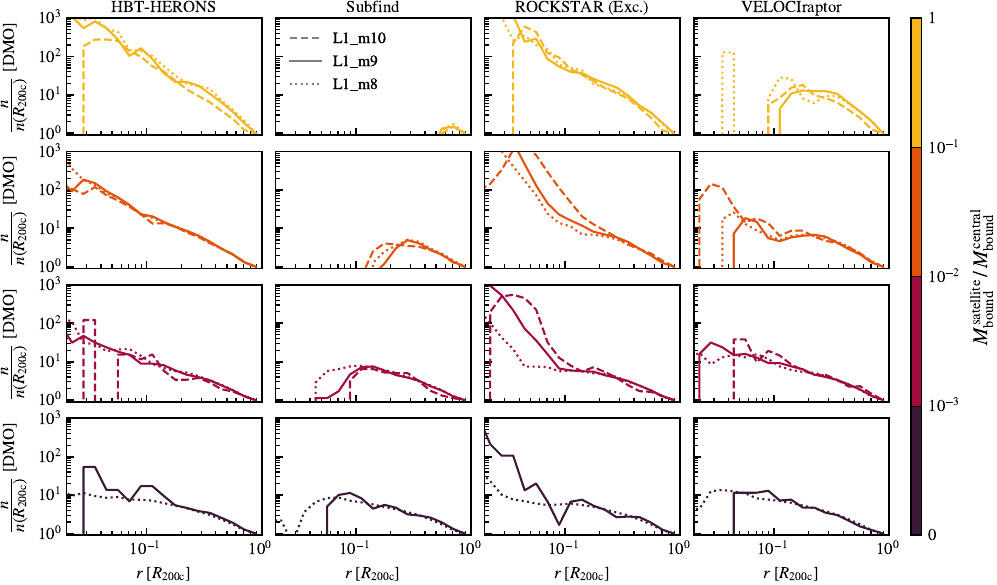}
    \caption{Similar to Fig.~\ref{figure:radial_distribution_satellite_mass_ratio_bins}, but the subhalo number density distributions are now shown across the three available DMO resolutions (different line styles). Different bins of satellite-to-host bound mass ratios are shown using different colours, as indicated in the colour bar. Note that the {\lowres} simulation has a very undersampled distribution in the lowest bin in $M_{\mathrm{satellite}} / M_{\mathrm{central}}$. For our cut of 100 bound particles, $M_{\mathrm{satellite}} \approx 10^{12} \,\mathrm{M}_{\odot}$. This means only very few satellites will contribute to our measurements, as the bin itself requires  $M_{\mathrm{satellite}} \leq 10^{-3} \times [10^{14.6}, 10^{15}] \,\Msun = [10^{11.6}, 10^{12}] \,\Msun$.}
    \label{figure:radial_distribution_satellite_mass_ratio_bins_convergence}
\end{figure*}

We now shift our attention to how the radial distributions change in the hydrodynamical simulations (bottom two rows of Fig.~\ref{figure:subhalo_number_density_distribution}). The shapes differ in several aspects from the DMO ones. The number density profiles become steeper across all subhalo finders and in all host $M_{\mathrm{200c}}$ bins. This results in normalised number densities that are higher by factors of two to tens at the smallest radii we explore here. Since the number of subhaloes near $R_{\mathrm{200c}}$ also increases (see Fig.~\ref{figure:subhalo_number_R200}), this implies there are more subhaloes throughout the virial regions of haloes. 

Previous studies found a similar change between DMO and hydrodynamical simulations \citep[e.g.][]{Haggar.2021, Riggs.2022}. Some explanations were put forward in \citet{Haggar.2021} to explain the cause behind the increase in abundance. One explanation was that their density-based subhalo finder ({\tt AHF}; \citealt{Gill.2004, Knollmann.2009}) finds subhaloes more effectively once baryons are included. Although we find this to be the case, e.g. {\SUBFIND} agrees more closely with {\HBTHERONS} in hydro than in DMO simulations, we see the increase across subhalo finders. Thus, the greater number of subhaloes occurs independently from changes due to the halo finding.

Alternative explanations attribute the difference to longer survival times of subhaloes. As discussed in \S\ref{section:number_haloes_r200}, subhaloes in {\tt FLAMINGO} have denser centres than in the DMO simulations, which presumably makes them more resilient to tidal stripping. \citet{Riggs.2022} matched subhaloes between their DMO and hydrodynamical counterparts, finding that many subhaloes in the hydrodynamical simulation were indeed missing in the DMO simulation. Interestingly, the subhaloes that were successfully matched were more spatially concentrated in the hydrodynamical simulation. They argued that the fact that the satellites were found closer to the centres of their hosts could also reflect changes in their orbits relative to the DMO simulations. The orbital changes could in principle be caused by a more concentrated gravitational potential of the host, or stronger dynamical friction due to the increased contrast between the density of the satellite and the background of its host. Determining how the orbits of subhaloes vary in clusters using the self-consistent trees of {\HBTHERONS} could help shed light on whether friction or the potential is the primary culprit. However, this is beyond the scope of this study.

Another change in the radial distributions is that the dependence on the host $M_{\mathrm{200c}}$ (i.e. resolution) is less pronounced for the hydrodynamical simulations. For {\VELOCIRAPTOR}, all mass bins are comparatively similar down to $0.05R_{\mathrm{200c}}$. {\SUBFIND} profile shapes also agree for all mass bins in the regions exterior to where it struggles to find subhaloes ($0.3R_{\mathrm{200c}}$ for the lowest $M_{\mathrm{200c}}$ bin). For {\HBTHERONS}, the shapes now agree across $M_{\mathrm{200c}}$ masses down to $0.3R_{\mathrm{200c}}$. This is likely due to more subhaloes surviving stripping due to their denser cores, and hence the distribution in the hydrodynamical simulation resembles that of a higher resolution DMO simulation.  

Another important difference is that, except for {\ROCKSTAR}, different subhalo finders find more consistent results down to smaller radii than for the DMO simulations. For the highest mass bin, {\VELOCIRAPTOR} and {\HBTHERONS} are within $30\%$ from each other at $0.02R_{\mathrm{200c}}$. For DMO, their values disagreed by $80\%$ at this radius. The radius at which {\SUBFIND} deviates from {\HBTHERONS} by $50\%$ is $0.02R_{\mathrm{200c}}$. In the DMO simulation, this already occurred at $0.1R_{\mathrm{200c}}$. The better agreement, beyond longer survival times for subhaloes, is due to the subhalo centres being denser, making it easier to identify density peaks. 

\subsubsection{Radial distributions as a function of satellite to central mass ratio}\label{Section:radial_distribution_mass_ratio_bins}

In the previous subsection we discussed how different subhalo finders result in different distributions of subhaloes within $R_{\mathrm{200c}}$. This was for the whole population of subhaloes with more bound mass than the equivalent of 100 DMO dark matter particles, regardless of how massive they are relative to the central. However, the distribution of subhaloes can depend on the mass ratio relative to their host, as the importance of various physical mechanisms depends on its value. For example, dynamical friction primarily affects massive satellites, making them sink towards the centre of the host. This can naturally lead to different radial distributions based on the relative mass between satellites and their host. 

In Fig.~\ref{figure:radial_distribution_satellite_mass_ratio_bins}, we explore how the subhalo population is distributed according to their satellite-to-central bound mass ratio. The subhalo number densities for each mass ratio bin are divided by the number density found by each corresponding subhalo finder near $R_{200\mathrm{c}}$. For this analysis, we only use the highest $M_{\mathrm{200c}}$ bin defined in  \S\ref{Section:radial_distribution_m200_bins} ($10^{14.6}\,\Msun \leq M_{\mathrm{200c}} \leq 10^{15}\,\Msun$), because it contains the largest number of particles within $R_{\mathrm{200c}}$ at a fixed resolution. We measure the relative mass between the satellite and its host based on the ratio of their $z=0$ bound masses, as measured by each subhalo finder. As we are interested in computing a satellite-to-host mass ratio whose upper value is at most 1, we change the definition of central to be the subhalo with the largest $M_{\mathrm{bound}}$ within $R_{\mathrm{200c}}$. When performing this re-centering, which results in spatial shifts that are typically of the order of $\sim \mathcal{O}(10)\,\mathrm{kpc}$, we keep the $R_{200\mathrm{c}}$ value estimated from the previous central. 

Differences between subhalo finders within a mass ratio bin can occur in two ways. A subhalo may be completely missing from the catalogue, in which case it does not contribute to any binned measurement. Alternatively, its bound mass may be over- or underestimated. When this happens, it will contribute to a different mass bin relative to its counterpart for a different subhalo finder. 

We confirm that the radial distribution functions depend on the satellite-to-host mass ratio, but the direction of the trend differs between subhalo finders. Both {\HBTHERONS} and {\ROCKSTAR} find high mass satellites to be more spatially concentrated than lower mass ones, as other studies have found in other simulation suites in the past (e.g. Fig.~8 of \citealt{Han.2018}). {\SUBFIND} and {\VELOCIRAPTOR} find the opposite, with a distinct lack of massive subhaloes within the central parts. {\VELOCIRAPTOR} does find larger numbers of high-mass subhaloes at smaller distances compared to {\SUBFIND}, likely thanks to its phase-space approach. However, as was shown in Fig. \ref{figure:image_example}, this is sometimes insufficient to separate closely spaced subhaloes. 

One may explain the correlation or anti-correlation between satellite mass and spatial concentration using dynamical friction. If the subhaloes remain self-bound as they sink, their number densities will increase at small radial distances, precisely as {\HBTHERONS} and {\ROCKSTAR} find. If they instead quickly disrupt and mix with their host, their number densities will be suppressed, as {\SUBFIND} and {\VELOCIRAPTOR} predict. 

Based on the evidence we have presented concerning the missing objects in {\SUBFIND} and {\VELOCIRAPTOR}, we think that the distribution of the most massive subhaloes is indeed more radially concentrated than for their lower mass counterparts. Previous studies that only used {\SUBFIND} to study the same relation \citep[e.g.][]{DeLucia.2004, Angulo.2009} reached the opposite conclusion. We attribute this to an incorrect interpretation of systematically biased subhalo catalogues, highlighting the importance of robustly finding subhaloes. Note that the high-mass subhalo population is less numerous than the lower mass one, so the former does not determine the overall distributions when all subhaloes are taken into account.

In Fig. \ref{figure:radial_distribution_satellite_mass_ratio_bins_convergence} we show that this dichotomy of interpretations is resilient to changes in the resolution. Few significant systematic differences with the resolution of the simulation exist in the normalised subhalo number density distribution across all but the lowest satellite-to-host mass ratios. The fact that the distribution does not become more concentrated as the resolution increases implies that the predictions from different subhalo finders converge to different answers. Based on visual inspection of many examples of missing massive satellites (e.g. \S\ref{Section:example}), and the well-known limitations of subhalo finders when subhaloes spatially overlap, we argue that the trends recovered by {\HBTHERONS} and {\ROCKSTAR} are the most physical.

Lastly, the fact that massive objects are not found means that they may be treated as orphans instead. Tracking orphaned subhaloes is done with the aim of improving the agreement between low- and high-resolution simulations, but tracking the evolution of orphans relies on assumptions that lead to different predictions \citep[e.g.][]{Pujol.2017}. It would be best to avoid treating massive objects that are resolvable as orphans, in an attempt to compensate for halo finding problems.

To give a sense of how many particles sample the massive, missing objects, we provide an order of magnitude estimate. The hosts we analyse here have $M_{\mathrm{200c}} > 10^{14.6}\,\Msun$. For the {\highres} simulation, this means they are sampled with $\approx 10^{6.6}$ particles within $R_{\mathrm{200c}}$. If we assume that the bound mass of the central is similar to its $M_{\mathrm{200c}}$, the highest mass ratio bin we use corresponds to satellites of $\approx 10^{5}$ particles. In other words, these are objects that should be very well resolved, and yet are missing from some catalogues.

We conclude this subsection by showing in the bottom panels of Fig.~\ref{figure:radial_distribution_satellite_mass_ratio_bins} how the number density as a function of satellite-to-host mass ratio bin compares across finders in the hydrodynamical simulations. Similar to what we found in the previous subsection, the agreement between {\SUBFIND} and {\HBTHERONS} depends on the mass bin under consideration. The smaller the mass ratio, the more likely the satellite is found by {\SUBFIND} at small distances. The inclusion of hydrodynamics greatly increases the number of objects found by both, particularly for {\SUBFIND}. The net effect is better agreement with {\HBTHERONS}, although {\SUBFIND} is still unable to find the massive satellites at all radii. For {\ROCKSTAR}, the addition of hydrodynamics makes the normalised subhalo number densities agree less well with {\HBTHERONS}  across all satellite-to-host mass bins, and suppresses the densities relative to {\ROCKSTAR}'s own results in the DMO simulations. For {\VELOCIRAPTOR}, the second lowest mass ratio bin ($10^{-3}$ to $10^{-2}$) agrees better in hydro than in DMO with the results found by {\HBTHERONS}, although the agreement for the second highest bin  ($10^{-2}$ to $10^{-1}$) worsens.

\subsection{Two-point correlation functions}\label{section:correlation_functions}

The last summary statistic we compare across subhalo finders is the two-point correlation function of subhaloes. As we have seen in the previous subsections, the choice of subhalo finder changes the halo virial mass functions, subhalo mass functions (based on $M_{\mathrm{bound}}$ and $V_{\mathrm{max}}$) and the radial distribution of subhaloes. One might thus expect that these differences are reflected in the predicted correlation functions, especially in the radial range where the one halo term dominates. 

To explore whether this is the case, we compute the auto-correlation function using the {\tt Corrfunc} package \citep{Sinha.2020}. We divide the whole subhalo population into four $V_{\mathrm{max}}$ bins, without taking into account whether the subhaloes are centrals or satellites. Our choice to use $V_{\mathrm{max}}$ reflects the fact that it is a mass proxy that is less sensitive to the chosen subhalo finder algorithm, compared to $M_{\mathrm{bound}}$. Additionally, observationally it is more accessible than the bound mass of subhaloes.

\begin{figure*}
    \centering
    \includegraphics{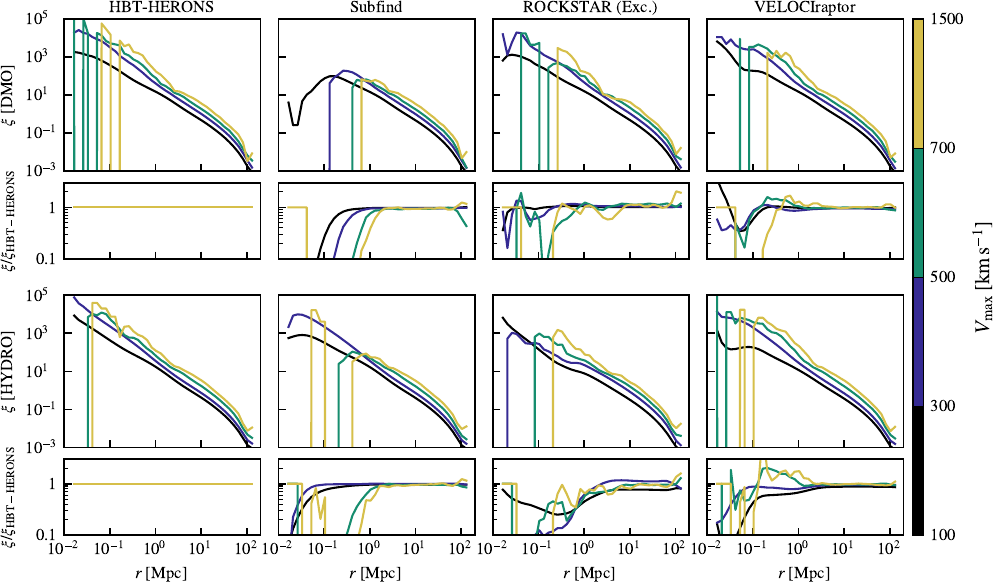}
    \caption{Auto-correlation function of subhaloes in the {\midres} simulation, measured in different bins of $V_{\mathrm{max}}$ (different colours). We select all subhaloes whose $V_{\mathrm{max}}$ is within the limits of the corresponding bin, regardless of whether they classified as satellites or centrals. \textit{Top panels}: the correlation functions found in the DMO version of the simulations. The secondary panel shows the ratio relative to the values found by {\HBTHERONS}. \textit{Bottom panels}: Same as the top panels, but for the hydrodynamical version of the simulations. Note that for the purposes of this comparison, we modified {\ROCKSTAR} to assign masses exclusively, which is the way mass is assigned by the other subhalo finders, but it is not {\ROCKSTAR}'s default behaviour.}
    \label{figure:correlation_functions}
\end{figure*}

The resulting correlation functions are shown in Fig.~\ref{figure:correlation_functions}. We recover the expected trend of more massive subhaloes being more strongly clustered across finders. However, the values and shapes of the correlation functions can vary. Looking at large scales, beyond $1$ to $10\, \mathrm{Mpc}$, the agreement is generally good. {\SUBFIND} and {\HBTHERONS} are the most consistent pair, but differences exist at the largest scales between the two highest $V_{\mathrm{max}}$ bins.

We note a small systematic shift in the values found at large scales between {\ROCKSTAR} and {\HBTHERONS}. This is driven by the fact that a cut in subhaloes according to their $V_{\mathrm{max}}$ results in slightly different population selections. As discussed in \S\ref{section:vmax_functions}, the bound mass of {\ROCKSTAR} centrals is suppressed relative to all other subhalo finders. This means that selecting on {\ROCKSTAR}'s $V_{\mathrm{max}}$ selects subhaloes with larger average $M_{\mathrm{200c}}$ than those selected using the same $V_{\mathrm{max}}$ found by the three other subhalo finders. As more massive subhaloes are more strongly clustered, this selection biases the correlation function to be slightly larger at fixed $V_{\mathrm{max}}$ for {\ROCKSTAR} relative to the other finders. 

On scales smaller than $\approx 2\,\mathrm{Mpc}$, where the one halo term contribution becomes important, the measured correlation functions differ significantly. The differences are due to how well the satellite systems are found by each subhalo finder. As such, most changes relative to {\HBTHERONS} are similar to those in the radial distribution functions. For example, the {\SUBFIND} correlation functions are increasingly suppressed at small distances, regardless of the $V_{\mathrm{max}}$ bin.

Comparing the correlation functions in the hydrodynamical simulations, in the bottom panels of Fig.~\ref{figure:correlation_functions}, we see that they change relative to DMO in a similar manner as the satellite radial distributions between DMO and hydro (see Fig.~\ref{figure:subhalo_number_density_distribution}). Subhaloes are typically more strongly clustered at smaller scales in the hydro than in DMO simulations, which reflects the fact that more are present within the virial regions of their centrals at $z=0$. As was argued before, this is likely due to a combination of longer survival times and easier identification of subhaloes thanks to their denser centres. For {\SUBFIND}, this results in the clustering of the lowest $V_{\mathrm{max}}$ subhaloes increasing at small scales relative to what is found in the DMO simulation. However, the two highest $V_{\mathrm{max}}$ bins remain unchanged for {\SUBFIND}, as the inclusion of hydro does not facilitate the identification of the most massive satellites.

The level of agreement between subhalo finders is different for the hydrodynamical simulations. {\ROCKSTAR} and {\VELOCIRAPTOR} become more discrepant relative to {\HBTHERONS} at small scales, which is caused by their worse performance in finding subhaloes once baryons are included (see \S\ref{section:halo_mass_functions}). On the other hand, {\SUBFIND} shows better agreement, as per the better identification of denser satellites within their associated haloes. However, the largest $V_{\mathrm{max}}$ bin shows no improvement. As an increase in the resolution does not improve the identification of subhaloes with large ratios of satellite to host bound mass (Fig.~\ref{figure:radial_distribution_satellite_mass_ratio_bins_convergence}), the population of merging high $V_{\mathrm{max}}$ subhaloes will not be found. We therefore posit that the {\SUBFIND} correlation functions in the highest $V_{\mathrm{max}}$ bin will remain suppressed relative to {\HBTHERONS} even for simulations that reach higher resolutions and larger volumes. 

\section{Conclusions}\label{section:conclusions}

Making inferences in the era of precision cosmology and large galaxy surveys requires quantifying how uncertain predictions are. One aspect that plays an important role is the identification of (sub)haloes and galaxies within simulations. The process of structure finding is non-trivial, as collapsed structures can be found in a variety of environments and evolutionary stages. In some instances, this results in the mischaracterisation of subhaloes and galaxies.

In this study, we have explored how the choice of subhalo finder influences basic summary statistics derived from a subset of the {\tt FLAMINGO} DMO and hydrodynamical simulations \citep{Schaye.2023, Kugel.2023}. The {\tt FLAMINGO} suite provides a good test bench to study how (sub)halo finding algorithms influence predictions relevant to large-scale surveys, as they provide a large statistical sample, include both DMO and hydrodynamical simulations spanning a large range of resolutions, and are tailored to reproduce observables relevant to such studies, e.g. cluster gas fractions and stellar mass functions. 

For our analysis, we chose to compare four representative subhalo finding algorithms, all of which rely on running the Friends-of-Friends algorithm first. Each of the subhalo finders we use are representative of commonly used methods: configuration- ({\SUBFIND}), phase- ({\ROCKSTAR} \& {\VELOCIRAPTOR}) and history-space ({\HBTHERONS}). {\SUBFIND} relies on identifying subhaloes based on finding spatial density peaks. {\VELOCIRAPTOR} instead finds peaks in position and velocity space. {\ROCKSTAR} is also based on phase-space finding, but runs a 6D FoF algorithm using the position and velocity of particles. {\HBTHERONS} uses the past evolution of subhaloes to tag associated particles. The association is propagated forward in time to identify potential subhalo candidates, based on grouping particles that share a subhalo progenitor.

As each finder makes different assumptions and uses different methods to compute subhalo properties, we used an external tool ({\SOAP}; \citealt{McGibbon.2025}) to measure subhalo properties in a consistent manner. Additionally, we also implemented exclusive mass assignment in {\ROCKSTAR} (i.e. a particle can only be bound to a single subhalo) so that it more closely resembles the manner in which the other three subhalo finders assign masses to subhaloes. To avoid trivial resolution effects, we only consider subhaloes with masses greater than 100 dark matter particles. Our main results can be summarised as follows:
\begin{enumerate}

    \item The computational cost and scalability of subhalo finders differs significantly. For the simulations we tested here, {\SUBFIND} and {\VELOCIRAPTOR} took $\mathcal{O}(10^{3})$~CPU-h to analyse the same output that took {\HBTHERONS} and {\ROCKSTAR} only $\mathcal{O}(10)$~CPU-h. As an example, the {\midres} hydrodynamical {\tt FLAMINGO} simulation took $\mathcal{O}(10^{6})$~CPU-h to complete and contains 79 outputs. The cost of using {\HBTHERONS} and {\ROCKSTAR} to find subhaloes across all outputs is negligible compared to the cost of running the simulation. In contrast, {\SUBFIND} and {\VELOCIRAPTOR} require a comparable investment to running the simulation itself. 

    \item During the development of subhalo finders, the effects of adding baryons to simulations is often overlooked, because the algorithms are initially developed for and tested with dark-matter-only simulations. As such, key assumptions used in the finding process may not hold for hydrodynamical simulations. This generally results in poorer performance, with the number of catastrophic failures during the identification dramatically increasing. For {\HBT}, we addressed these issues by implementing a series of improvements in a new version named {\HBTHERONS}. These were primarily made with the objective of improving its subhalo finding and time-tracking performance in hydrodynamical simulations (\S\ref{Appendix:incorrect_tracers}). However, the new additions also propagate to better tracking in dark-matter-only simulations (\S\ref{appendix:host_finding}, \S\ref{appendix:merging_calculations}, \S\ref{appendix:orphan_tracking}, \S\ref{appendix:disrupted_descendants}, \S\ref{appendix:unbinding_subsampling_effects}). Additionally, a number of other longstanding issues, such as particle duplication in the catalogues (\S\ref{appendix:duplicate_particles}) and inadequate particle partitioning, have also been solved.
    
    \item All finders except {\SUBFIND} and {\HBTHERONS} perform worse in hydrodynamical than in DMO runs. As a matter of fact, {\SUBFIND} performs better for hydro than for DMO by virtue of more concentrated subhalo centres facilitating the identification of density peaks. However, for higher resolution simulations, e.g. with cold and dense gas clumps within the ISM or CGM, we expect {\SUBFIND} to identify those clumps as separate subhaloes from the galaxy that hosts them, which would require additional processing to construct a galaxy catalogue.

    \item Basic summary statistics that should ideally be relatively insensitive to the choice of subhalo finder, such as the $M_{\mathrm{200c}}$ mass functions, change by up to 10\% (Fig.~\ref{figure:M200_mass_function}) and the differences are insensitive to resolution (Fig.~\ref{figure:M200_mass_function_convergence}). Only {\HBTHERONS} and {\SUBFIND} agree within 1\%. The discrepancies in the values we found are due to halo miscentering and differences in how central subhaloes are defined. Miscentering primarily affects {\VELOCIRAPTOR}, worsens for hydro, and causes $3\%$-level differences for well-resolved haloes. The definition of which subhalo is the central primarily affects {\ROCKSTAR} and is the more problematic source of disagreement. This is because, beyond its induced 10\%-level differences, it highlights that the current freedom in defining centrals causes systematic shifts. Although some authors may argue for or against a given definition, this effect should be accounted for when comparing with observations.

    \item Following from iv), an implicit choice made when running subhalo finders that assign a single central per FoF group ({\HBTHERONS}, {\SUBFIND} and {\VELOCIRAPTOR}) is the FoF linking length. Given that there is no a priori physically correct choice when it comes to comparisons to observations, as it is impossible to run a 3D FoF algorithm on the underlying DM distribution around observed galaxies, we explored how strongly the choice of linking length affects the resulting z = 0 $M_{\mathrm{200c}}$ mass functions of {\HBTHERONS}. We did this by varying the fiducial linking length of 0.2 to 0.16 and 0.25 times the mean interparticle separation. These linking lengths roughly correspond to a factor of two lower (0.25) and higher (0.16) overdensities relative to the commonly used choice. We found steeper and shallower mass functions for the shorter and longer linking lengths, respectively. This can change the predicted abundance of Milky Way-mass haloes by up to 6\%, and the difference only drops below 1\% for the most massive clusters of galaxies    (Fig.~\ref{figure:fof_linking_length_effect}).

    \item The bound mass functions differ by larger amounts ($\approx 50\%$) than the $M_{\mathrm{200c}}$ mass functions (Fig.~\ref{figure:bound_mass_function}). {\SUBFIND} assigns less mass to satellites than {\HBTHERONS}, a well-known shortcoming which boosts the bound mass of the central subhaloes it finds. {\VELOCIRAPTOR} and {\HBTHERONS} agree well, although there are slightly more massive satellites in {\VELOCIRAPTOR}. This is likely due to the central definition, as {\VELOCIRAPTOR} does not require the most massive subhalo to be the central of a FoF group. {\ROCKSTAR} satellites have a lot less bound mass than for the other finders, and the associated number densities also change more strongly with resolution (Fig.~\ref{figure:convergence_bound_mass_function}). Differences remain even when counting subhaloes as a function of $V_{\mathrm{max}}$ instead of $M_{\mathrm{bound}}$ (Fig.~\ref{figure:vmax_mass_function_exclusive}), although it improves the agreement between {\SUBFIND} and {\HBTHERONS}.

    \item The normalisation and shape of the subhalo radial distributions are both sensitive to which subhalo finder is used. Close to $R_{\mathrm{200c}}$, where the difficulties associated with analysing the dense core of haloes are less important, the number density of subhaloes, $n(r)$, can vary by as much as $20\%$ between subhalo finders (Fig.~\ref{figure:subhalo_number_R200}). The shapes of the distribution within $R_{\mathrm{200c}}$ can also change dramatically, with $10\%$ agreement for $n(r) / n(R_{\mathrm{200c}})$ only reached beyond $0.1R_{\mathrm{200c}}$ (Fig.~\ref{figure:subhalo_number_density_distribution}). Out of the fours subhalo finders we compare, {\SUBFIND} struggles the most to find subhaloes near the central regions of haloes, as it exhibits a clear suppression in the number density of subhaloes at small radii regardless of $M_{200\mathrm{c}}$. The problem with the identification of subhaloes by {\SUBFIND} worsens as the mass of the satellite subhalo relative to its host increases, meaning that either the most massive satellite subhaloes are typically missing or their masses are severely suppressed relative to the values found in the other subhalo finders (Fig.~\ref{figure:radial_distribution_satellite_mass_ratio_bins}). This occurs because the a significant fraction of the satellite mass is artificially assigned to the chosen central subhalo, leaving a subhalo with an otherwise comparable mass severally depleted in particle content.

    \item The inclusion of hydrodynamics increases the number of subhaloes near $R_{\mathrm{200c}}$ and steepens their number density profiles. This is likely driven by the cores of subhaloes becoming denser, both due to baryons being more concentrated than the DM and the increase in the central DM density in response to the baryons. As denser subhaloes are more resilient to disruption and stripping, more survive to $z=0$. Additionally, the agreement between subhalo finders tends to improve, particularly for {\SUBFIND}. This is because the more prominent density peaks are easier to identify against the backdrop of the halo they are embedded in. 
    
    \item Satellites are segregated by their mass relative to that of the host, with the more massive ones being more concentrated towards the centre of the host halo (Fig.~\ref{figure:radial_distribution_satellite_mass_ratio_bins}). This is opposite to the trend found in previous studies based on {\SUBFIND}, which claimed that more massive satellites are \textit{less} concentrated towards the centre of their host. We think the interpretation offered in those studies suffered from systematically biased subhalo catalogues. This major discrepancy across subhalo finders remains across resolution levels (Fig.~\ref{figure:radial_distribution_satellite_mass_ratio_bins_convergence}). In other words, different subhalo finders converge to qualitatively different answers.

    \item The auto-correlation functions of all subhaloes, when selected in bins of different $V_{\mathrm{max}}$, can differ at both large and small distances between different subhalo finders (Fig.~\ref{figure:correlation_functions}). The disagreement between subhalo finders depends sensitively on the chosen $V_{\mathrm{max}}$ bin, and worsens at small scales for larger values of $V_{\mathrm{max}}$, i.e. for subhaloes that should be well resolved. The primary difference in the two-point correlation function at large scales is a systematic offset between {\ROCKSTAR} and the rest of the subhalo finders, caused by a different $V_{\mathrm{max}}$ dependence on mass. Applying the same $V_{\mathrm{max}}$ cut selects subhaloes of different average masses, and hence clustering strength. This reflects the fact that the $V_{\mathrm{max}}$ of a subhalo is calculated using only the particles bound to it, so the unbinding procedure that each algorithm uses can influence its value.  At small scales, the differences are driven by how well satellite systems can be recovered. As such, all subhalo finders under predict the correlation function relative to {\HBTHERONS} within $\approx 2~\mathrm{Mpc}$.

\end{enumerate}

In short, (sub)halo finding remains an open problem, and the choice of which algorithm to use needs to be made carefully. As it is unlikely, and probably undesirable, for the whole community to adopt a single approach to finding subhaloes and to use the same definitions -- and none of the (sub)halo finding algorithms can be directly applied to observations -- part of the uncertainty budget in making predictions should be devoted to the manner in which (sub)haloes are found. A related, but slightly different aspect, is how the properties of galaxies change across subhalo finders. As they are located in the centres of dark matter subhaloes, and the identification of a subhalo core is typically easier than the identification of its edge, the properties of galaxies may not differ as much as what we have discussed in this study.

Indeed, answering the question of which subhalo finder is the best to use is complicated. The difficulty arises primarily from the fact that there are no clear, definitive answers to the question of what the properties of any given subhalo population should be. Any realistic prediction to compare against is inevitably based on simulations, and hence on a series of choices of how the subhaloes were found and defined. As we have shown here, many differences exist depending on the chosen subhalo finder, so trying to `calibrate' the performance of a subhalo finder to a reference solution is moot.

Nevertheless, some useful insights can still be gained. Imaging many haloes to identify whether any visually obvious subhaloes are missing from the catalogues reveals catastrophic failures (e.g. Fig.~\ref{figure:image_example}). The convergence of the properties of subhalo populations with increasing resolution, and whether the recovered trends are sensible, also provide a handle with which to identify better performing algorithms. However, this is always with the caveat that trends can often be explained by a variety of plausible mechanisms (e.g. the radial distribution of the most massive subhaloes) and that the finders seem to converge towards different answers as the resolution is increased. Lastly, an extremely powerful test is to examine the time evolution of the subhalo population, as failures difficult to identify at a fixed time can become evident when integrated over a Hubble time (e.g. \citealt{Chandro-Gomez.2025}).

Based on the above methodology, we conclude that {\HBTHERONS} provides the most robust subhalo catalogues out of the four subhalo finders we have explored in this work. There are generally no visibly obvious subhaloes missing from its catalogues based on extensive imaging, and the trends and convergence appear sensible. For time-integrated tests, \citet{Chandro-Gomez.2025} find that {\HBTHERONS} drastically suppresses the number of unphysical results in the time evolution of subhalo populations compared to several combinations of alternative subhalo finders ({\VELOCIRAPTOR} and {\SUBFIND}) and merger tree algorithms ({\tt D-Trees+DHalo}, \citealt{Jiang.2014}; and {\tt TreeFrog}, \citealt{Elahi.2019b}). As such, we make {\HBTHERONS} the default subhalo finder for the {\tt FLAMINGO} simulations and make it publicly available.

\section*{Acknowledgements}

We thank Peter Behroozi, Pascal Elahi and Volker Springel for their useful comments and discussion during the preparation of this manuscript.  We also thank the referee for a constructive report that improved the presentation of our work. JCH thanks Shaun Cole and Carlos Frenk for their discussions and suggestions. This work has benefited from using {\tt swiftsimio} \citep{Borrow.2020}, both to handle particle and (sub)halo data, and to image (sub)haloes. We thank Jeger Broxteman and Will McDonald for providing a suitable bird-related acronym.

This work used the DiRAC@Durham facility managed by the Institute for Computational Cosmology on behalf of the STFC DiRAC HPC Facility (www.dirac.ac.uk). The equipment was funded by BEIS capital funding via STFC capital grants ST/K00042X/1, ST/P002293/1, ST/R002371/1 and ST/S002502/1, Durham University and STFC operations grant ST/R000832/1. DiRAC is part of the National e-Infrastructure. VJFM acknowledges support by NWO through the Dark Universe Science Collaboration (OCENW.XL21.XL21.025). JXH acknowledges support from National Key R\&D Program of China (2023YFA1607800, 2023YFA1607801), China Manned Space Project (No.\ CMS-
CSST-2021-A03), and 111 project (No.\ B20019). YMB acknowledges support from UK Research and Innovation through a Future Leaders Fellowship (grant agreement MR/X035166/1) and financial support from the Swiss National Science Foundation (SNSF) under project 200021\_213076

%%%%%%%%%%%%%%%%%%%%%%%%%%%%%%%%%%%%%%%%%%%%%%%%%%
\section*{Data Availability}

The data used in this study can be made available upon reasonable request to the corresponding author. The {\HBTHERONS} algorithm (\href{https://github.com/SWIFTSIM/HBT-HERONS}{https://github.com/SWIFTSIM/HBT-HERONS}) and {\SOAP} package ({\href{https://github.com/SWIFTSIM/SOAP}{https://github.com/SWIFTSIM/SOAP}}) are publicly available and can handle the data output of {\tt Swift} simulations.

%%%%%%%%%%%%%%%%%%%% REFERENCES %%%%%%%%%%%%%%%%%%

\bibliographystyle{mnras}
\bibliography{references}

\begin{thebibliography}{}
\makeatletter
\relax
\def\mn@urlcharsother{\let\do\@makeother \do\$\do\&\do\#\do\^\do\_\do\%\do\~}
\def\mn@doi{\begingroup\mn@urlcharsother \@ifnextchar [ {\mn@doi@} {\mn@doi@[]}}
\def\mn@doi@[#1]#2{\def\@tempa{#1}\ifx\@tempa\@empty \href {http://dx.doi.org/#2} {doi:#2}\else \href {http://dx.doi.org/#2} {#1}\fi \endgroup}
\def\mn@eprint#1#2{\mn@eprint@#1:#2::\@nil}
\def\mn@eprint@arXiv#1{\href {http://arxiv.org/abs/#1} {{\tt arXiv:#1}}}
\def\mn@eprint@dblp#1{\href {http://dblp.uni-trier.de/rec/bibtex/#1.xml} {dblp:#1}}
\def\mn@eprint@#1:#2:#3:#4\@nil{\def\@tempa {#1}\def\@tempb {#2}\def\@tempc {#3}\ifx \@tempc \@empty \let \@tempc \@tempb \let \@tempb \@tempa \fi \ifx \@tempb \@empty \def\@tempb {arXiv}\fi \@ifundefined {mn@eprint@\@tempb}{\@tempb:\@tempc}{\expandafter \expandafter \csname mn@eprint@\@tempb\endcsname \expandafter{\@tempc}}}

\bibitem[\protect\citeauthoryear{{Abbott} et~al.,}{{Abbott} et~al.}{2022}]{Abbott.2022}
{Abbott} T.~M.~C.,  et~al., 2022, \mn@doi [\prd] {10.1103/PhysRevD.105.023520}, \href {https://ui.adsabs.harvard.edu/abs/2022PhRvD.105b3520A} {105, 023520}

\bibitem[\protect\citeauthoryear{{Angulo}, {Lacey}, {Baugh}  \& {Frenk}}{{Angulo} et~al.}{2009}]{Angulo.2009}
{Angulo} R.~E.,  {Lacey} C.~G.,  {Baugh} C.~M.,   {Frenk} C.~S.,  2009, \mn@doi [\mnras] {10.1111/j.1365-2966.2009.15333.x}, \href {https://ui.adsabs.harvard.edu/abs/2009MNRAS.399..983A} {399, 983}

\bibitem[\protect\citeauthoryear{{Aubert}, {Pichon}  \& {Colombi}}{{Aubert} et~al.}{2004}]{Aubert.2004}
{Aubert} D.,  {Pichon} C.,   {Colombi} S.,  2004, \mn@doi [\mnras] {10.1111/j.1365-2966.2004.07883.x}, \href {https://ui.adsabs.harvard.edu/abs/2004MNRAS.352..376A} {352, 376}

\bibitem[\protect\citeauthoryear{{Bah{\'e}} et~al.,}{{Bah{\'e}} et~al.}{2019}]{Bahe.2019}
{Bah{\'e}} Y.~M.,  et~al., 2019, \mn@doi [\mnras] {10.1093/mnras/stz361}, \href {https://ui.adsabs.harvard.edu/abs/2019MNRAS.485.2287B} {485, 2287}

\bibitem[\protect\citeauthoryear{{Bah{\'e}} et~al.,}{{Bah{\'e}} et~al.}{2022}]{Bahe.2022}
{Bah{\'e}} Y.~M.,  et~al., 2022, \mn@doi [\mnras] {10.1093/mnras/stac1339}, \href {https://ui.adsabs.harvard.edu/abs/2022MNRAS.516..167B} {516, 167}

\bibitem[\protect\citeauthoryear{{Behroozi}, {Wechsler}  \& {Wu}}{{Behroozi} et~al.}{2013a}]{Behroozi.2013}
{Behroozi} P.~S.,  {Wechsler} R.~H.,   {Wu} H.-Y.,  2013a, \mn@doi [\apj] {10.1088/0004-637X/762/2/109}, \href {https://ui.adsabs.harvard.edu/abs/2013ApJ...762..109B} {762, 109}

\bibitem[\protect\citeauthoryear{{Behroozi}, {Wechsler}, {Wu}, {Busha}, {Klypin}  \& {Primack}}{{Behroozi} et~al.}{2013b}]{Behroozi.2013_consistent_trees}
{Behroozi} P.~S.,  {Wechsler} R.~H.,  {Wu} H.-Y.,  {Busha} M.~T.,  {Klypin} A.~A.,   {Primack} J.~R.,  2013b, \mn@doi [\apj] {10.1088/0004-637X/763/1/18}, \href {https://ui.adsabs.harvard.edu/abs/2013ApJ...763...18B} {763, 18}

\bibitem[\protect\citeauthoryear{{Behroozi} et~al.,}{{Behroozi} et~al.}{2015}]{Behroozi.2015}
{Behroozi} P.,  et~al., 2015, \mn@doi [\mnras] {10.1093/mnras/stv2046}, \href {https://ui.adsabs.harvard.edu/abs/2015MNRAS.454.3020B} {454, 3020}

\bibitem[\protect\citeauthoryear{{Booth} \& {Schaye}}{{Booth} \& {Schaye}}{2009}]{Booth.2009}
{Booth} C.~M.,  {Schaye} J.,  2009, \mn@doi [\mnras] {10.1111/j.1365-2966.2009.15043.x}, \href {https://ui.adsabs.harvard.edu/abs/2009MNRAS.398...53B} {398, 53}

\bibitem[\protect\citeauthoryear{Borrow \& Borrisov}{Borrow \& Borrisov}{2020}]{Borrow.2020}
Borrow J.,  Borrisov A.,  2020, \mn@doi [Journal of Open Source Software] {10.21105/joss.02430}, 5, 2430

\bibitem[\protect\citeauthoryear{{Borrow}, {Schaller}, {Bower}  \& {Schaye}}{{Borrow} et~al.}{2022}]{Borrow.2022}
{Borrow} J.,  {Schaller} M.,  {Bower} R.~G.,   {Schaye} J.,  2022, \mn@doi [\mnras] {10.1093/mnras/stab3166}, \href {https://ui.adsabs.harvard.edu/abs/2022MNRAS.511.2367B} {511, 2367}

\bibitem[\protect\citeauthoryear{{Bryan} \& {Norman}}{{Bryan} \& {Norman}}{1998}]{Bryan.1998}
{Bryan} G.~L.,  {Norman} M.~L.,  1998, \mn@doi [\apj] {10.1086/305262}, \href {https://ui.adsabs.harvard.edu/abs/1998ApJ...495...80B} {495, 80}

\bibitem[\protect\citeauthoryear{{Chaikin}, {Schaye}, {Schaller}, {Bah{\'e}}, {Nobels}  \& {Ploeckinger}}{{Chaikin} et~al.}{2022}]{Chaikin.2022}
{Chaikin} E.,  {Schaye} J.,  {Schaller} M.,  {Bah{\'e}} Y.~M.,  {Nobels} F. S.~J.,   {Ploeckinger} S.,  2022, \mn@doi [\mnras] {10.1093/mnras/stac1132}, \href {https://ui.adsabs.harvard.edu/abs/2022MNRAS.514..249C} {514, 249}

\bibitem[\protect\citeauthoryear{{Chaikin}, {Schaye}, {Schaller}, {Ben{\'\i}tez-Llambay}, {Nobels}  \& {Ploeckinger}}{{Chaikin} et~al.}{2023}]{Chaikin.2023}
{Chaikin} E.,  {Schaye} J.,  {Schaller} M.,  {Ben{\'\i}tez-Llambay} A.,  {Nobels} F. S.~J.,   {Ploeckinger} S.,  2023, \mn@doi [\mnras] {10.1093/mnras/stad1626}, \href {https://ui.adsabs.harvard.edu/abs/2023MNRAS.523.3709C} {523, 3709}

\bibitem[\protect\citeauthoryear{{Chandro-G{\'o}mez} et~al.,}{{Chandro-G{\'o}mez} et~al.}{2025}]{Chandro-Gomez.2025}
{Chandro-G{\'o}mez} {\'A}.,  et~al., 2025, \mn@doi [arXiv e-prints] {10.48550/arXiv.2501.07677}, \href {https://ui.adsabs.harvard.edu/abs/2025arXiv250107677C} {p. arXiv:2501.07677}

\bibitem[\protect\citeauthoryear{{Dalla Vecchia} \& {Schaye}}{{Dalla Vecchia} \& {Schaye}}{2008}]{DallaVecchia.2008}
{Dalla Vecchia} C.,  {Schaye} J.,  2008, \mn@doi [\mnras] {10.1111/j.1365-2966.2008.13322.x}, \href {https://ui.adsabs.harvard.edu/abs/2008MNRAS.387.1431D} {387, 1431}

\bibitem[\protect\citeauthoryear{{De Lucia}, {Kauffmann}, {Springel}, {White}, {Lanzoni}, {Stoehr}, {Tormen}  \& {Yoshida}}{{De Lucia} et~al.}{2004}]{DeLucia.2004}
{De Lucia} G.,  {Kauffmann} G.,  {Springel} V.,  {White} S.~D.~M.,  {Lanzoni} B.,  {Stoehr} F.,  {Tormen} G.,   {Yoshida} N.,  2004, \mn@doi [\mnras] {10.1111/j.1365-2966.2004.07372.x}, \href {https://ui.adsabs.harvard.edu/abs/2004MNRAS.348..333D} {348, 333}

\bibitem[\protect\citeauthoryear{{Diemer}, {Behroozi}  \& {Mansfield}}{{Diemer} et~al.}{2024}]{Diemer.2024}
{Diemer} B.,  {Behroozi} P.,   {Mansfield} P.,  2024, \mn@doi [\mnras] {10.1093/mnras/stae2007}, \href {https://ui.adsabs.harvard.edu/abs/2024MNRAS.533.3811D} {533, 3811}

\bibitem[\protect\citeauthoryear{{Driver} et~al.,}{{Driver} et~al.}{2022}]{Driver.2022}
{Driver} S.~P.,  et~al., 2022, \mn@doi [\mnras] {10.1093/mnras/stac472}, \href {https://ui.adsabs.harvard.edu/abs/2022MNRAS.513..439D} {513, 439}

\bibitem[\protect\citeauthoryear{{Drlica-Wagner} et~al.,}{{Drlica-Wagner} et~al.}{2019}]{Drlica-Wagner.2019}
{Drlica-Wagner} A.,  et~al., 2019, \mn@doi [arXiv e-prints] {10.48550/arXiv.1902.01055}, \href {https://ui.adsabs.harvard.edu/abs/2019arXiv190201055D} {p. arXiv:1902.01055}

\bibitem[\protect\citeauthoryear{{Efstathiou}, {Sutherland}  \& {Maddox}}{{Efstathiou} et~al.}{1990}]{Efstathiou.1990}
{Efstathiou} G.,  {Sutherland} W.~J.,   {Maddox} S.~J.,  1990, \mn@doi [\nat] {10.1038/348705a0}, \href {https://ui.adsabs.harvard.edu/abs/1990Natur.348..705E} {348, 705}

\bibitem[\protect\citeauthoryear{{Elahi}, {Thacker}  \& {Widrow}}{{Elahi} et~al.}{2011}]{Elahi.2011}
{Elahi} P.~J.,  {Thacker} R.~J.,   {Widrow} L.~M.,  2011, \mn@doi [\mnras] {10.1111/j.1365-2966.2011.19485.x}, \href {https://ui.adsabs.harvard.edu/abs/2011MNRAS.418..320E} {418, 320}

\bibitem[\protect\citeauthoryear{{Elahi}, {Ca{\~n}as}, {Poulton}, {Tobar}, {Willis}, {Lagos}, {Power}  \& {Robotham}}{{Elahi} et~al.}{2019a}]{Elahi.2019}
{Elahi} P.~J.,  {Ca{\~n}as} R.,  {Poulton} R. J.~J.,  {Tobar} R.~J.,  {Willis} J.~S.,  {Lagos} C. d.~P.,  {Power} C.,   {Robotham} A. S.~G.,  2019a, \mn@doi [\pasa] {10.1017/pasa.2019.12}, \href {https://ui.adsabs.harvard.edu/abs/2019PASA...36...21E} {36, e021}

\bibitem[\protect\citeauthoryear{{Elahi}, {Poulton}, {Tobar}, {Ca{\~n}as}, {Lagos}, {Power}  \& {Robotham}}{{Elahi} et~al.}{2019b}]{Elahi.2019b}
{Elahi} P.~J.,  {Poulton} R. J.~J.,  {Tobar} R.~J.,  {Ca{\~n}as} R.,  {Lagos} C. d.~P.,  {Power} C.,   {Robotham} A. S.~G.,  2019b, \mn@doi [\pasa] {10.1017/pasa.2019.18}, \href {https://ui.adsabs.harvard.edu/abs/2019PASA...36...28E} {36, e028}

\bibitem[\protect\citeauthoryear{{Elbers}, {Frenk}, {Jenkins}, {Li}  \& {Pascoli}}{{Elbers} et~al.}{2021}]{Elbers.2021}
{Elbers} W.,  {Frenk} C.~S.,  {Jenkins} A.,  {Li} B.,   {Pascoli} S.,  2021, \mn@doi [\mnras] {10.1093/mnras/stab2260}, \href {https://ui.adsabs.harvard.edu/abs/2021MNRAS.507.2614E} {507, 2614}

\bibitem[\protect\citeauthoryear{{Elbers}, {Frenk}, {Jenkins}, {Li}  \& {Pascoli}}{{Elbers} et~al.}{2022}]{Elbers.2022}
{Elbers} W.,  {Frenk} C.~S.,  {Jenkins} A.,  {Li} B.,   {Pascoli} S.,  2022, \mn@doi [\mnras] {10.1093/mnras/stac2365}, \href {https://ui.adsabs.harvard.edu/abs/2022MNRAS.516.3821E} {516, 3821}

\bibitem[\protect\citeauthoryear{{Euclid Collaboration} et~al.,}{{Euclid Collaboration} et~al.}{2023}]{Euclid.2023}
{Euclid Collaboration} et~al., 2023, \mn@doi [\aap] {10.1051/0004-6361/202244674}, \href {https://ui.adsabs.harvard.edu/abs/2023A&A...671A.100E} {671, A100}

\bibitem[\protect\citeauthoryear{{Euclid Collaboration} et~al.,}{{Euclid Collaboration} et~al.}{2025a}]{Euclid.2024}
{Euclid Collaboration} et~al., 2025a, \mn@doi [\aap] {10.1051/0004-6361/202450810}, \href {https://ui.adsabs.harvard.edu/abs/2025A&A...697A...1E} {697, A1}

\bibitem[\protect\citeauthoryear{{Euclid Collaboration} et~al.,}{{Euclid Collaboration} et~al.}{2025b}]{EuclidCollaboration.2024}
{Euclid Collaboration} et~al., 2025b, \mn@doi [\aap] {10.1051/0004-6361/202450853}, \href {https://ui.adsabs.harvard.edu/abs/2025A&A...697A...5E} {697, A5}

\bibitem[\protect\citeauthoryear{{Garc{\'\i}a} \& {Rozo}}{{Garc{\'\i}a} \& {Rozo}}{2019}]{Garcia.2019}
{Garc{\'\i}a} R.,  {Rozo} E.,  2019, \mn@doi [\mnras] {10.1093/mnras/stz2458}, \href {https://ui.adsabs.harvard.edu/abs/2019MNRAS.489.4170G} {489, 4170}

\bibitem[\protect\citeauthoryear{{Garrison-Kimmel} et~al.,}{{Garrison-Kimmel} et~al.}{2017}]{Garrison-Kimmel.2017}
{Garrison-Kimmel} S.,  et~al., 2017, \mn@doi [\mnras] {10.1093/mnras/stx1710}, \href {https://ui.adsabs.harvard.edu/abs/2017MNRAS.471.1709G} {471, 1709}

\bibitem[\protect\citeauthoryear{{Genel}, {Genzel}, {Bouch{\'e}}, {Naab}  \& {Sternberg}}{{Genel} et~al.}{2009}]{Genel.2009}
{Genel} S.,  {Genzel} R.,  {Bouch{\'e}} N.,  {Naab} T.,   {Sternberg} A.,  2009, \mn@doi [\apj] {10.1088/0004-637X/701/2/2002}, \href {https://ui.adsabs.harvard.edu/abs/2009ApJ...701.2002G} {701, 2002}

\bibitem[\protect\citeauthoryear{{Gill}, {Knebe}  \& {Gibson}}{{Gill} et~al.}{2004}]{Gill.2004}
{Gill} S. P.~D.,  {Knebe} A.,   {Gibson} B.~K.,  2004, \mn@doi [\mnras] {10.1111/j.1365-2966.2004.07786.x}, \href {https://ui.adsabs.harvard.edu/abs/2004MNRAS.351..399G} {351, 399}

\bibitem[\protect\citeauthoryear{{Giocoli}, {Tormen}, {Sheth}  \& {van den Bosch}}{{Giocoli} et~al.}{2010}]{Giocoli.2010}
{Giocoli} C.,  {Tormen} G.,  {Sheth} R.~K.,   {van den Bosch} F.~C.,  2010, \mn@doi [\mnras] {10.1111/j.1365-2966.2010.16311.x}, \href {https://ui.adsabs.harvard.edu/abs/2010MNRAS.404..502G} {404, 502}

\bibitem[\protect\citeauthoryear{{G{\'o}mez}, {Padilla}, {Helly}, {Lacey}, {Baugh}  \& {Lagos}}{{G{\'o}mez} et~al.}{2022}]{Gomez.2022}
{G{\'o}mez} J.~S.,  {Padilla} N.~D.,  {Helly} J.~C.,  {Lacey} C.~G.,  {Baugh} C.~M.,   {Lagos} C.~D.~P.,  2022, \mn@doi [\mnras] {10.1093/mnras/stab3661}, \href {https://ui.adsabs.harvard.edu/abs/2022MNRAS.510.5500G} {510, 5500}

\bibitem[\protect\citeauthoryear{{Griffen}, {Ji}, {Dooley}, {G{\'o}mez}, {Vogelsberger}, {O'Shea}  \& {Frebel}}{{Griffen} et~al.}{2016}]{Griffen.2016}
{Griffen} B.~F.,  {Ji} A.~P.,  {Dooley} G.~A.,  {G{\'o}mez} F.~A.,  {Vogelsberger} M.,  {O'Shea} B.~W.,   {Frebel} A.,  2016, \mn@doi [\apj] {10.3847/0004-637X/818/1/10}, \href {https://ui.adsabs.harvard.edu/abs/2016ApJ...818...10G} {818, 10}

\bibitem[\protect\citeauthoryear{{Guo} \& {White}}{{Guo} \& {White}}{2014}]{Guo.2014}
{Guo} Q.,  {White} S.,  2014, \mn@doi [\mnras] {10.1093/mnras/stt2116}, \href {https://ui.adsabs.harvard.edu/abs/2014MNRAS.437.3228G} {437, 3228}

\bibitem[\protect\citeauthoryear{{Haggar}, {Pearce}, {Gray}, {Knebe}  \& {Yepes}}{{Haggar} et~al.}{2021}]{Haggar.2021}
{Haggar} R.,  {Pearce} F.~R.,  {Gray} M.~E.,  {Knebe} A.,   {Yepes} G.,  2021, \mn@doi [\mnras] {10.1093/mnras/stab064}, \href {https://ui.adsabs.harvard.edu/abs/2021MNRAS.502.1191H} {502, 1191}

\bibitem[\protect\citeauthoryear{{Hahn}, {Martizzi}, {Wu}, {Evrard}, {Teyssier}  \& {Wechsler}}{{Hahn} et~al.}{2017}]{Hahn.2017}
{Hahn} O.,  {Martizzi} D.,  {Wu} H.-Y.,  {Evrard} A.~E.,  {Teyssier} R.,   {Wechsler} R.~H.,  2017, \mn@doi [\mnras] {10.1093/mnras/stx001}, \href {https://ui.adsabs.harvard.edu/abs/2017MNRAS.470..166H} {470, 166}

\bibitem[\protect\citeauthoryear{{Hahn}, {Rampf}  \& {Uhlemann}}{{Hahn} et~al.}{2021}]{Hahn.2021}
{Hahn} O.,  {Rampf} C.,   {Uhlemann} C.,  2021, \mn@doi [\mnras] {10.1093/mnras/staa3773}, \href {https://ui.adsabs.harvard.edu/abs/2021MNRAS.503..426H} {503, 426}

\bibitem[\protect\citeauthoryear{{Han}, {Jing}, {Wang}  \& {Wang}}{{Han} et~al.}{2012}]{Han.2012}
{Han} J.,  {Jing} Y.~P.,  {Wang} H.,   {Wang} W.,  2012, \mn@doi [\mnras] {10.1111/j.1365-2966.2012.22111.x}, \href {https://ui.adsabs.harvard.edu/abs/2012MNRAS.427.2437H} {427, 2437}

\bibitem[\protect\citeauthoryear{{Han}, {Cole}, {Frenk}, {Benitez-Llambay}  \& {Helly}}{{Han} et~al.}{2018}]{Han.2018}
{Han} J.,  {Cole} S.,  {Frenk} C.~S.,  {Benitez-Llambay} A.,   {Helly} J.,  2018, \mn@doi [\mnras] {10.1093/mnras/stx2792}, \href {https://ui.adsabs.harvard.edu/abs/2018MNRAS.474..604H} {474, 604}

\bibitem[\protect\citeauthoryear{{Hu{\v{s}}ko}, {Lacey}, {Schaye}, {Schaller}  \& {Nobels}}{{Hu{\v{s}}ko} et~al.}{2022}]{Husko.2022}
{Hu{\v{s}}ko} F.,  {Lacey} C.~G.,  {Schaye} J.,  {Schaller} M.,   {Nobels} F. S.~J.,  2022, \mn@doi [\mnras] {10.1093/mnras/stac2278}, \href {https://ui.adsabs.harvard.edu/abs/2022MNRAS.516.3750H} {516, 3750}

\bibitem[\protect\citeauthoryear{{Jiang}, {Helly}, {Cole}  \& {Frenk}}{{Jiang} et~al.}{2014}]{Jiang.2014}
{Jiang} L.,  {Helly} J.~C.,  {Cole} S.,   {Frenk} C.~S.,  2014, \mn@doi [\mnras] {10.1093/mnras/stu390}, \href {https://ui.adsabs.harvard.edu/abs/2014MNRAS.440.2115J} {440, 2115}

\bibitem[\protect\citeauthoryear{{Jung} et~al.,}{{Jung} et~al.}{2024}]{Jung.2024}
{Jung} M.,  et~al., 2024, \mn@doi [\apj] {10.3847/1538-4357/ad245b}, \href {https://ui.adsabs.harvard.edu/abs/2024ApJ...964..123J} {964, 123}

\bibitem[\protect\citeauthoryear{{Kitzbichler} \& {White}}{{Kitzbichler} \& {White}}{2008}]{Kitzbichler.2008}
{Kitzbichler} M.~G.,  {White} S.~D.~M.,  2008, \mn@doi [\mnras] {10.1111/j.1365-2966.2008.13873.x}, \href {https://ui.adsabs.harvard.edu/abs/2008MNRAS.391.1489K} {391, 1489}

\bibitem[\protect\citeauthoryear{{Klypin}, {Gottl{\"o}ber}, {Kravtsov}  \& {Khokhlov}}{{Klypin} et~al.}{1999}]{Klypin.1999}
{Klypin} A.,  {Gottl{\"o}ber} S.,  {Kravtsov} A.~V.,   {Khokhlov} A.~M.,  1999, \mn@doi [\apj] {10.1086/307122}, \href {https://ui.adsabs.harvard.edu/abs/1999ApJ...516..530K} {516, 530}

\bibitem[\protect\citeauthoryear{{Knebe} et~al.,}{{Knebe} et~al.}{2011}]{Knebe.2011}
{Knebe} A.,  et~al., 2011, \mn@doi [\mnras] {10.1111/j.1365-2966.2011.18858.x}, \href {https://ui.adsabs.harvard.edu/abs/2011MNRAS.415.2293K} {415, 2293}

\bibitem[\protect\citeauthoryear{{Knollmann} \& {Knebe}}{{Knollmann} \& {Knebe}}{2009}]{Knollmann.2009}
{Knollmann} S.~R.,  {Knebe} A.,  2009, \mn@doi [\apjs] {10.1088/0067-0049/182/2/608}, \href {https://ui.adsabs.harvard.edu/abs/2009ApJS..182..608K} {182, 608}

\bibitem[\protect\citeauthoryear{{Kong}, {Boylan-Kolchin}  \& {Bullock}}{{Kong} et~al.}{2025}]{Kong.2025}
{Kong} H.,  {Boylan-Kolchin} M.,   {Bullock} J.~S.,  2025, \mn@doi [arXiv e-prints] {10.48550/arXiv.2503.10766}, \href {https://ui.adsabs.harvard.edu/abs/2025arXiv250310766K} {p. arXiv:2503.10766}

\bibitem[\protect\citeauthoryear{{Kugel} et~al.,}{{Kugel} et~al.}{2023}]{Kugel.2023}
{Kugel} R.,  et~al., 2023, \mn@doi [\mnras] {10.1093/mnras/stad2540}, \href {https://ui.adsabs.harvard.edu/abs/2023MNRAS.526.6103K} {526, 6103}

\bibitem[\protect\citeauthoryear{{Kugel}, {Schaye}, {Schaller}, {Moreno}  \& {McGibbon}}{{Kugel} et~al.}{2025}]{Kugel.2025}
{Kugel} R.,  {Schaye} J.,  {Schaller} M.,  {Moreno} V. J.~F.,   {McGibbon} R.~J.,  2025, \mn@doi [\mnras] {10.1093/mnras/staf111}, \href {https://ui.adsabs.harvard.edu/abs/2025MNRAS.tmp...98K} {}

\bibitem[\protect\citeauthoryear{{Mansfield}, {Darragh-Ford}, {Wang}, {Nadler}, {Diemer}  \& {Wechsler}}{{Mansfield} et~al.}{2024}]{Mansfield.2024}
{Mansfield} P.,  {Darragh-Ford} E.,  {Wang} Y.,  {Nadler} E.~O.,  {Diemer} B.,   {Wechsler} R.~H.,  2024, \mn@doi [\apj] {10.3847/1538-4357/ad4e33}, \href {https://ui.adsabs.harvard.edu/abs/2024ApJ...970..178M} {970, 178}

\bibitem[\protect\citeauthoryear{McGibbon, Helly, Schaye, Schaller  \& Vandenbroucke}{McGibbon et~al.}{2025}]{McGibbon.2025}
McGibbon R.,  Helly J.~C.,  Schaye J.,  Schaller M.,   Vandenbroucke B.,  2025, \mn@doi [Journal of Open Source Software] {10.21105/joss.08252}, 10, 8252

\bibitem[\protect\citeauthoryear{{More}, {Kravtsov}, {Dalal}  \& {Gottl{\"o}ber}}{{More} et~al.}{2011}]{More.2011}
{More} S.,  {Kravtsov} A.~V.,  {Dalal} N.,   {Gottl{\"o}ber} S.,  2011, \mn@doi [\apjs] {10.1088/0067-0049/195/1/4}, \href {https://ui.adsabs.harvard.edu/abs/2011ApJS..195....4M} {195, 4}

\bibitem[\protect\citeauthoryear{{Necib}, {Lisanti}, {Garrison-Kimmel}, {Wetzel}, {Sanderson}, {Hopkins}, {Faucher-Gigu{\`e}re}  \& {Kere{\v{s}}}}{{Necib} et~al.}{2019}]{Necib.2019}
{Necib} L.,  {Lisanti} M.,  {Garrison-Kimmel} S.,  {Wetzel} A.,  {Sanderson} R.,  {Hopkins} P.~F.,  {Faucher-Gigu{\`e}re} C.-A.,   {Kere{\v{s}}} D.,  2019, \mn@doi [\apj] {10.3847/1538-4357/ab3afc}, \href {https://ui.adsabs.harvard.edu/abs/2019ApJ...883...27N} {883, 27}

\bibitem[\protect\citeauthoryear{{Onions} et~al.,}{{Onions} et~al.}{2012}]{Onions.2012}
{Onions} J.,  et~al., 2012, \mn@doi [\mnras] {10.1111/j.1365-2966.2012.20947.x}, \href {https://ui.adsabs.harvard.edu/abs/2012MNRAS.423.1200O} {423, 1200}

\bibitem[\protect\citeauthoryear{{Pe{\~n}arrubia}, {Benson}, {Walker}, {Gilmore}, {McConnachie}  \& {Mayer}}{{Pe{\~n}arrubia} et~al.}{2010}]{Penarrubia.2010}
{Pe{\~n}arrubia} J.,  {Benson} A.~J.,  {Walker} M.~G.,  {Gilmore} G.,  {McConnachie} A.~W.,   {Mayer} L.,  2010, \mn@doi [\mnras] {10.1111/j.1365-2966.2010.16762.x}, \href {https://ui.adsabs.harvard.edu/abs/2010MNRAS.406.1290P} {406, 1290}

\bibitem[\protect\citeauthoryear{{Pizzati} et~al.,}{{Pizzati} et~al.}{2024}]{Pizzati.2024}
{Pizzati} E.,  et~al., 2024, \mn@doi [\mnras] {10.1093/mnras/stae2307}, \href {https://ui.adsabs.harvard.edu/abs/2024MNRAS.534.3155P} {534, 3155}

\bibitem[\protect\citeauthoryear{{Ploeckinger} \& {Schaye}}{{Ploeckinger} \& {Schaye}}{2020}]{Ploeckinger.2020}
{Ploeckinger} S.,  {Schaye} J.,  2020, \mn@doi [\mnras] {10.1093/mnras/staa2172}, \href {https://ui.adsabs.harvard.edu/abs/2020MNRAS.497.4857P} {497, 4857}

\bibitem[\protect\citeauthoryear{{Press} \& {Davis}}{{Press} \& {Davis}}{1982}]{Press.1982}
{Press} W.~H.,  {Davis} M.,  1982, \mn@doi [\apj] {10.1086/160183}, \href {https://ui.adsabs.harvard.edu/abs/1982ApJ...259..449P} {259, 449}

\bibitem[\protect\citeauthoryear{{Pujol} et~al.,}{{Pujol} et~al.}{2014}]{Pujol.2014}
{Pujol} A.,  et~al., 2014, \mn@doi [\mnras] {10.1093/mnras/stt2446}, \href {https://ui.adsabs.harvard.edu/abs/2014MNRAS.438.3205P} {438, 3205}

\bibitem[\protect\citeauthoryear{{Pujol} et~al.,}{{Pujol} et~al.}{2017}]{Pujol.2017}
{Pujol} A.,  et~al., 2017, \mn@doi [\mnras] {10.1093/mnras/stx913}, \href {https://ui.adsabs.harvard.edu/abs/2017MNRAS.469..749P} {469, 749}

\bibitem[\protect\citeauthoryear{{Riggs}, {Loveday}, {Thomas}, {Pillepich}, {Nelson}  \& {Holwerda}}{{Riggs} et~al.}{2022}]{Riggs.2022}
{Riggs} S.~D.,  {Loveday} J.,  {Thomas} P.~A.,  {Pillepich} A.,  {Nelson} D.,   {Holwerda} B.~W.,  2022, \mn@doi [\mnras] {10.1093/mnras/stac1591}, \href {https://ui.adsabs.harvard.edu/abs/2022MNRAS.514.4676R} {514, 4676}

\bibitem[\protect\citeauthoryear{{Samuel} et~al.,}{{Samuel} et~al.}{2020}]{Samuel.2020}
{Samuel} J.,  et~al., 2020, \mn@doi [\mnras] {10.1093/mnras/stz3054}, \href {https://ui.adsabs.harvard.edu/abs/2020MNRAS.491.1471S} {491, 1471}

\bibitem[\protect\citeauthoryear{{Sawala}, {Pihajoki}, {Johansson}, {Frenk}, {Navarro}, {Oman}  \& {White}}{{Sawala} et~al.}{2017}]{Sawala.2017}
{Sawala} T.,  {Pihajoki} P.,  {Johansson} P.~H.,  {Frenk} C.~S.,  {Navarro} J.~F.,  {Oman} K.~A.,   {White} S. D.~M.,  2017, \mn@doi [\mnras] {10.1093/mnras/stx360}, \href {https://ui.adsabs.harvard.edu/abs/2017MNRAS.467.4383S} {467, 4383}

\bibitem[\protect\citeauthoryear{{Schaller} et~al.,}{{Schaller} et~al.}{2024}]{Schaller.2024}
{Schaller} M.,  et~al., 2024, \mn@doi [\mnras] {10.1093/mnras/stae922}, \href {https://ui.adsabs.harvard.edu/abs/2024MNRAS.530.2378S} {530, 2378}

\bibitem[\protect\citeauthoryear{{Schaye} \& {Dalla Vecchia}}{{Schaye} \& {Dalla Vecchia}}{2008}]{Schaye.2008}
{Schaye} J.,  {Dalla Vecchia} C.,  2008, \mn@doi [\mnras] {10.1111/j.1365-2966.2007.12639.x}, \href {https://ui.adsabs.harvard.edu/abs/2008MNRAS.383.1210S} {383, 1210}

\bibitem[\protect\citeauthoryear{{Schaye} et~al.,}{{Schaye} et~al.}{2015}]{Schaye.2015}
{Schaye} J.,  et~al., 2015, \mn@doi [\mnras] {10.1093/mnras/stu2058}, \href {https://ui.adsabs.harvard.edu/abs/2015MNRAS.446..521S} {446, 521}

\bibitem[\protect\citeauthoryear{{Schaye} et~al.,}{{Schaye} et~al.}{2023}]{Schaye.2023}
{Schaye} J.,  et~al., 2023, \mn@doi [\mnras] {10.1093/mnras/stad2419}, \href {https://ui.adsabs.harvard.edu/abs/2023MNRAS.526.4978S} {526, 4978}

\bibitem[\protect\citeauthoryear{{Sharma} \& {Steinmetz}}{{Sharma} \& {Steinmetz}}{2006}]{Sharma.2006}
{Sharma} S.,  {Steinmetz} M.,  2006, \mn@doi [\mnras] {10.1111/j.1365-2966.2006.11043.x}, \href {https://ui.adsabs.harvard.edu/abs/2006MNRAS.373.1293S} {373, 1293}

\bibitem[\protect\citeauthoryear{{Sinha} \& {Garrison}}{{Sinha} \& {Garrison}}{2020}]{Sinha.2020}
{Sinha} M.,  {Garrison} L.~H.,  2020, \mn@doi [\mnras] {10.1093/mnras/stz3157}, \href {https://ui.adsabs.harvard.edu/abs/2020MNRAS.491.3022S} {491, 3022}

\bibitem[\protect\citeauthoryear{{Springel}, {White}, {Tormen}  \& {Kauffmann}}{{Springel} et~al.}{2001}]{Springel.2001}
{Springel} V.,  {White} S. D.~M.,  {Tormen} G.,   {Kauffmann} G.,  2001, \mn@doi [\mnras] {10.1046/j.1365-8711.2001.04912.x}, \href {https://ui.adsabs.harvard.edu/abs/2001MNRAS.328..726S} {328, 726}

\bibitem[\protect\citeauthoryear{{Springel} et~al.,}{{Springel} et~al.}{2005}]{Springel.2005}
{Springel} V.,  et~al., 2005, \mn@doi [\nat] {10.1038/nature03597}, \href {https://ui.adsabs.harvard.edu/abs/2005Natur.435..629S} {435, 629}

\bibitem[\protect\citeauthoryear{{Springel}, {Pakmor}, {Zier}  \& {Reinecke}}{{Springel} et~al.}{2021}]{Springel.2021}
{Springel} V.,  {Pakmor} R.,  {Zier} O.,   {Reinecke} M.,  2021, \mn@doi [\mnras] {10.1093/mnras/stab1855}, \href {https://ui.adsabs.harvard.edu/abs/2021MNRAS.506.2871S} {506, 2871}

\bibitem[\protect\citeauthoryear{Villaescusa-Navarro et~al.,}{Villaescusa-Navarro et~al.}{2023}]{VillaescusaNavarro.2023}
Villaescusa-Navarro F.,  et~al., 2023, \mn@doi [The Astrophysical Journal Supplement Series] {10.3847/1538-4365/acbf47}, 265, 54

\bibitem[\protect\citeauthoryear{{Vogelsberger} et~al.,}{{Vogelsberger} et~al.}{2009}]{Vogelsberger.2009}
{Vogelsberger} M.,  et~al., 2009, \mn@doi [\mnras] {10.1111/j.1365-2966.2009.14630.x}, \href {https://ui.adsabs.harvard.edu/abs/2009MNRAS.395..797V} {395, 797}

\bibitem[\protect\citeauthoryear{{Wang}, {Frenk}, {Navarro}, {Gao}  \& {Sawala}}{{Wang} et~al.}{2012}]{Wang.2012}
{Wang} J.,  {Frenk} C.~S.,  {Navarro} J.~F.,  {Gao} L.,   {Sawala} T.,  2012, \mn@doi [\mnras] {10.1111/j.1365-2966.2012.21357.x}, \href {https://ui.adsabs.harvard.edu/abs/2012MNRAS.424.2715W} {424, 2715}

\bibitem[\protect\citeauthoryear{{White}, {Frenk}  \& {Davis}}{{White} et~al.}{1983}]{White.1983}
{White} S.~D.~M.,  {Frenk} C.~S.,   {Davis} M.,  1983, \mn@doi [\apjl] {10.1086/184139}, \href {https://ui.adsabs.harvard.edu/abs/1983ApJ...274L...1W} {274, L1}

\bibitem[\protect\citeauthoryear{{Wiersma}, {Schaye}, {Theuns}, {Dalla Vecchia}  \& {Tornatore}}{{Wiersma} et~al.}{2009}]{Wiersma.2009}
{Wiersma} R. P.~C.,  {Schaye} J.,  {Theuns} T.,  {Dalla Vecchia} C.,   {Tornatore} L.,  2009, \mn@doi [\mnras] {10.1111/j.1365-2966.2009.15331.x}, \href {https://ui.adsabs.harvard.edu/abs/2009MNRAS.399..574W} {399, 574}

\makeatother
\end{thebibliography}
%%%%%%%%%%%%%%%%%%%%%%%%%%%%%%%%%%%%%%%%%%%%%%%%%%

\appendix

\section{Differences between HBT-HERONS and HBT+}\label{section:hbt_improvements}
As mentioned in \S\ref{section:halo_finders}, several improvements were made to the {\HBT} algorithm, resulting in a version of the subhalo finder named {\HBTHERONS}. These changes primarily target the tracking of subhaloes, and hence improve the capabilities of {\HBT} in both DMO and hydrodynamical simulations. However, some changes were specifically tailored for hydrodynamical simulations (\S\ref{Appendix:incorrect_tracers}). In this Appendix, we explain in detail what all the changes are and their motivation.

An important step that we refer to in several occasions below is particle `masking'. This is done internally within {\HBT} and {\HBTHERONS} before analysing any given FoF group, and the aim is to ensure that no particles are shared across subhaloes that belong to the same hierarchical branch. The process proceeds as follows. First, the IDs of particles that are associated to subhaloes without any children subhaloes are logged in a list. Next, their parent subhaloes check whether any of the IDs of their respective particles are already in the list. The particles with a match are removed from the source of the parent subhalo, i.e. the particles are masked. The parent subhalo retains the particles without any matches, and adds their IDs to the list. This process is repeated for each subhalo going up the hierarchy until the central subhalo is reached. The end result is that the particle distribution of each subhalo has the particles associated to other subhaloes `masked' out. As masking only occurs between within a given hierarchical branch, there is no masking across FoF groups.

\subsection{Collisionless, time-persistent particle tracers}\label{Appendix:incorrect_tracers}

The original {\HBT} algorithm relies on identifying the most bound particle of each subhalo as the tracer of its location. The FoF group membership of the tracer particle is subsequently used to determine which FoF group hosts a given subhalo. As this is used internally to determine the hierarchical relations of subhaloes within a FoF group, the tracer particle plays an important role. Additionally, the tracer particle is also used to identify the position and velocity of orphaned subhaloes. {\HBT} does not take into account the particle type of the tracer, meaning it can choose a gas or black hole particle if these are the most bound ones. If the chosen tracer particle is a gas particle then it can be ejected from the halo by feedback, and if the tracer is a black hole particle, then it can merge with another black hole and cease to exist.

\begin{figure}
\centering
    \input{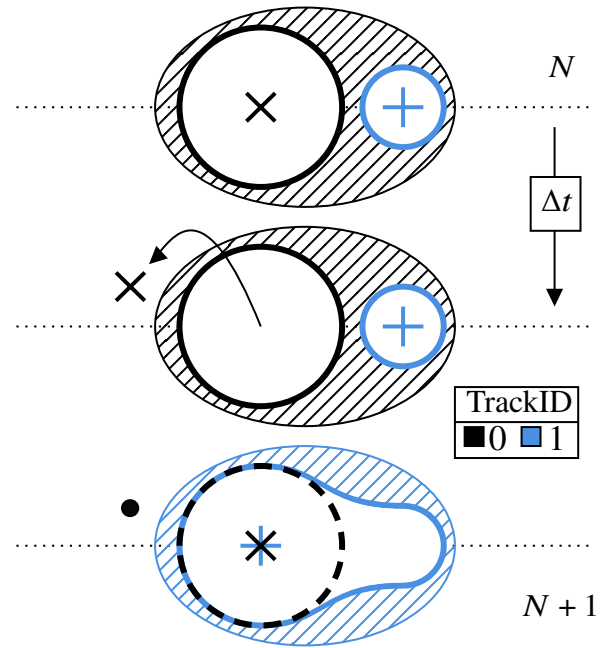}
    \caption{Schematic representation of how using gas as particle tracers of subhaloes may lead to the loss of satellites in {\HBT}. Each column shows the time evolution of a unique subhalo, whose bound component at a given time is represented by a circle. Different colours correspond to different evolutionary branches, whose TrackIDs are given by the legend. The shaded ellipses are FoF groups, coloured as their corresponding central subhalo. The crosses and pluses represent the centre of each subhalo, which is the position of its most bound tracer particle. The filled circle represents an individual particle. Initially, Tracks 0 and 1 are contained in the same group. The gas tracer of the central subhalo, Track 0, is blown out of the FoF group. This makes {\HBT} reassign Track 1 as the new central, as Track 0 is now a hostless subhalo. This mixes the particles of the two existing subhaloes together, making {\HBT} unable to identify the actual subhalo that Track 1 was meant to track. Since TrackId 0 retains its previous particles, the two subhaloes now share a large number of particles. Thus, in output $N+1$, both subhaloes are likely to have the same position in phase-space, but no merging checks are applied as {\HBT} incorrectly thinks they are in separate FoF groups.}
    \label{figure:incorrect_tracers}
\end{figure}

If the tracer is indeed blown out from a FoF group, the subhalo will be deemed hostless, even if it the majority of its constituent particles remain part of a FoF group. The inappropriate choice of particle tracer has important ramifications, as it leads to the sudden loss of well-resolved substructure. We illustrate how this happens in the diagram shown in Fig.~\ref{figure:incorrect_tracers}. In this example, we start with two subhaloes in the same FoF group, with Track 0 being the central. The most bound particle of each subhalo is located at its centre, but crucially, the tracer of Track 0 is a gas particle. 

Between outputs $N$ and $N+1$, the galaxy experiences a gas outflow, e.g. resulting from AGN activity. Consequently, the gas tracer particle is ejected from the FoF group. In output $N+1$, the FoF group has no central, as the algorithm has identified Track 0 as hostless. Therefore, the most massive satellite becomes its central, i.e. Track 1. In doing so, the particles that belonged to Track 0 and 1 get grouped together. As {\HBT} bases its subhalo identification on the past membership of particles, it will be unable to identify the satellite subhalo as a separate entity from the central throughout the remainder of the simulation. 

Another problem resulting from this example is that, since {\HBT} allows the existence of hostless subhaloes, Track 0 retains its previously associated particles. This means that Tracks 0 and 1 share large numbers of bound particles in output $N+1$. Their most bound particles will thus be located close in phase-space. Despite the significant overlap between both subhaloes, no merger condition is triggered because they are formally not in the same FoF group. In output $N+2$, if no other blowouts occur, both subhaloes will be located in the same FoF group. However, {\HBT} assigns particles within a FoF group to subhaloes in an exclusive manner by applying a mask. The significant overlap in particle content between subhaloes means that one of the two subhaloes is masked out of existence, leading to the sudden loss of a subhalo.  

The chain of events described above can happen several times to a subhalo throughout a simulation, potentially affecting more than one satellite. Since the occurrence of the problem only requires the misidentification of the FoF group based on a single particle, it can happen to subhaloes regardless of how well resolved they are. To address this issue, {\HBTHERONS} can constrain which particle type is allowed to be a tracer of a subhalo. Ideal particle tracers are collisionless and time persistent. In the {\tt FLAMINGO} simulations, only star and dark matter particles types satisfy these two requirements simultaneously. To reflect this, {\HBTHERONS} defaults to only allowing star or dark matter particles to be subhalo particle tracers.

\tikzset{every picture/.style={line width=0.75pt}} %set default line width to 0.75pt        

Our preference towards using time-persistent and collisionless particles as tracers is reflected in other {\HBTHERONS} routines relying on a subset of the most bound particles of a subhalo. For example, orphans are constrained to use the most bound star or dark matter particle when they were last resolved. However, in {\HBT}, they could have been assigned to a gas or black hole particle. This change solves issues stemming from using gas and black hole tracers, since they can disappear from the simulation, leading to the loss of the orphan.

\subsection{Weighted host finding}\label{appendix:host_finding}

Once a tracer particle has been assigned to a subhalo, its FoF membership is used at the next output time to identify which FoF group hosts the subhalo it is associated to. In {\HBT} this approach thus relies on a single particle, and is hence prone to discreteness noise. For host finding, this noise primarily manifests itself as the tracer straying away from the core of a subhalo. This produces a clear mismatch between the FoF group assigned to a subhalo, and the actual FoF group where the majority of the core particles of the subhalo end up.

\begin{figure}
    \centering
    \input{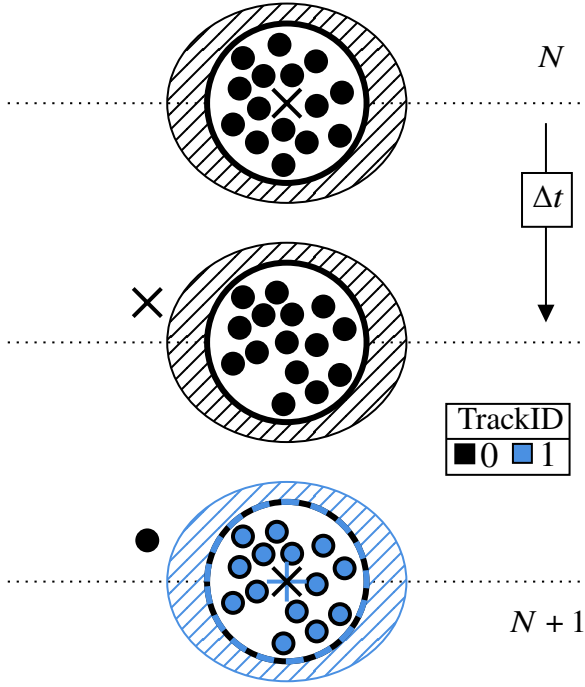}
    \caption{Schematic representation (as in Fig.~\ref{figure:incorrect_tracers}) of how noise in the particle distribution may lead to misidentification of the FoF group hosting a central subhalo in {\HBT}. Initially, there is a single subhalo (Track 0). Subsequently, its tracer particle is not part of any FoF group, and hence {\HBT} makes Track 0 hostless. This is in contrast to the bulk of its particles, which remain within the FoF group. Since the FoF group has no subhalo associated to it, a new subhalo (Track 1) is created. Similar to the example shown in Fig.~\ref{figure:incorrect_tracers}, the two subhaloes share a large number of particles.}
    \label{figure:weighted_host_finding}
\end{figure}

Similar to the issue discussed in \S\ref{Appendix:incorrect_tracers}, the incorrect assignment of the FoF group of a subhalo has important ramifications. We explain the issues that arise as a consequence with the example shown in Fig.~\ref{figure:weighted_host_finding}. In output $N$, we have a single central subhalo with TrackId 0. At the next output, its tracer has strayed away from the FoF group and the subhalo is deemed hostless. This is in contrast to the majority of its associated particles, as they remain within a FoF group. In this example, since said FoF group has no associated central, a new subhalo is spawned with TrackId 1. The subhaloes now share a large fraction of their particles, as masking is not done across FoF groups, and will likely end up with the same tracer in output $N+1$. Eventually, if the tracer of TrackId 0 is found in the same FoF as TrackId 1 in output $N+2$, one of the two will be masked.

Contrary to the problem discussed in \S\ref{Appendix:incorrect_tracers}, this issue only arises for subhaloes that are poorly sampled by particles. This is because the spatial extent of FoF groups with $\mathcal{O}(10^{2})$ particles is much smaller than groups with many more particles. Consequently, a particle that is near the centre of the FoF group can more easily make it beyond its outskirts in the time difference between two consecutive time outputs. As the number density of subhaloes increases with decreasing subhalo mass, this issue results in many short-lived spuriously-created subhaloes. We note that the subhaloes prone to experience this appear to be more elongated than subhaloes of similar mass that do not exhibit the same issue. Presumably, this facilitates the tracer particle straying away from the FoF group along the minor axis of its particle distribution.

In order to not rely on a single particle, the host finding in {\HBTHERONS} is done through a weighted estimate of the $N_{\mathrm{host-finding}}$ (10) most bound collisionless tracers. The weight assigned to each particle, $w_{i}$, reflects their boundness ranking in the previous output, $r_{i}$. The functional form is given by equation (\ref{host-finding-equation}) \citep[e.g.][]{Springel.2021}.

This improves the tracing of subhaloes in hydrodynamical and DMO simulations, as the host is not determined by a single particle. Given the reliance of the algorithm on using more than one particle tracers, we require at least $N^{\mathrm{min}}_{\mathrm{tracer}}$ particles of the accepted tracer type (equal to $N_{\mathrm{host-finding}}$ by default) to be present in any subhalo for it to be deemed resolved. In other words, a resolved subhalo requires at least $N^{\mathrm{min}}_{\mathrm{bound}}$ total bound particles \textit{and} $N^{\mathrm{min}}_{\mathrm{tracer}}$ bound particles that are valid tracer particle types.

We note that there are still subhaloes deemed as hostless even with this improved host finding method. Their existence is not necessarily unphysical, as it simply reflects the fact that linking particles close to the resolution limit of the simulation is noisy. In other words, a FoF group with a self-bound subhalo may lose one particle and dip below the threshold to be found as a FoF group. This makes the subhalo hostless, even though not much might have changed in terms of its self-boundness.  

\subsection{Phase-space calculations}\label{appendix:merging_calculations}

An important step in {\HBT} that is not generally required for other subhalo finders is to check whether resolved subhaloes overlap in position and velocity space. The reason is that {\HBT} is able to find two separate resolved subhaloes even if they have coalesced following a major merger. This is not a problem for alternative algorithms, like those relying on density peaks, as they stop being able to distinguish coalescing subhaloes long before the merger is complete. Failing to manually merge overlapping subhaloes in {\HBT} would lead to a spurious increase in the satellite number density at extremely close distances from the centre of haloes.

Computing the offset in phase-space between two pairs of subhaloes requires assigning a phase-space location to each (equation \ref{equation:phase_space_offset}) and an associated dispersion. In {\HBT}, the higher mass subhalo of a given hierarchical pair uses the centre of mass position and velocity of its $N^{\rm min}_{\rm core}$ (20 by default) most bound tracers, which are also used to estimate the position and velocity dispersion. The lower mass subhalo only uses its most bound particle tracer to locate its centre in position and velocity space. 

However, using only one particle to estimate the phase-space location of subhaloes can be prone to noise. Therefore, {\HBTHERONS} estimates the phase-space position of both subhaloes in a hierarchical pair using the same number of particles (10 most bound tracers by default) as the most massive of the pair. Only when the least massive subhalo of the pair is an orphan, i.e. it only has one tracer particle, do we revert to using one particle to estimate its position and velocity. Additionally, we apply equation (\ref{equation:phase_space_offset}) symmetrically between the two subhaloes under consideration. We compute two phase-space offsets, each using the dispersion measured from one of the two subhaloes, and we take the minimum value as the offset. This change removes the assumption that the most massive subhalo of the pair has the largest dispersion of the two subhaloes, which is not always the case for poorly resolved systems. 

\begin{figure}
\centering
    \input{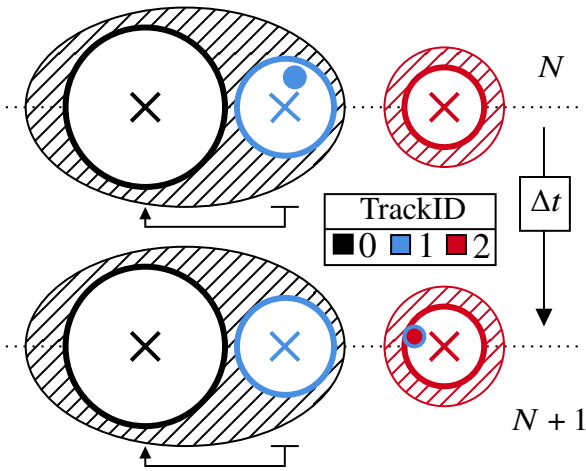}
    \caption{Schematic representation (as in Fig.~\ref{figure:incorrect_tracers}) of how a stray particle originally associated to a satellite subhalo in a different FoF can lead to its duplication in {\HBT}. Initially, the particle is bound and associated to Track 1. In the next output, the particle is found in the FoF group whose central is Track 2. This means that the source subhalo of Track 2 acquires the particle, and it is found to be bound. However, the source subhalo of Track 1 still contains the particle in question, which can also be found to be bound. Hence, the particle is bound to two different subhaloes and appears duplicated in the catalogues. {\HBTHERONS} constrains particles to be part of the same FoF group as the Track they are associated to, if the particle is found in any FoF group. For this example, Track 1 would lose the particle since they are no longer located in the same FoF group.}
    \label{figure:duplication_explanation_1}
\end{figure}

\subsection{Exclusive particle membership}\label{appendix:duplicate_particles}

An inconsistency present in {\HBT} is that particles can be shared among subhaloes, and hence appear duplicated in the catalogues. This is clearly unintended behaviour, as it would occur even when mass was supposed to be assigned to subhaloes in an exclusive manner. There are three main reasons that would result in particle duplication, which are facilitated by the fact that there is no particle masking for exclusive mass assignment across separate FoF groups. We explain each below, with an accompanying diagram when applicable.

\begin{figure}
\centering
    \input{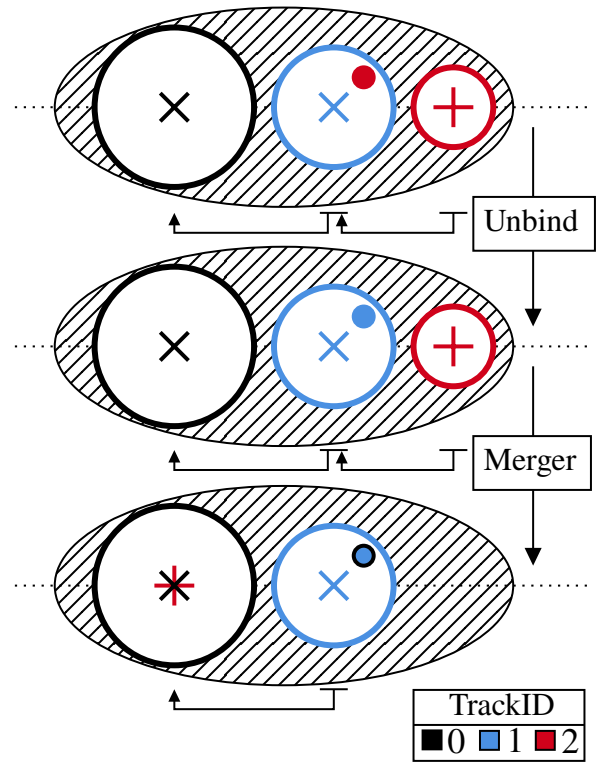}
    \caption{Schematic representation (as in Fig.~\ref{figure:incorrect_tracers}) of how the previous way of dealing with subhalo mergers can lead to duplicate particles in {\HBT}. The rows represent distinct steps done when analysing the subhaloes in a FoF group during a single time output. The particle is originally associated to the source subhalo of Track 2, but is not bound to it. It is instead bound to Track 1, but it remains in the source of Track 2. After analysing all Tracks in the FoF, merging checks happen, and Track 2 merges with Track 0. All the particles previously associated to Track 2 are now part of the source subhalo of Track 0. A new unbinding iteration is triggered for Track 0, to account for the newly accreted mass. The original particle is now also bound to Track 0, and the duplication occurs because it is bound to Tracks 0 and 1. {\HBTHERONS} only allows the subhalo that accretes particles during a merger to acquire the bound particles of the merged subhalo. For this example, the particle will only be bound to Track 1.}
    \label{figure:duplication_explanation_2}
\end{figure}

The first reason is that particles associated to a satellite subhalo can end up in a different FoF group than its host. We illustrate how this can happen in Fig.~\ref{figure:duplication_explanation_1}. The particle retains its association to the satellite that it was originally bound to, reflecting its past history. However, the central in the FoF group that the particle is currently in also adds it to its source subhalo. Since no particle masking occurs across FoF groups, the particle can be found to be bound to both subhaloes. {\HBTHERONS} addresses this by required that, if a particle is found in a FoF group, any previous association to subhaloes in a different FoF group are removed.  

The second cause is two related choices in how the re-unbinding following a subhalo merger is done. {\HBT} only triggers the re-unbinding after analysing every subhalo in a given FoF group. Additionally, the unbinding is only done for the subhaloes that accreted another subhalo through a merger. As mergers can happen between subhaloes of different depths, this can lead to duplication of particles. We show an example of how this can happen in Fig. ~\ref{figure:duplication_explanation_2}. This issue is fixed in {\HBTHERONS} by doing the merging checks immediately after the unbinding of a given subhalo, and only passing the particles bound to the now-merged subhalo onto the subhalo it merged with. 

\begin{figure}
\centering
    \input{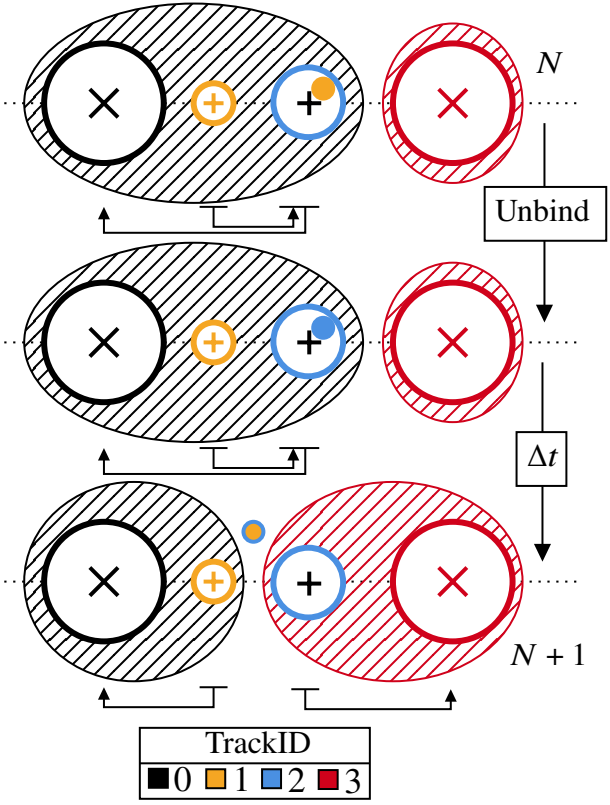}
    \caption{Schematic representation (as in Fig.~\ref{figure:incorrect_tracers}) of how the incorrect assignment of particles to more than one source subhalo can lead to particle duplication in {\HBT}. The filled circle corresponds to an individual particle that is originally associated to Track 1. After unbinding is done, the particle is now bound to Track 2. Similar to the example of Fig.~\ref{figure:duplication_explanation_2}, this means the particle now belongs to the source subhalo of both Track 1 and Track 2. In the next output, Track 1 and 2 are satellites in different FoF groups. As there is no particle masking across different FoF groups, the particle can be found as bound to both Tracks and hence be duplicated. This duplication process only occurs if the particle in question is not part of a FoF group, as the fix for the Fig.~\ref{figure:duplication_explanation_1} would otherwise be able to handle it. {\HBTHERONS} constrains particles to belong to a single source subhalo after unbinding every subhalo in the FoF group. The particles are preferentially assigned to original source subhalo from which they originated. For this example, the particle will only be bound to Track 1, as it is the subhalo the particle was originally associated with.}
    \label{figure:duplication_explanation_3}
\end{figure}

The final cause is the specific implementation of how unbound particles are internally passed between child-parent subhalo pairs. Specifically, if a particle is not bound to a subhalo, it is passed onto its parent subhalo to check if it is bound to it instead. Internally, this is equivalent to adding the particle to the source subhalo of the parent. However, the particle may still remain in the source of its original subhalo if it was not trimmed from it (see \S\ref{unbinding_step}). This is not an issue if both subhaloes remain in the same FoF group in the next output, as inter-FoF particle masking will remove the duplicate particle from the relevant source. However, if the subhaloes are found in different FoFs, the particle can be bound to both, as shown in Fig.~\ref{figure:duplication_explanation_3}. Note that an additional condition for this to occur is that the particle is not found in a FoF group. Otherwise, the fix shown for case Fig.~\ref{figure:duplication_explanation_1} would handle this case. To remedy this problem, {\HBTHERONS} applies a mask on the source of subhaloes after the analysis of a whole FoF group is done. The masking preferentially gives particles to more deeply nested subhaloes, to reflect the fact that those particles always originate from deeper subhaloes.

\subsection{Orphan tracking and mass assignment}\label{appendix:orphan_tracking}

Orphan subhaloes are subhaloes that were self-bound in the past, but whose particles have been stripped causing the subhaloes to fall below the resolution limit of the simulation. In {\HBT} and {\HBTHERONS}, their position and velocities are subsequently tracked using the most bound particle tracer when they were last resolved. In {\HBT}, the orphan tracer particles are explicitly associated to orphan subhaloes. This means that, as mass is assigned in an exclusive manner, subhalos with many particles associated to orphans would have less mass. 

Another closely related problem is that a particle already associated with an orphan subhalo could well be the most bound particle of a resolved subhalo. Thus, in {\HBT}, the resolved subhalo would be unable to acquire its `true' most bound particle and would hence be traced by an incorrect particle if it subsequently became an orphan. As these situations can happen several times over the course of a simulation, the most bound core of resolved subhaloes could potentially become dominated by orphan subhaloes and not be explicitly bound to resolved subhaloes.

In {\HBTHERONS}, particle tracers are no longer explicitly assigned to orphan subhaloes, relying instead on locating the tracer particle properties based on ID matching. This means that orphan subhaloes can be identified, whilst their bound mass can be correctly assigned to a resolved subhalo. This change also allows for the possibility of more than one orphan being assigned to a single tracer particle.

\subsection{Descendants of disrupted subhaloes}\label{appendix:disrupted_descendants}

In {\HBT}, resolved subhalos can cease to exist in two different ways: disruption and merging. Disruption occurs when the number of bound particles drops below a given threshold, which is set by default to $N^{\mathrm{min}}_{\mathrm{bound}} = 20$. Mergers happen between subhaloes that are identified as self-bound but overlap sufficiently in phase space. Only a minority of subhaloes undergo actual mergers while still being resolved. Most fall below the resolution limit before sinking to the centre of their host subhalo.

{\HBT} only provides information about the descendants of resolved mergers, i.e. which subhalo overlapped in phase space with the merged one. Thus, the merger trees built by {\HBT} do not incorporate information about the descendants of (unresolved) mergers resulting from subhalo disruption. {\HBTHERONS} provides the required information to build complete merger trees, by outputting descendant information for both merged and disrupted subhaloes. To assign a descendant subhalo to each disrupted subhalo (`DescendantTrackId'), {\HBTHERONS} uses the $N_{\mathrm{descendant}}$ (10 by default) most bound particles from the previous output of the now-disrupted subhalo. The subhalo that contains the majority of this set of $N_{\mathrm{descendant}}$ particles is the descendant of the disrupted subhalo.  

\begin{figure}
    \centering
    \includegraphics[width=0.45\textwidth,keepaspectratio]{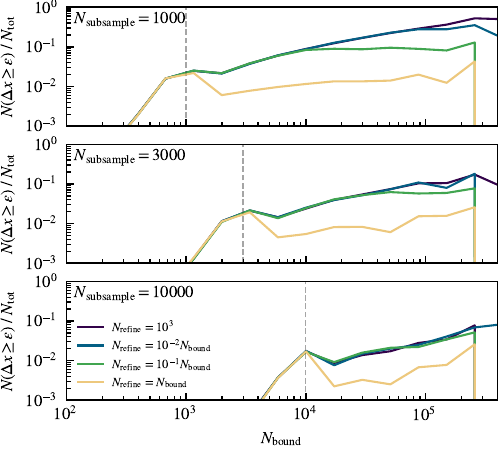}
    \caption{The fraction of resolved subhaloes whose centre is offset from its counterpart in a different {\HBTHERONS} run by more than the softening length of the DMO {\midres} simulation ($\epsilon = 5.7\,\mathrm{kpc}$). Each panel corresponds to using a different choice for the number of particles used to subsample the potential calculation when unbinding subhaloes. The line colour reflects how many particles are used to refine the centre of the subhalo, with a constant value of $10^{3}$ being the choice used in {\HBT}.}
    \label{figure:centering_influence_particles}
\end{figure}

\subsection{Refinement of the centres of subhaloes}\label{appendix:unbinding_subsampling_effects}

{\HBT} randomly subsamples the particle distribution of subhaloes whose source contains more than $N_{\mathrm{subsample}}$ particles (1000 by default) to compute the gravitational potential during unbinding. If the source subhalo contains more particles than this threshold, it randomly selects $N_{\mathrm{subsample}}$ particles, and if it contains fewer, it uses all particles for potential estimates. The subsampling is done with the aim of speeding up the unbinding iterations, which are often one of the most expensive calculations within halo finding. Although this approximate method works well to identify which particles are bound, their relative binding energies are affected by the choice of subsampled particles. Thus, this can influence the estimated subhalo centre from run to run, as it is defined based on its most bound particle. 

To address this, {\HBT} does a `centre refinement' after completing the unbinding of a subhalo. This consists of computing the self-binding energy of the $N_{\mathrm{refine}}$ = $N_{\mathrm{subsample}}$ most bound particles identified during the subsampled unbinding. The most bound particle of that subset is then the centre of the subhalo. This approach works well, except if the subhalo in question has $N_{\mathrm{bound}}\gg N_{\mathrm{refine}}$. This is because the chosen subset becomes a very small fraction of the overall number of bound particles, and randomness in the subsampled unbinding step becomes increasingly important. For example, if $N_{\mathrm{bound}} = 10^{10}$, one must correctly identify the centre based on $10^{-7}$ of bound particles. In practice, this is almost impossible to do, and hence the centre is likely significantly offset from the true centre found using all particles during unbinding.    

In order to reduce the effect of subsampling on the identification of subhalo centres, {\HBTHERONS} scales $N_{\mathrm{refine}}$ with $N_{\mathrm{bound}}$ whenever $N_{\mathrm{bound}} > N_{\mathrm{subsample}}$:
\begin{equation}\label{equation:scaling_Nrefine}
    N_{\mathrm{refine}} = 0.1 N_{\mathrm{bound}}\,.
\end{equation}
The proportionality factor was chosen based on the tests shown in Fig.~\ref{figure:centering_influence_particles}. To perform these tests, we first ran {\HBTHERONS} using the {\HBT} configuration until the second to last snapshot of the DMO {\midres} simulation. We then re-ran {\HBTHERONS} on the last snapshot using different choices for $N_{\mathrm{subsample}}$ and $N_{\mathrm{refine}}$, and independently run each combination 5 times to quantify the run-to-run scatter. We then compared the offset between pairs of matched subhaloes across runs that use the same configuration. 

For the parameter choices of {\HBT} ($N_{\mathrm{subsample}} = N_{\mathrm{refine}} = 10^{3}$), the fraction of matched counterparts whose centre deviates by more than the gravitational softening length is $0.53$. For the same value of $N_{\mathrm{subsample}}$, making $N_{\mathrm{refine}}$ scale as in equation \ref{equation:scaling_Nrefine} reduces this fraction to $0.13$. This scaling is accompanied by an increased computational cost, although it amounts to an average of only +2\% for this simulation. In contrast, doing the unbinding without any subsampling of the particles takes +54\% extra. We note that the additional cost of the better refinement will depend on the number of objects with $N_{\mathrm{bound}} \geq N_{\mathrm{subsample}}$, i.e. it will depend on the resolution and box size. 

Exploring alternative choices for the scaling of $N_{\mathrm{refine}}$ and $N_{\mathrm{subsample}}$, we identify several trends. There is a sharp decrease in the fraction of offset subhaloes whose particle number is below $N_{\mathrm{subsample}}$, indicating the approximate threshold where particle subsampling no longer takes place. However, there is no sudden truncation, as the subsampling is based on the number of particles associated to the subhalo, rather than its bound particles. Thus, even if a subhalo has $N_{\mathrm{bound}} = 500$, it would have been subject to subsampling if its source (see \S\ref{unbinding_step}) had more than $10^{3}$ particles. Additionally, for a fixed value/scaling of $N_{\mathrm{refine}}$, the spatial offsets are smaller if $N_{\mathrm{subsample}}$ increases. This is because the (approximate) relative binding energies are estimated more accurately, so picking the correct most bound particle subset is less prone to run-to-run randomness. Indeed, doing no subsampling would mean no centre refinement would be required. Lastly, the runs where we set $N_{\mathrm{refine}} = N_{\mathrm{bound}}$ highlight the effect of identifying different sets of bound particles due to potential subsampling. In other words, if a subhalo with the same set of bound particles would have its centre refined, the same centre would always be chosen for $N_{\mathrm{refine}} = N_{\mathrm{bound}}$. The fact that there is still a non-zero offset indicates that the same subhalo in different runs has a similar yet-different set of particles bound to it. 

\subsection{Re-attachment of gas particles}\label{appendix:gas_reattaching}

A fundamental assumption used by {\tt HBT+} is that mass can only be transferred from lower mass subhaloes to their more massive parent subhaloes, which are connected in a way that reflects their past accretion history. This assumption works well for collisionless particle species, but may not hold for collisional particles like gas. For example, a pair of closely interacting galaxies may experience some transfer of gas particles between them, which does not necessarily follow the flow of mass as assumed possible by {\tt HBT+}. This situation can lead to the misassignment of gas mass (or even stellar mass if those particles subsequently become star forming) in the central regions of galaxies, as those particles are unable to change the subhalo they are associated to.  

To prevent this from happening, {\HBTHERONS} performs a gas re-attachment step before subjecting subhaloes to unbinding. First, we find for each gas particle in a FoF group the nearest 10 tracer particles. We then check whether any of those neighbours share the same TrackId as the gas particle in question. If none of them do, it indicates the gas particle is well beyond the confines of the subhalo it is associated to, and hence that it should be re-attached to a more suitable subhalo. We identify the subhalo the gas particle should belong to based on the subhalo that the majority of the neighbours are associated to. This change effectively removes  the assumption that gas transfer is always be hierarchical in nature.

\section{Differences between HBT+ and SUBFIND-HBT}\label{Appendix:VersionsOfHBT}

\begin{figure}
    \centering
    \includegraphics[width=0.45\textwidth,keepaspectratio]{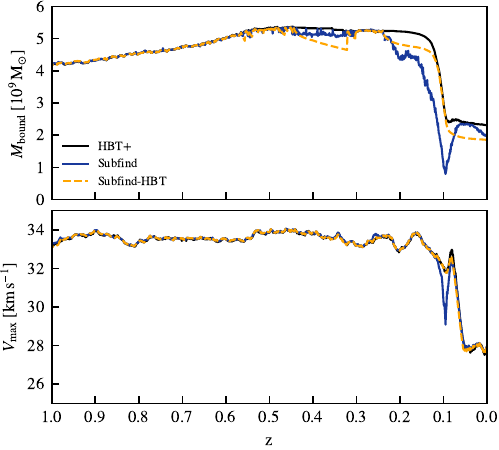}
    \caption{The bound mass (top panel) and $V_{\mathbf{max}}$ (bottom panel) time evolution of a $z = 0$ subhalo, measured by {\HBT} (solid black), {\SUBFIND} (solid blue) and {\tt Subfind-HBT} (dashed orange).}
    \label{figure:subfind_hbt}
\end{figure}

The similar names of the {\HBT} and {\tt Subfind-HBT} subhalo finders may suggest that they are the same algorithm. Although both rely on history-space, the details of how this approach is implemented differs. This can lead to small, yet systematic differences when comparing the properties and evolution of subhaloes matched between catalogues. As such, they should be thought of as two independent halo finding algorithms, rather than {\HBT} running on-the-fly within {\tt GADGET-4}. 

To illustrate the differences, we show how the bound mass of a $z = 0$ satellite subhalo evolves with time according to {\HBT} and {\tt Subfind-HBT} in Fig~\ref{figure:subfind_hbt}. Note that this example subhalo was taken from a {\tt GADGET-4} zoom-in dark-matter-only simulation of a $M_{\mathrm{200c}} \approx 10^{12}\, \Msun$ halo. The particle mass is $m_{\mathrm{DM}} = 4.3 \times10^{5}\Msun$, and the time cadence between outputs below $z = 2$ is 10~Myr. 

Their bound masses agree relatively well at early times, to within $\approx 5\%$. However, their masses diverge after the subhalo becomes a satellite shortly after $z \approx 0.5$. Whereas {\HBT} finds little to no change in its bound mass, the subhalo continuously loses mass in {\tt Subfind-HBT}. This eventually leads to a $20\%$ difference between both finders. The bound mass agrees between finders once it becomes a central again ($z \approx 0.3$), with the mass shooting up in just 10~Myr for {\tt Subfind-HBT}. As soon as it becomes a satellite again ($z\approx0.2$), the masses begin to diverge once more. It then undergoes a pericentre passage and mass is lost according to both finders, as expected from the effect of tidal stripping. By $z = 0$, the bound mass differs by $20\%$. 

Through the evolution, the value of $V_{\max}$ agrees much better between the subhalo finders. It remains constant after the inital growth of the subhalo, and only changes after the mass loss induced by tides at pericentre. As this quantity probes the inner part of the subhalo, the dichotomy between $V_{\mathrm{max}}$ and $M_{\mathrm{bound}}$ tells us that mass is lost from the outskirts of the {\tt Subfind-HBT} subhalo.

Indeed, this highlights one of the main differences between the two algorithms. {\tt Subfind-HBT} finds (satellite) subhalo candidates within a FoF group by grouping particles that were bound to the same subhalo in the previous snapshot. The key difference relative to {\HBT} is that {\tt Subfind-HBT} does not account for particles that are re-accreted by satellites. In other words, particles may be found to be momentarily unbound at one output and then bound in the next. In {\HBT}, this is dealt with by associating more particles to a subhalo than those that are bound (see \citealt{Han.2012} for a discussion of why this is required). The approach that {\tt Subfind-HBT} uses means that any particle that is found momentarily unbound will no longer be associated to that subhalo. Over time, this results into the outer layers of mass `evaporating'. As soon as the subhalo becomes a central, the method of candidate finding changes, and the missing mass is found once again. Despite the disagreement in the bound mass between {\tt Subfind-HBT} and {\HBT}, it is easy to see that the tracking of both is better than {\SUBFIND}, particularly during the pericentric passage at $z \approx 0.1$. 

\begin{figure}
    \centering
    \includegraphics{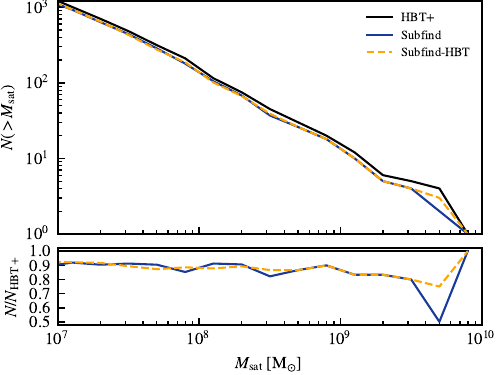}
    \caption{The cumulative bound mass function of satellite subhalos around a $M_{\mathrm{200c}} = 10^{12}\,\Msun$ halo. It was measured using {\HBT}, {\SUBFIND} and {\tt Subfind-HBT}. The ratios relative to {\HBT} are shown in the bottom panel. Despite the history-based approach of {\tt Subfind-HBT}, it is in closer agreement with {\SUBFIND} than with {\HBT}.}
    \label{figure:subfind_hbt_mass_function}
\end{figure}

The differences in the bound mass evolution propagate to summary statistics. We show the cumulative bound mass function of satellites for the zoom-in simulation in Fig.~\ref{figure:subfind_hbt_mass_function}. {\SUBFIND} and {\tt Subfind-HBT} agree well with each other, as found in the Gadget-4 release paper \citep{Springel.2021}. However, they are both suppressed relative to what {\HBT} finds. There is a weak mass dependence, with a 90\% agreement at the lowest masses and 80\% at the highest masses. We note that these differences should not be directly compared to those found in Fig.~\ref{figure:bound_mass_function}. In the version discussed in the main text, a fixed $M_{\mathrm{bound}}$ is made up by contributions from the satellite systems of centrals with different $M_{\mathrm{200c}}$ values. Here, we only consider a single FoF group of $M_{\mathrm{200c}} \approx 10^{12}\,\Msun$. Since {\SUBFIND} struggles with major mergers (Fig.~\ref{figure:radial_distribution_satellite_mass_ratio_bins}), the current dynamical state of haloes is clearly important in determining how well it recovers the satellite population. As zoom-in simulations generally target isolated haloes at $z = 0$, their selection may not be representative of the merger rates of haloes with similar masses \citep[e.g.][]{Genel.2009}.

Finally, other differences exist between {\HBT} and {\tt Subfind-HBT}. There is no hierarchy of subhaloes within FoF groups in {\tt Subfind-HBT}. This helps improve parallelisation, as unbinding does not need to be done in a depth first approach. However, this means that mass stripped from satellites always ends up in the central subhalo, rather than possibly being accreted by its immediate parent. No phase-space checks are done between subhaloes, no hostless subhaloes exist and {\tt Subfind-HBT} does not subsample particles during the unbinding of subhaloes and iterates until the number of bound particles has converged (similar to {\SUBFIND}). In summary, these similarly named algorithms should be treated as different approaches to history-based halo finding. 

\section{Differences in virial masses}\label{Appendix:VirialMassDifferences}

In \S\ref{Section:M200_mass_functions} we showed the $M_{\mathrm{200c}}$ mass functions measured using different subhalo finders in the DMO and hydrodynamical versions of the simulations we analyse. Compared to {\HBTHERONS}, {\VELOCIRAPTOR} and {\ROCKSTAR} show systematic differences beyond the 1\%-level across all the resolutions we consider in this study. We attributed these differences to miscentering in {\VELOCIRAPTOR} and the different central definition in {\ROCKSTAR}.

To show that this is the case, we match central subhaloes across subhalo finders. This approach allows us to identify whether any systematic differences exist in the properties of centrals, identified by any two subhalo finders. We perform the bijective matching  as follows. We first match central subhaloes in catalogue A by identifying which subhalo in catalogue $B$ contains the largest share of its bound particles. We then do the matching in reverse, i.e. from catalogue $B$ to catalogue A, and only keep consistent matches. For the purposes of this discussion, catalogue A refers to the {\HBTHERONS} catalgoues

\begin{figure*}
    \centering
    \includegraphics{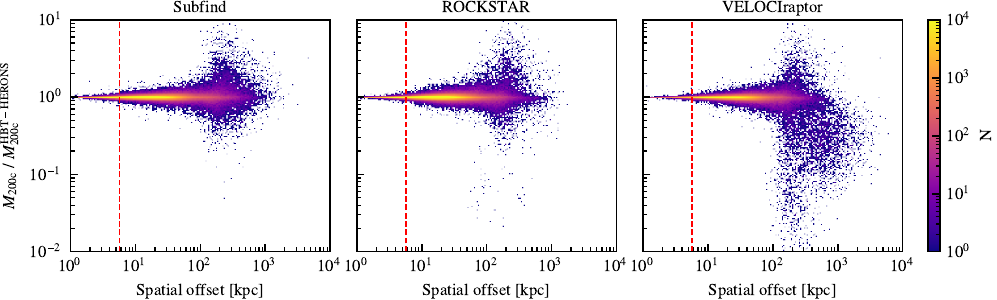}
    \caption{The $M_{\mathrm{200c}}$ values of haloes in the DMO {\midres} simulation, measured in different subhalo finders and expressed relative to their matched counterparts found by {\HBTHERONS}. Only matched subhaloes identified as centrals by {\HBTHERONS} and with more than 100 particles enclosed within $R_{\mathrm{200c}}$ are shown. We limit the radial range to only show spatial offsets that are larger than those attributable to rounding errors. The vertical red line indicates the  value of the gravitational softening length of the simulation at $z = 0$.}
    \label{figure:mass_ratio_M200_matched}
\end{figure*}

In Fig.~\ref{figure:mass_ratio_M200_matched} we show the $M_{\mathrm{200c}}$ ratio of matched central subhaloes in the DMO {\midres} simulation, as well as the spatial offset between their centres. The majority of centrals have similar $M_{\mathrm{200c}}$ values, even when the spatial offset is large. By visually inspecting several examples, we conclude that these occur whenever two similar mass subhaloes are interchangeably identified as centrals. Leaving aside these cases, there are outliers when matching across subhalo finders. 

The most prominent outlier population occurs when matching to {\VELOCIRAPTOR}, whose $M_{\mathrm{200c}}$ are underestimated by up to one order of magnitude. By imaging the matches which disagreed the most in their $M_{\mathrm{200c}}$ values, we see that the underestimate is due to the centre being placed on a satellite, rather than on a central. As the centre of the aperture is not adequately positioned on the true peak of the overdensity, the $R_{\mathrm{200c}}$ is smaller and encloses less mass.

Some differences still remain when comparing to {\SUBFIND} and {\ROCKSTAR}. In particular, there  are central subhaloes with large spatial offsets ($\geq 200\,\mathrm{kpc}$) that have more or less mass within $R_{\mathrm{200c}}$. The distribution is very slightly skewed towards higher $M_{\mathrm{200c}}$ values in {\SUBFIND}, but appears more symmetric in {\ROCKSTAR}. We imaged several examples and saw that the differences are caused by different central identifications within the same FoF group. Despite the disagreement by factors of $4$ in some individual cases, {\SUBFIND} and {\HBT} agree within 1\% in their $M_{\mathrm{200c}}$ mass function. As {\ROCKSTAR} has a similar distribution as {\SUBFIND}, the $M_{\mathrm{200c}}$ mass function would agree to the same level, were it not for the different number of centrals resulting from its different definition.

\section{Linking only DM or all particles}\label{Appendix:LinkingParticleTypes}

Running {\ROCKSTAR} on hydrodynamical simulations is typically done in one of two ways. The linking can be done using all particle types or just the dark matter ones. As explained in \S\ref{section:halo_finders}, we link all particle types for the main study. To provide an idea of how this choice affects the results, we repeat the comparison of mass functions as shown \S\ref{section:halo_mass_functions}.

\begin{figure}
    \centering\includegraphics{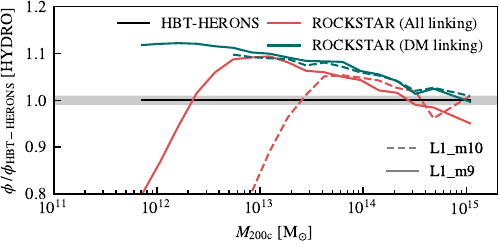}
    \caption{$M_{\mathrm{200c}}$ mass functions relative to those found in {\HBTHERONS}, for the hydrodynamical simulations used in this work. Two {\ROCKSTAR} lines are present, which show how results change depending on how the 6D FoF algorithm is run. The red line is the method we used to run {\ROCKSTAR} for the results shown in the main text, which does the 6D FoF finding using all particle types except for black holes. The teal line is an alternative approach that just uses the DM particles for 6D FoF finding and ignores the other particle types.}
    \label{figure:m200_rockstar_dm_linking_vs_not}
\end{figure}

We show the $M_{\mathrm{200c}}$ mass function found in both choices of how to run {\ROCKSTAR} in Fig.~\ref{figure:m200_rockstar_dm_linking_vs_not}, expressed relative to the value found by {\HBTHERONS}. The choice of linking using all particles \textit{vs} only DM ones can alter the measured mass function between the two {\ROCKSTAR} variations by 5\% at high and intermediate masses. The overall lower masses when running on all the particles is due to the issues that were identified in the main text: {\ROCKSTAR} is unable to find subhaloes, and the centres are offset from the true density peaks. Thus, similar to what happens in the {\VELOCIRAPTOR} miscentering, this lowers the measured $M_{\mathrm{200c}}$. The differences become increasingly large at the lowest masses. 

\section{Different mass definitions}\label{Appendix:MassDefinitions}

The main results of this work depend on being able to compute the properties of (sub)haloes as consistently as possible across subhalo finders. To do this for the $M_{\mathrm{200c}}$ masses of haloes, we load the centres of central subhaloes as given by each subhalo finder into {\SOAP}, so that we use the same routine to find the corresponding spherical overdensity mass. However, each subhalo finder outputs their own internally-computed $M_{200\mathrm{c}}$, which may not be the same as the {\SOAP}-computed value we use in this work. As past studies have likely used the values provided by the subhalo finders themselves, we explore by how much the value of $M_{200\mathrm{c}}$ changes once we compute it within {\SOAP}. 

\begin{figure}
    \centering\includegraphics{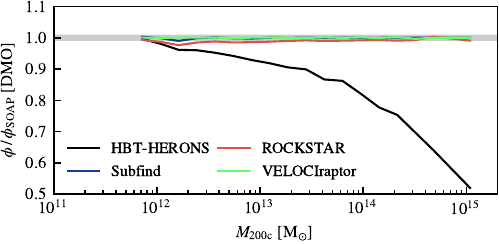}
    \caption{$M_{\mathrm{200c}}$ mass functions based on the values computed internally by each subhalo finder, relative to those computed by {\SOAP}. The results are shown for the DMO {\midres} simulation.}
    \label{figure:m200_soap_vs_internal}
\end{figure}

We show in Fig.~\ref{figure:m200_soap_vs_internal} the $M_{200\mathrm{c}}$ mass functions based on the internally-computed values for each subhalo finder, and we normalise each by the corresponding mass function based on the $M_{200\mathrm{c}}$ computed within {\SOAP}. {\SUBFIND} and {\VELOCIRAPTOR} are well within the 1\% difference, which reflects the fact that they use a very similar approach to {\SOAP} when computing $M_{200\mathrm{c}}$. All three include all mass, bound and unbound, within a given aperture. 

The mass function derived from the {\ROCKSTAR}-computed $M_{200\mathrm{c}}$ is also in good agreement with the one based on using {\SOAP}. The agreement is not expected a priori, as {\ROCKSTAR} only includes particles that are bound to the central, and so one might expect lower values relative to those measured in {\SOAP}. However, a key choice made by {\ROCKSTAR} reduces the differences between total and bound-only mass calculations. The bound mass is assigned inclusively, so the mass that is bound to satellite subhaloes may also be bound to central subhaloes. In practice this means that a large fraction of the particles within $R_{200\mathrm{c}}$ end up formally bound to the central, at least according to the unbinding criterion used by {\ROCKSTAR}. Hence, the differences in $M_{200\mathrm{c}}$ are minimal. 

The $M_{200\mathrm{c}}$ mass function that changes the most is that based on the values provided by {\HBTHERONS}. Similar to {\ROCKSTAR}, it only includes bound particles in its internal calculation. However, mass is assigned exclusively. This entails that mass bound to resolved satellites is not included within a given aperture. Consequently, the radius at which the mean enclosed density equals $200\rho_{\mathrm{crit}}$ is smaller, and the mass enclosed within that physical radius is also less than when including the total mass. The extent of the effect depends on how many satellites are resolved, which is why the differences between {\HBTHERONS} and {\SOAP} become larger as $M_{200\mathrm{c}}$ increases (i.e. as more particles sample the virial region of the halo). 

\begin{figure}
    \centering
    \includegraphics{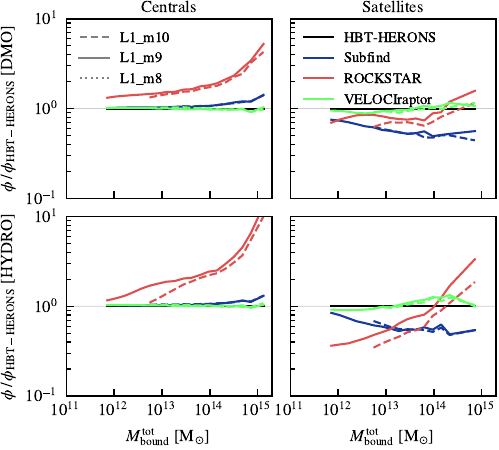}
    \caption{Same as in Fig.~\ref{figure:bound_mass_function}, but showing the results as provided by the default version of {\ROCKSTAR}. As the masses computed by {\ROCKSTAR} are inclusive, there is a systematic upwards shift in the number density at a fixed $M_{\mathrm{bound}}$, as the other subhalo finders use exclusive mass definitions.}
    \label{figure:bound_mass_function_inclusive}
\end{figure}

\begin{figure}
    \centering
    \includegraphics{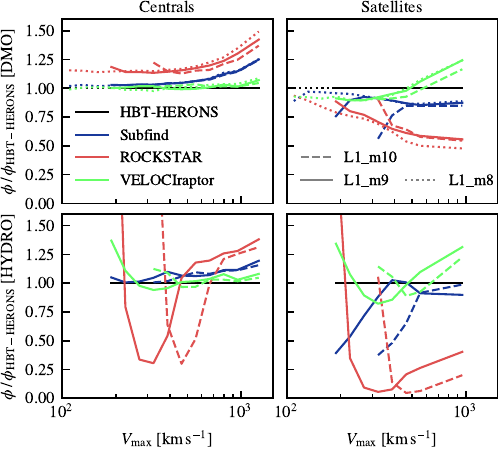}
    \caption{Same as in Fig.~\ref{figure:vmax_mass_function_exclusive}, but showing the results as provided by the default version of {\ROCKSTAR}. As the masses computed by {\ROCKSTAR} are inclusive, there is a systematic upwards shift in the number density of central subhaloes at a fixed $V_{\mathrm{max}}$, as the other subhalo finders use exclusive mass definitions.}
    \label{figure:vmax_function_inclusive}
\end{figure}

The second change we made to ensure a more consistent comparison is to make {\ROCKSTAR} assign masses to subhaloes in an exclusive manner. We do this because {\HBTHERONS}, {\SUBFIND} and {\VELOCIRAPTOR} use an exclusive mass assignment, so {\ROCKSTAR}-derived values for $M_{\mathrm{bound}}$ and $V_{\mathrm{max}}$ are systematically higher just because of this operational difference. The effect of this difference is clearly seen in Fig.~\ref{figure:bound_mass_function_inclusive}, where {\ROCKSTAR} has factors of 5 to 10 more bound mass for the highest mass central subhaloes compared to the exclusive mass version of Fig.~\ref{figure:bound_mass_function}. This change in definition also propagates to changes in $V_{\mathrm{max}}$, as seen in Fig.~\ref{figure:vmax_function_inclusive}.

To make {\ROCKSTAR} work with exclusive masses, we modified its code so that particles can only be bound to a single subhalo. We make the internal change to the algorithm because including particles bound to other subhaloes during the unbinding step changes the overall gravitational potential, and hence the binding energies of particles. A simpler fix based on subtracting the mass bound to satellites from a given parent subhalo in post-processing would not account for the removal of additional gravitation potential contributed by particles already bound to satellites.

Leaving aside the practicalities of implementing exclusive or inclusive mass assignment, one may argue which of the two mass definitions is `best'. The dynamics in gravitational systems depend on the gravitational potential, regardless of whether the mass that sources it is formally associated to one object or another (or if it is not bound to any object, a possibility that none of the four subhalo finders we use here contemplate). One may therefore be inclined to use inclusive masses for $V_{\mathrm{max}}$. However, inclusive masses can lead to double counting mass, if for example the stars of a galaxy are bound to both a satellite subhalo and its parent subhalo. This highlights that the choice of one definition should reflect the question one aims to answer. In any case, when making predictions to compare against the real Universe, this discussion is moot. Any, if not all, metrics we usually use to assign masses to (sub)haloes in simulations are theory-motivated and are not realistically measurable in the real Universe.

% Don't change these lines
\bsp	% typesetting comment
\label{lastpage}
\end{document}